\documentclass[twocolumn]{aastex631}
\usepackage{bm}
\usepackage{amsmath}
\usepackage{booktabs}
\usepackage{threeparttable}

\newcommand{\Ha}{H$\alpha$\,$\lambda$6563}
\newcommand{\Hb}{H$\beta$\,$\lambda$4861}
\newcommand{\OIIII}{{\rm [O\texorpdfstring{\,}{}{\sc iii}]\texorpdfstring{\,}{}\texorpdfstring{$\lambda$}{}5007}}
\newcommand{\NII}{[N\,{\sc ii}]\,$\lambda$6583}

\begin{document}

\title{Radial Profiles of Binary Fraction in Elliptical Galaxies}

\correspondingauthor{Xiejin Li}
\email{lixiejin@ynao.ac.cn, zhangfh@ynao.ac.cn}

\author[0009-0009-4030-5208]{Xiejin Li}
\affiliation{Yunnan Observatories, Chinese Academy of Sciences, Kunming 650216, China}
\affiliation{University of Chinese Academy of Sciences, Beijing 100049, China}

\author{Fenghui Zhang}
\affiliation{Yunnan Observatories, Chinese Academy of Sciences, Kunming 650216, China}
\affiliation{International Centre of Supernovae, Yunnan Key Laboratory, Kunming 650216, China}
\affiliation{Key Laboratory for the Structure and Evolution of Celestial Objects, Chinese Academy of Sciences, Kunming 650216, China}

\author[0000-0002-9128-818X]{Yinghe Zhao}
\affiliation{Yunnan Observatories, Chinese Academy of Sciences, Kunming 650216, China}
\affiliation{State Key Laboratory of Radio Astronomy and Technology, National Astronomical Observatories, Chinese Academy of Sciences, Beijing 100101, China}

\author{Cheng Li}
\affiliation{Department of Astronomy, Tsinghua University, Beijing 100084, China}

\author{Zhanwen Han}
\affiliation{Yunnan Observatories, Chinese Academy of Sciences, Kunming 650216, China}
\affiliation{International Centre of Supernovae, Yunnan Key Laboratory, Kunming 650216, China}
\affiliation{Key Laboratory for the Structure and Evolution of Celestial Objects, Chinese Academy of Sciences, Kunming 650216, China}

\author{Yunkun Han}
\affiliation{Yunnan Observatories, Chinese Academy of Sciences, Kunming 650216, China}
\affiliation{Key Laboratory for the Structure and Evolution of Celestial Objects, Chinese Academy of Sciences, Kunming 650216, China}
\affiliation{Center for Astronomical Mega-Science, Chinese Academy of Sciences, 20A Datun Road, Chaoyang District, Beijing 100012, China}

\author{Xiaoyu Kang}
\affiliation{Yunnan Observatories, Chinese Academy of Sciences, Kunming 650216, China}
\affiliation{Key Laboratory for the Structure and Evolution of Celestial Objects, Chinese Academy of Sciences, Kunming 650216, China}
\affiliation{Center for Astronomical Mega-Science, Chinese Academy of Sciences, 20A Datun Road, Chaoyang District, Beijing 100012, China}



\begin{abstract}

The radial profile of binary fraction may vary with environment and is of significant importance for studying the formation mechanisms of binary stars and their dynamical evolution within globular clusters (GCs) and galaxies. However, existing studies remain limited to the Milky Way and its neighboring galaxies. Leveraging the method proposed by Zhang et al.\ for estimating the variation of binary fraction from integrated spectral features, we analyze a sample of 513 elliptical galaxies drawn from the Mapping Nearby Galaxies at Apache Point Observatory (MaNGA) survey to measure their radial binary fraction profiles.
Our results show that after accounting for the effect induced by radial variations in the stellar population (SP), the median SP-subtracted binary fraction, $r_{\rm b,sub}^{\rm med}$, becomes approximately flat. For nearly all elliptical galaxies in our sample, the variation in binary fraction relative to the galaxy center at $1R_e$ is less than 5\%. No clear correlation is found between the binary fraction gradient and the gradients of SP properties. Moreover, we also compare differences between ultraviolet (UV) upturn and non-UV upturn galaxies. The overall binary fraction profiles and SP properties of the non-UV upturn galaxies in our sample are comparable to those of the UV upturn galaxies. This similarity may arise from the presence of residual star formation (RSF) in the non-UV upturn systems.

\end{abstract}



\section{Introduction} \label{sec:Introduction}

Most stars reside in binary systems \citep{Duch2013ARA&A..51..269D}. The binary fraction and its spatial distribution not only trace the initial conditions of binary formation but also evolve in response to varying external environments.

The binary fraction differs between dense \citep{Fregeau2009ApJ...707.1533F} and sparse \citep{Li2018ApJ...859...36L} stellar environments and changes with the dynamical evolution of globular clusters (GCs).
Both observations \citep{Ji2015ApJ...807...32J, Giesers2019A&A...632A...3G, Albrow2024MNRAS.528.6211A} and numerical simulations \citep{Fregeau2009ApJ...707.1533F, Hurley2007ApJ...665..707H, Geller2013AJ....145....8G} consistently show that the binary fraction generally decreases with increasing radius in the Galactic GCs as a result of mass segregation where the massive stars tend to sink toward the center of clusters due to the dynamical friction. 
\citet{Milone2012A&A...540A..16M} found in a survey of 59 Galactic GCs that the fraction of main-sequence binaries with mass ratios $q > 0.5$ typically declines by a few percent to tens of percent from the cluster center to regions beyond the half-mass radius. Flat \citep{Li2013MNRAS.436.1497L} and even radially increasing \citep{deGrijs2013ApJ...765....4D} binary fraction profiles have also been reported in star clusters of the Large Magellanic Cloud, and this discrepancy likely arises because they experience different stages along a similar dynamical evolutionary sequence \citep{Geller2015ApJ...805...11G}.


For the Milky Way, the binary fraction also depends on the specific environment and dynamical process. Binary fraction in the field is approximately 50\% \citep{Raghavan2010ApJS..190....1R}, usually higher than that in star clusters \citep{Milone2012A&A...540A..16M}, indicating field binaries may experience distinct dynamical evolution processing \citep{Liu2019MNRAS.490..550L}. 
The binary fraction in the Galactic halo is comparable to that of the thick disk, and both are higher than that in the thin disk \citep{Gao2014ApJ...788L..37G, Hwang2022MNRAS.513..754H}, implying that binaries in halo and thick disk may not migrate radially like the thin disk stars \citep{Sellwood2002MNRAS.336..785S, Liu2012MNRAS.425.2144L}.
The Galactic tidal field could also create/destroy wide binaries \citep{Jiang2010MNRAS.401..977J, Pe2021MNRAS.501.3670P, Grishin2022MNRAS.512.4993G}, thereby affecting binay fraction. Recently, \citet{Gautam2024ApJ...964..164G} discovered that, within a region of about 0.4\,pc from the Galactic Center, the binary fraction of young stars increases with radius---opposite to the trend seen in GCs---suggesting that frequent strong gravitational encounters in the extreme, dense environment near the supermassive black hole lead to significant depletion of binaries in the central region through evaporation or coalescence. 

The radial profile of binary fraction, therefore, offers valuable insights into the formation mechanism of binaries, dynamical evolution of clusters, radial migration of stars within galaxies, and is essential for understanding the mass assembly or merger history, as well as the star formation history of galaxies. However, studies on radial distribution of binary fractions in the galaxies, especially extragalactic systems, remain extremely limited due to the constraints of directed measurements that require resolvable photometric or spectroscopic observations of individual stars. 

On the other hand, a rising flux toward shorter wavelengths in the far-ultraviolet (UV) band is observed in some elliptical galaxies \citep{OConnell1999ARA&A..37..603O}, which are generally considered old, red systems that are not expected to produce such strong UV emission. This UV upturn phenomenon (also called UV excess) is commonly accepted to be caused by a hot horizontal branch (HB) star population, whose origin remains controversial with several formation channels proposed: (i) hot subdwarf stars (sdB) produced through binary-star evolution \citep{Han2003MNRAS.341..669H, Han2007MNRAS.380.1098H}; (ii) populations formed via different single star evolution, could be metal-poor, metal-rich, or helium-enhanced HB stars \citep{Lee2005ApJ...621L..57L, Yi2008ASPC..392....3Y, Chung2013ApJS..204....3C, Chung2017ApJ...842...91C}. In addition, residual star formation (RSF) could also result in this phenomenon \citep{Yi2005ApJ...619L.111Y, Schawinski2007ApJS..173..512S, Kaviraj2007ApJS..173..619K, Pipino2009MNRAS.395..462P, Salim2010ApJ...714L.290S}. 
In a recent study, \citet{Jiang2025MNRAS.540.3770J} applied a semi-analytic model of galaxy formation to investigate all related physical processes, including dust attenuation, stellar population (SP) properties, and binary evolution. They found that UV upturn galaxies can be divided into two categories: old, metal-rich, quenched elliptical galaxies for which the UV upturn is induced by hot sdB in binary evolution (explanation (i) above); and dusty star-forming galaxies, which may be identified as UV upturn galaxies due to dust attenuation.

Systematic investigations of UV upturn galaxies---such as measurements of the radial distribution of the binary fraction or SP properties---hold great promise for revealing the physical nature of their stellar constituents and providing critical clues to resolve the origin of the UV upturn.

Therefore, in this paper, we leverage the empirical relationship between the binary fraction and spectral absorption feature indices (SAFIs) established by \citet{Zhang2024MNRAS.531.3468Z} to construct, for the first time, radial binary fraction profiles for a sample of 513 elliptical galaxies. 
We compare binary fraction profiles between the UV upturn and the non-UV upturn subsample, and examine differences in their SP properties. Furthermore, to gain deeper insight into the physical nature of UV upturn galaxies, we also investigate how different evolutionary population synthesis (EPS) models affect the inferred SP properties.

The paper is organized as follows: Section~\ref{sec:Sample, Data and Methods} describes the selection of the elliptical galaxy sample and the methodology for measuring the radial binary fraction profiles. The results of the radial binary fraction profiles are presented in Section~\ref{sec:Binary Fraction Profiles}. Comparison of the properties between UV and non-UV upturn galaxies is given in Section~\ref{sec:UV Upturn vs non-UV Upturn Galaxies}. Connections between the binary fraction gradients and the gradients in SP properties, and the impact of the EPS models and signal-to-noise ratio (S/N) on derived binary fraction profiles are discussed in Section~\ref{sec:Discussion}. Finally, we summarize our main conclusions in Section~\ref{sec:Summary}.

\section{Sample, Data and Methods} \label{sec:Sample, Data and Methods}

\begin{deluxetable}{lll}
\tablenum{1}
\tablewidth{0pt}
\tablecaption{Criteria for Sample Selection.\label{tab1}}
\tablehead{\colhead{Criterion} & \colhead{Condition} & \colhead{Number}}
\startdata
Parent & --- & 10125 \\
BPT & \citet{Kewley2001ApJ...556..121K} & 6622 \\
Morphology & {\tt Unsure=0} \& & 4308 \\
 & {\tt Tidal=0} \& & \\
 & {\tt cas\_flag=1} & \\
Elliptical & {\tt T-Type} = -5 & 521 \\
\hline
\hline
Final Sample & Condition & Number \\
\hline
All $^a$ & --- & 513 \\
UV & ${\rm NUV}-r>5.4$ \& & 57 \\
   & ${\rm FUV-NUV}<0.9$ \& & \\
   & ${\rm FUV}-r<6.6$ & \\
Non-UV & --- & 456 \\
\enddata
\tablecomments{All$^a$: Eight galaxies without available $R_e$, position angle measurement, or the number of elliptical annuli $\leqslant3$ are excluded.}
\end{deluxetable}

\begin{deluxetable*}{lcccccccccc}
\tablenum{2}
\tablewidth{0pt}
\tablecaption{Calibration Coefficients $a_k$ and Weight $w_k$ of Binary Fraction-Sensitive SAFIs.\label{tab2}}
\tablehead{\colhead{SAFIs} & \colhead{Ca4455} & \colhead{C$_2$4668} & \colhead{TiO$_1$} & \colhead{O$_{\rm III}$-1} & \colhead{O$_{\rm III}$-2} & \colhead{H$\beta$} & \colhead{H$\rm \delta_A$} & \colhead{H$\rm \gamma_A$} & \colhead{H$\rm \delta_F$} & \colhead{H$\rm  \gamma_F$} \\
\colhead{} & \colhead{(1)} & \colhead{(2)} & \colhead{(3)} & \colhead{(4)} & \colhead{(5)} & \colhead{(6)} & \colhead{(7)} & \colhead{(8)} & \colhead{(9)} & \colhead{(10)} }
\startdata
$a_k$   &   $-1.30$  & $-76.87$  & $-2.08$ & $-40.92$ &  $-35.14$ & $10.88$ & $125.64$ & $108.58$ & $101.35$ & $94.57$ \\
$w_k$ & 0.040 & 0.096  & 0.155  & 0.203  & 0.166  & 0.041 & 0.07 & 0.04  & 0.1052  & 0.079\\
\enddata
\end{deluxetable*}

\begin{deluxetable*}{llll}
\tablenum{3}
\tablewidth{0pt}
\tablecaption{Specific Parameter Configurations Used in This Work for Different EPS Models.\label{tab3}}
\tablehead{\colhead{EPS Model} & \colhead{GALAXEV} & \colhead{Yunnan II} & \colhead{BPASS}}
\startdata
Age $t$ & 1 Myr -- 15.85 Gyr  & 1 Myr -- 15 Gyr & 1 Myr -- 15.85 Gyr \\
Metallicity $[Z/H]$ &  -1.74 -- 0.47 &  -1.3 -- 0.2 & -1.3 -- 0.3  \\
IMF &\citet{Chabrier2003PASP..115..763C} & \citet{Miller1979ApJS...41..513M} & \citet{Byrne2022MNRAS.512.5329B} \\
Isochrone$^a$ & Padova & BSE & STARS \\
Spectra library$^b$  & MILES & BLUERED \& BaSeL-2.0 & C3K \\
\enddata
\tablecomments{Isochrone$^a$: Padova \citep{Bertelli1994AAS..106..275B}, BSE \citep{Hurley2000MNRAS.315..543H, Hurley2002MNRAS.329..897H}, STARS \citep{Eggleton1971MNRAS.151..351E,Eldridge2008MNRAS.384.1109E}. Spectra library$^b$: MILES \citep{Sanchez2006MNRAS.371..703S,Falcon2011AA...532A..95F}, BLUERED \citep{Bertone2008AA...485..823B} \& BaSeL-2.0 \citep{Lejeune1997AAS..125..229L, Lejeune1998AAS..130...65L}, C3K \citep{Choi2016ApJ...823..102C, Conroy2018ApJ...854..139C}.}
\end{deluxetable*}

\begin{figure}[htb!]
\centering
\includegraphics[width=0.35\textwidth]{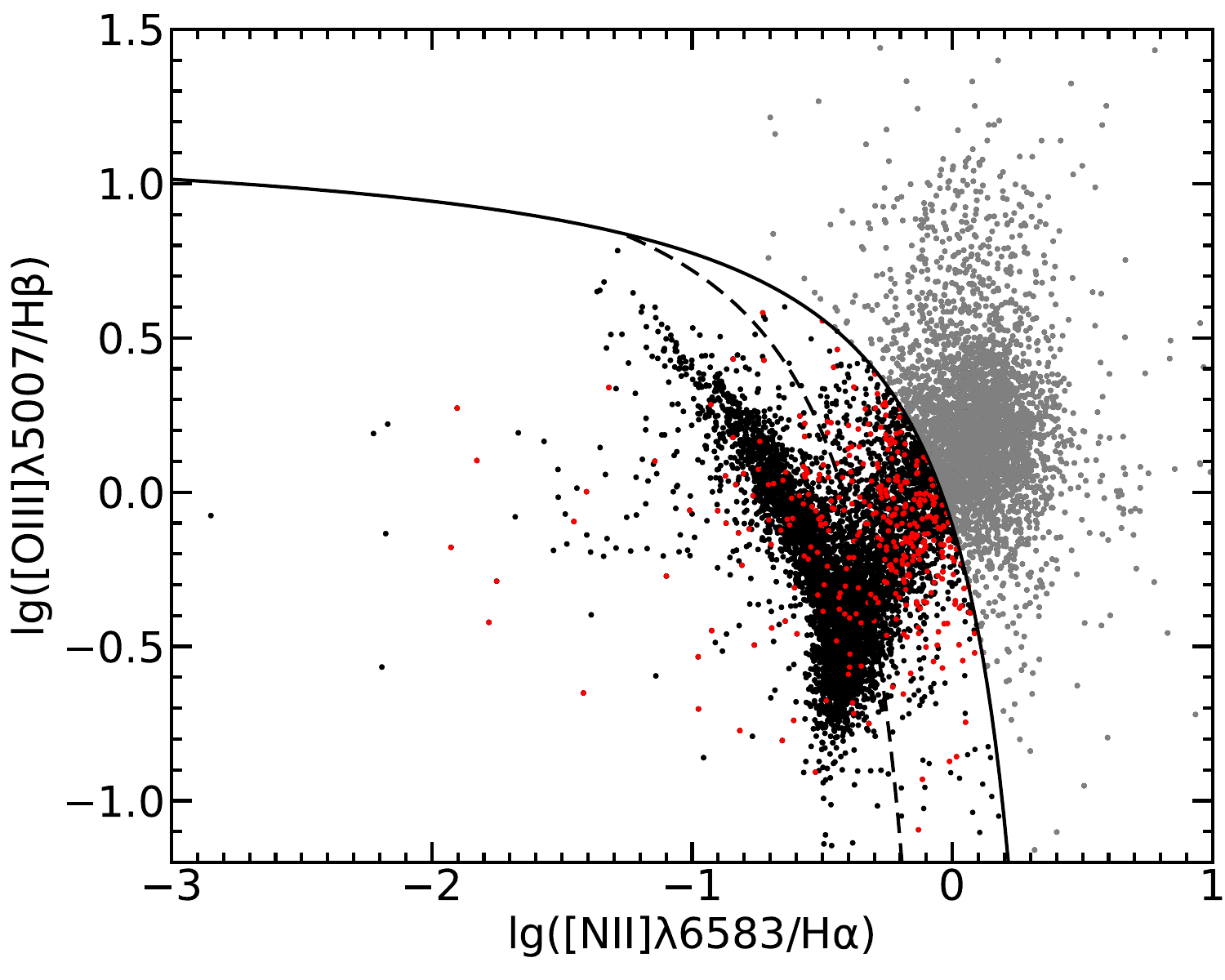}
\caption{The BPT diagram for 9,500 galaxies that can be drawn with valid emission lines measurements from the parent sample. The gray dots correspond to the AGNs that lie above the theoretical maximum starburst line (solid curve) defined by \citet{Kewley2001ApJ...556..121K} and are excluded from our sample. The red dots indicate 363 elliptical galaxies selected from the remaining 5,997 galaxies (black dots). The dashed curve plots the demarcation between star formation and composite region defined by \citet{Kauffmann2003MNRAS.346.1055K}.
\label{fig1}}
\end{figure}

\begin{figure}[htb!]
\centering
\includegraphics[width=0.35\textwidth]{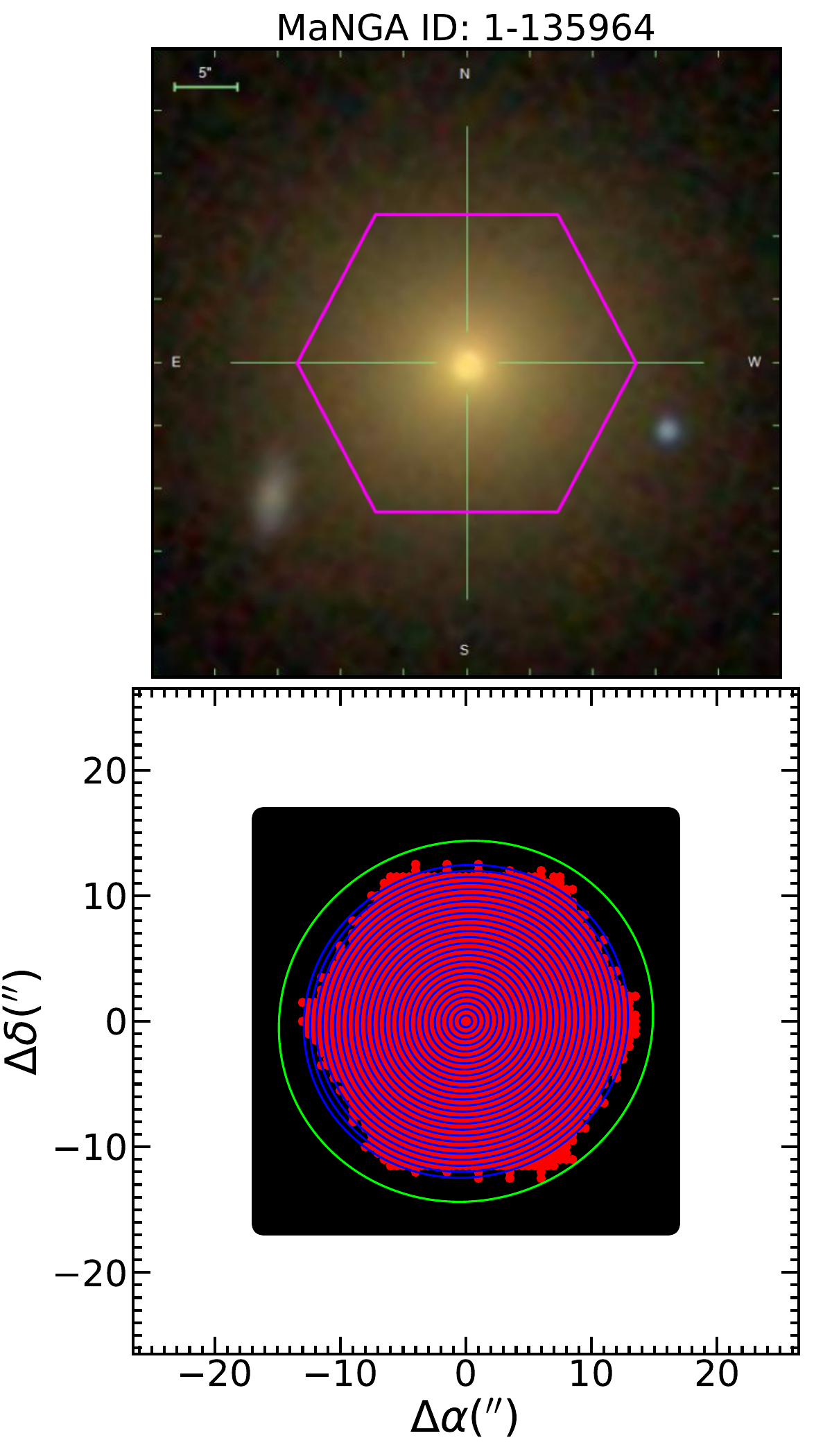}
\caption{Image mosaic (top panel) and annular binned spaxels (red dots, bottom panel) of galaxy with MaNGA ID: 1-135964. The blue elliptical annuli represent that the S/N is greater than 10, while the green elliptical annulus indicates excluded external regions where the S/N of the stacked spectra is below 10.
\label{fig2}}
\end{figure}

In Section~\ref{subsec:Data and Sample Selection}, we present the observational data used in this study and describe the selection of our sample of elliptical galaxies, including the identification of UV and non-UV upturn galaxies. The methodology for constructing radial spectra of elliptical galaxies is detailed in Section~\ref{subsec:Radial Spectra Construction}. Section~\ref{subsec:Measurement of Binary Fraction Profiles} provides a comprehensive description of the procedure for measuring the binary fraction profiles, along with an overview of the spectral fitting code and the EPS models employed. In Section~\ref{subsec:SP-Induced and Subtracted Binary Fraction Variations and Additional Clarifications}, we introduce the SP-induced and -subtracted relative binary fraction variations, and provide additional clarifications regarding the spectra used in Sections 2.2 and 2.3.

\subsection{Data and Sample Selection} \label{subsec:Data and Sample Selection}
To derive the radial profiles of the binary fraction in elliptical galaxies, we first construct their radial spectra. Therefore, this study adopts as its base sample galaxies from the Mapping Nearby Galaxies at Apache Point Observatory (MaNGA) survey \citep{Bundy2015ApJ...798....7B, Yan2016AJ....152..197Y}, which provides spatially resolved integral field unit (IFU) spectroscopy. 
The sample is drawn from version v1\_0\_1 of the NASA-Sloan Atlas (NSA) catalog\footnote{\url{https://www.sdss.org/dr16/manga/manga-target-selection/nsa}} which comprises approximately 10,000 nearby galaxies with redshifts in the range $0.01$--$0.15$, and an approximately flat stellar mass distribution spanning $\rm 5\times10^8$--$3\times10^{11}\,M_{\odot}\,h^{-2}$. The MaNGA spectra cover a wavelength range of 3600--10,300\,\AA\ at a typical spectral resolution of $R \sim 2000$. Observations are carried out using five IFU sizes with diameters ranging from $12\arcsec$ to $32\arcsec$ (19-127 fibers, \citealt{Drory2015AJ....149...77D}). These IFUs are assigned to ensure coverage out to $2.5\,R_e$ (the $r$-band effective radius) for the Primary and Primary+ samples, 
which constitute $\sim$67\% of the targets, and out to $1.5\,R_e$ for the Secondary sample, making up the remaining $\sim$33\% \citep{Law2015AJ....150...19L, Wake2017AJ....154...86W}. 

Secondly, to construct a reliable elliptical galaxy sample, we implement a multi-step selection procedure, summarized in Table~\ref{tab1}:

---We selected galaxies with morphological classifications by utilizing the MaNGA value-added catalog of visual morphological classifications provided by \citet{Vazquez-Mata2022MNRAS.512.2222V}. This catalog classifies all galaxies in MaNGA DR17 according to the Hubble scheme and provides type values ({\tt T-Type}) using digitally enhanced image mosaics that combine the Sloan Digital Sky Survey (SDSS) and the Dark Energy Spectroscopy Instrument (DESI) Legacy Survey $r$-band imaging data to more accurately identify internal structures (e.g., bars, spiral arms) and external features (e.g., tidal debris). We cross-matched this catalog with the MaNGA Data Analysis Pipeline (DAP) summary Table~{\tt DAPall}\footnote{{\tt DAPall} provides spectral measurements, while the NSA catalog supplies physical parameters such as redshift; the two are linked via the MaNGA ID.}, obtaining 10,125 matched galaxies as our initial parent sample.

---We removed galaxies potentially contaminated by active galactic nuclei (AGNs). Since AGN activity can produce strong emission lines that may contaminate the SAFIs used to derive binary fractions, we utilized central ($2\arcsec .5$ aperture) emission line fluxes from {\tt DAPall} ({\tt EMLINE\_SFLUX\_CEN}) and applied the BPT diagnostic diagram \citep{Baldwin1981PASP...93....5B, Veilleux1987ApJS...63..295V} with the criterion defined by \citet{Kewley2001ApJ...556..121K}. (see Fig.~\ref{fig1}; the boundary from \citealt{Kauffmann2003MNRAS.346.1055K} is also shown). We note that only 9,500 galaxies with all four (\Ha, \Hb, \OIIII, \NII) emission lines available are used in the BPT classification, while the remaining 625 galaxies without available emission line measurements are also retained for the following selection. After this step, the sample was reduced to 6,622 galaxies (including 625 galaxies).

---We further excluded galaxies with uncertain morphological classifications (e.g., diffuse, faint, or compact galaxies). We chose sources flagged as determined morphological ({\tt Unsure=0}), without tidal debris ({\tt Tidal=0}) as well as reliable concentration-asymmetry-clumpiness parameter estimates ({\tt cas\_flag=1}) in the \citet{Vazquez-Mata2022MNRAS.512.2222V} catalog. This resulted in a final sample of 4,308 galaxies with unambiguous morphological types.

---Among these, 521 galaxies were classified as ellipticals with galaxy type {\tt T-Type} = $-5$. Of these, 513 galaxies had sufficient S/N and spatial coverage for subsequent spectral stacking analysis (see Section~\ref{subsec:Measurement of Binary Fraction Profiles}), forming our final sample of elliptical galaxies for analysis, and 363 out of 513 galaxies with valid emission line measurements were also plotted on Fig.~\ref{fig1}.

According to the color criteria established by \citet[][see Table~\ref{tab1}]{Yi2011ApJS..195...22Y}, we further identified 57 galaxies as UV upturn systems (hereafter UV subsample) and the remaining 456 are designated as non-UV upturn galaxies (hereafter non-UV subsample) using photometric data from the NSA catalog---which re-analyzed co-added image mosaics from the Galaxy Evolution Explorer and SDSS using the methodology described in \citet{Blanton2011AJ....142...31B}.

\subsection{Radial Spectra Construction}\label{subsec:Radial Spectra Construction}
The MaNGA provides spectra for each spatial pixel (spaxel), and we construct the radial spectra via spectral stacking to derive the radial profile of the binary fraction. Before stacking, we correct both spectra and photometry for Milky Way foreground extinction using the extinction law of \citet{Cardelli1989ApJ...345..245C} and \citet{ODonnell1994ApJ...422..158O} with $R_V = 3.1$, adopting the visual extinction $A_V$ values from \citet{Schlegel1998ApJ...500..525S} as provided in the NASA/IPAC Extragalactic Database, and shift all spectra to the rest frame using redshift from NSA and stellar velocity measurements for each spaxel from \citet{Li2023ChPhB..32c9801L}. Additionally, during the stacking process, we mask spaxels contaminated by foreground stars to avoid contamination in the final stacked spectra.

After preprocessing, we stack the spectra in elliptical annuli, following a method similar to that of \citet{Cai2021RAA....21..204C}. Taking MaNGA ID 1-135964 as an example (see Fig.~\ref{fig2}), we construct elliptical annuli from the center of the galaxy and iteratively expand their radial extent until the stacked spectrum within each annulus achieves an S/N greater than 10. Here, S/N is defined as the ratio of the mean flux to its error in a relatively clean continuum window between 4730 and 4780\,\AA. The ellipticity of the annuli adopted the {\tt ELPETRO\_BA} from the NSA catalog, and the initial annular width is set to $0\arcsec .5$, matching the size of a single spaxel in {\tt LOGCUBE} data cube.

\subsection{Measurement of Binary Fraction Profiles}\label{subsec:Measurement of Binary Fraction Profiles}

\subsubsection{Method}\label{subsubsec:Method}
In this paper, we derive relative binary fraction variation according to \citet{Zhang2023A&A...679A..27Z, Zhang2024MNRAS.531.3468Z}. \citet{Zhang2023A&A...679A..27Z} investigated the correlation between the known binary fractions of Galactic GCs and their SAFIs in different observed regions. They identified ten Lick/IDS indices---O$_{\rm III}$-1, O$_{\rm III}$-2, H$\gamma_{\rm F}$, H$\delta_{\rm F}$, H$\gamma_{\rm A}$, H$\delta_{\rm A}$, H$\beta$, Ca4455, C$_2$4668, and TiO$_1$---as being sensitive to the binary fraction with mass ratio $q > 0.5$. Subsequently, \citet{Zhang2024MNRAS.531.3468Z} established an empirical relation between variations in SAFIs and variations in binary fraction, proposing a method to determine the relative binary fraction variation based on these index changes. Using this method, they derived binary fraction variations between different observed regions for some GCs and validated its reliability by comparing the results with Hubble Space Telescope (HST) observations.

The reasons these indices can trace the binary fraction are that the Balmer and O$_{\rm III}$ indices are temperature-sensitive \citep{Worthey1994ApJS...94..687W}, some binaries would undergo or evolve toward a system comprising high-temperature components or single stars (e.g., blue straggler, and helium main sequence stars or extreme HB stars) due to the Roche lobe overflow process. The Ca4455, C$_2$4668, and TiO$_1$ are Fe-like indices, and are dominated by the C abundance \citep{Trager1998ApJS..116....1T}. The C will be produced during the third dredge-up process of the thermally-pulsing asymptotic giant branch phase of the primary star and might be accreted then by the secondary star, resulting in the C overabundance in the secondary. As the increase/decrease of the binary fraction, the number of high-temperature or C overabundance components will also increase/decrease, leading to the variation in spectral indices. A more detailed explanation can be found in \citet{Zhang2023A&A...679A..27Z}.

The reason why no absolute binary fraction was measured is that for two or multiple clusters with different ages, metallicities, and binary fractions, their spectral indices may be similar. Therefore, we don't have a reference value or benchmark that can directly calibrate the spectral indices to the binary fraction for these different age and metallicity clusters. A possible way to exclude the influence of age and metallicity on spectral indices and find out which index may be sensitive to the binary fraction is to compare the binary fraction changes and the index changes among different regions within the same cluster (considering it as a single star population, but the binary fraction is different at different radii). Accordingly, a linear empirical relationship between the variation of binary fraction and the variation of spectral indices is established, and we only obtain relative changes in the binary fraction based on the spectral indices changes, rather than the absolute binary fraction. And the binary fraction variation presented in this paper refers to the full population of surviving binaries with various ages, metallicities, and masses, but with a mass ratio $q>0.5$, consistent with the analysis in \citet{Zhang2024MNRAS.531.3468Z}.

Following the method of \citet{Zhang2024MNRAS.531.3468Z}, we degrade the resolution of the stacked pure continuum spectrum of each galaxy to 5\,\AA, and then compute the difference in the $k$-th spectral index between the $i$-th elliptical annulus, ${\rm I}_{k,i}$, and the central (first) annulus, ${\rm I}_{k,i=1}$, as
$\Delta {\rm I}_{k,i} = {\rm I}_{k,i} - {\rm I}_{k,i=1}$,
according to the definitions of the SAFIs \citep{Worthey1994ApJS...94..687W, Trager1998ApJS..116....1T}. 
The binary fraction variation in the $i$-th annulus relative to the galaxy center (i.e., the first annulus), denoted by $r_{{\rm b},i}$, can then be obtained as follows:
\begin{equation}
r_{{\rm b},i} = \Sigma_{k=1}^{10} \frac{\Delta {\rm I}_{k,i}}{a_k} w_k\times100\%, \ k=1...10,
\label{Eq:method-1}
\end{equation}
where $a_k$ and $w_k$ are the calibration coefficient and weight, respectively, for the $k$-th SAFI, with their specific values listed in Table~\ref{tab2}. Once the derived binary fraction variations for all elliptical annuli are obtained, the radial profile of the binary fraction can be constructed.

To compare and quantify the trends and variations in the radial profile of the binary fraction, we further compute binary fraction gradient, $\nabla r_{\rm b}$, and the binary fraction variation at 1\,$R_e$ relative to the center, $r_{{\rm b}1R_e}$. These quantities are derived from a linear fit to the final radial profile: the outermost annulus is excluded, and only data within 2\,$R_e$ are used in the fit to avoid large fluctuations caused by declining signal-to-noise ratios in the galaxy outskirts.

\begin{figure*}[htb!]
\plotone{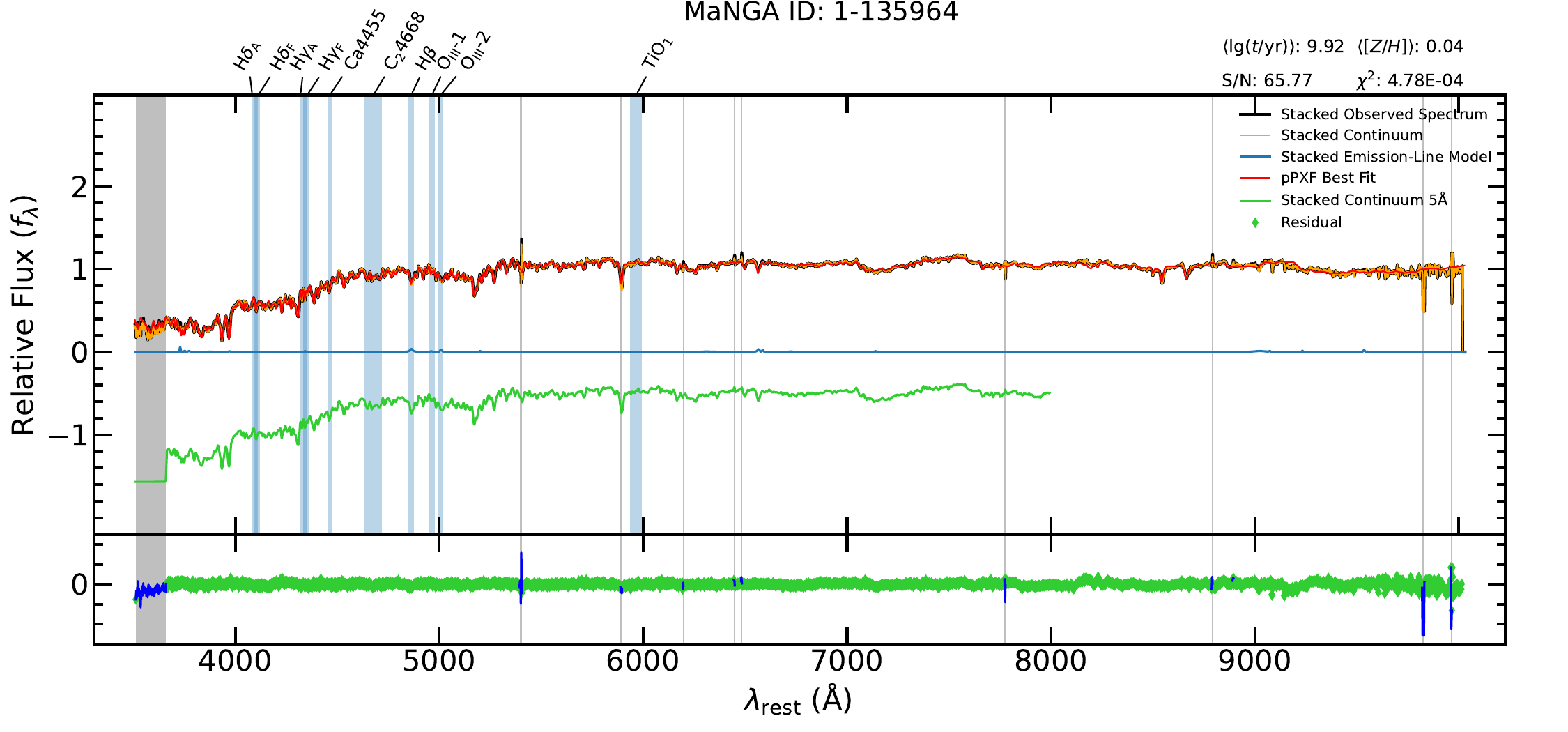}
\caption{An example of stacked spectrum process and fitting of the first (innermost) elliptical annulus for the galaxy with MaNGA ID: 1-135964. The black line indicates the stacked observed spectrum, the orange line indicates the stacked pure continuum whose emission-line has been subtracted by the stacked emission-line model (light blue line), and the red line is the best-fitted stellar template with the pPXF. The stacked pure continuum is degraded to the 5 \AA\ resolution (green line, truncated at 8000 \AA\ due to the low resolution) with an arbitrary offset. The residuals (green diamonds) and the masked pixels (blue lines, also represented by grey vertical bands on the spectrum) are shown in the bottom panel. All spectra are galactic extinction-corrected and shifted to the rest frame. The head of the figure labeled the ten spectral indices (plotted with vertical light blue bands) measured in this work and the fitting results.
\label{fit_illustration}}
\end{figure*}

\subsubsection{Spectral Fitting Code and EPS Models}\label{subsubsec:Spectral Fitting Code and EPS Models}
This study also requires an analysis of SP properties (e.g., comparing differences in age and metallicity between UV and non-UV upturn galaxies and assessing how these differences affect the measurement of the binary fraction variation), which necessitates the use of spectral fitting code and EPS models.

We employ the penalized Pixel-Fitting \citep[pPXF;][]{Cappellari2004PASP..116..138C, Cappellari2017MNRAS.466..798C, Cappellari2023MNRAS.526.3273C}, a robust and efficient SP synthesis full-spectrum fitting code, to derive the light-fractional contribution of each simple stellar population (SSP). Studies have shown that at signal-to-noise ratios S/N\,$>$\,10, the pPXF recovers mean mass-weighted ages or metallicities with uncertainties below $\sim$0.1\,dex \citep{Woo2024MNRAS.530.4260W}. The pPXF can fit photometric fluxes and spectra simultaneously---a capability we exploit in Section~\ref{sec:UV Upturn vs non-UV Upturn Galaxies} to compare SP properties between the UV and non-UV subsamples.

We compute the light-weighted mean stellar age $\langle \lg (t/{\rm yr}) \rangle$ and metallicity $\langle [Z/H]\rangle$ following Eq.39 in \citet{Cappellari2023MNRAS.526.3273C}:
\begin{subequations}\label{Eq.2}
        \begin{align}
        &\langle \lg (t/{\rm yr})\rangle=\sum\limits_{j=1}^{N_{\star}} x_{j}\lg (t_{j}/{\rm yr}),\\
        &\langle [Z/H]\rangle=\sum\limits_{j=1}^{N_{\star}} x_{j}[Z/H]_{j}.
        \end{align}
\end{subequations}
where $x_{j}$ is the fractional contribution of the $j$-th SSP to the total stellar light, with corresponding age $t_{j}$ and metallicity $[Z/H]_j$, and $N_{\star}$ is the total number of SSP templates. To estimate the uncertainty of light-weighted mean stellar age and metallicity, we perform 30 Monte Carlo simulations by adding Gaussian noise (same level as the observed spectrum) to the best-fit spectrum. The final age and metallicity are averaged over the fitting results of all simulations.
Furthermore, by performing linear fits to the radial profiles of SP properties for each galaxy, we derive the gradients of the mean stellar age ($\nabla \langle \lg (t/{\rm yr}) \rangle$), metallicity ($\nabla \langle [Z/H]\rangle$), and stellar velocity dispersion ($\nabla \sigma_\star$).

Since the inferred stellar population properties could be affected by the fitting code \citep{Pacifici2023ApJ...944..141P} and spectral template libraries \citep{Jones2025MNRAS.543..167J}. In this work, we employ three independent SSP templates based on EPS models, with their specific parameter configurations listed in Table~\ref{tab3}. All fluxes of SSP templates are normalized over the wavelength range 5070--5950\,\AA. We note that the SSP templates used here are to predict possible changes in spectral indices induced by radial stellar population variation, rather than to measure their binary fraction variations. Therefore, the choice of the SSP template is irrelevant to whether it includes binaries. Unless otherwise stated, our conclusions are based on the GALAXEV model.

The GALAXEV model \citep{Bruzual2003MNRAS.344.1000B} is an EPS model without binaries. It is based on the Padova 1994 evolutionary tracks \citep{Bertelli1994AAS..106..275B, Bruzual2003MNRAS.344.1000B} with lower and upper stellar mass cutoffs of 0.1$M_\odot$ and 100$M_\odot$. We use the Fortran code provided by it to extract SSPs constructed by the Chabrier \citep{Chabrier2003PASP..115..763C} initial mass function (IMF) and the MILES stellar spectral library \citep{Sanchez2006MNRAS.371..703S, Falcon2011AA...532A..95F}. The SSPs include 5 metallicities [Z/H] = -1.74, -0.73, -0.42, 0., 0.47, and 43 ages spaced logarithmically by 0.1 dex from 1 Myr to 15.85 Gyr.

The Yunnan II model \citep{Zhang2004A&A...415..117Z, Zhang2005MNRAS.357.1088Z} is developed in both single and binary stars evolution models with the binary-star evolution (BSE) algorithm of \citet{Hurley2000MNRAS.315..543H, Hurley2002MNRAS.329..897H}. Both single and binary star EPS models use an approximation IMF of \citet{Miller1979ApJS...41..513M}, also described in \citet{Eggleton1989ApJ...347..998E}, with the initial (primary) stellar mass range from 0.1$M_\odot$ to 100$M_\odot$. The binary star evolution model assumes a uniform form of mass ratio and eccentricity. The distribution of orbital separation falls smoothly at close ($\leqslant$ 10 $R_\odot$) separation, and is constant for wide binaries per logarithmic interval, making approximately 50\% of stellar systems are binary systems with orbital periods less than 100 yr. Both the single and binaries SSPs adopted here consist of 43 ages, and 5 metallicities [Z/H] = -1.3, -0.7, -0.3, 0., 0.2, compiled by the BLUERED \citep{Bertone2008AA...485..823B} stellar atmosphere models in the optical (3500-7500 \AA) region and the BaSeL-2.0 spectral library \citep{Lejeune1997AAS..125..229L, Lejeune1998AAS..130...65L} in the rest region.

The BPASS v2.3 \citep{Byrne2022MNRAS.512.5329B} also provides the EPS models with and without binary stars. Both of them built upon the Cambridge {\tt STARS} stellar evolution model \citep{Eggleton1971MNRAS.151..351E, Eldridge2008MNRAS.384.1109E} with (primary) stellar mass range from 0.1$M_\odot$ to 300$M_\odot$. They use a default IMF with a slope of -1.3 between 0.1$M_\odot$-0.5$M_\odot$ and -2.35 at stellar mass $M_{\star}>0.5M_\odot$, and the C3K stellar spectra library \citep{Choi2016ApJ...823..102C, Conroy2018ApJ...854..139C}. The binary template is modeled with the parameters (binary fraction, mass ratio, and orbital period) from the statistical results summarized in \citet{Moe2017ApJS..230...15M}, in which the binary fraction varies from about 40\% to 90\% as a function of mass of the primary star. Both single and binary SSPs adopted here consist of 43 ages spaced logarithmically by 0.1 dex from 1 Myr to 15.85 Gyr, same as the BC03 used above, and 10 metallicities [Z/H] = -1.3, -1., -0.8, -0.7, -0.5, -0.4, -0.3, 0., 0.2, 0.3.

\subsection{SP-Induced and Subtracted Binary Fraction Variations and Additional Clarifications}\label{subsec:SP-Induced and Subtracted Binary Fraction Variations and Additional Clarifications}

\subsubsection{SP-Induced and Subtracted Binary Fraction Variations}
\label{subsec:SP-Induced and Subtracted Binary Fraction Variations}
The binary fraction variations are derived from variations in SAFIs (see Eq.~\ref{Eq:method-1}), which are known to be sensitive to SP age and metallicity \citep{Worthey1994ApJS...95..107W}. As a result, the radial profile of the binary fraction inferred from SAFIs may be contaminated by radial SP variations.

To account for this, we define the SP-induced binary fraction variation $r_{{\rm b,sp},i}$---representing the pseudo-variation of relative binary fraction caused purely by SP variations. It was obtained by measuring the spectral indices variation of the intrinsic stellar population spectra generated by interpolating the SSP spectral templates with the radial light-weighted mean stellar age $\langle \lg (t/{\rm yr}) \rangle$ and metallicity $\langle [Z/H]\rangle$ derived above. We also take the uncertainty of age and metallicity estimated above into account to generate upper- and lower-limited (the oldest and most metal-rich, and the youngest and most metal-poor) intrinsic stellar population spectra and consider these as an uncertainty of stellar population-induced indices variation, thereby to maximize isolating the influence of changes in radial stellar population on spectral indices and derived binary fraction variations.

$r_{{\rm b,sp},i}$ is obtained in the same way as Eq.~\ref{Eq:method-1}, and the SP-subtracted binary fraction variation $r_{{\rm b,sub},i} = r_{{\rm b},i} - r_{{\rm b,sp},i}$.
The radial profiles, gradients, and values at $1R_e$ of SP-subtracted binary fraction variations are processed identically to those of $r_{{\rm b},i}$ (see the final two paragraphs of Section~\ref{subsubsec:Method}).

\subsubsection{Additional Clarifications for Spectrum}
\label{subsec:Additional Clarifications for Spectrum}
To minimize the impact of emission lines on our results as much as possible, we use a pure continuum spectrum---containing only stellar components---for both SAFI measurements and pPXF fitting (Sections~\ref{subsubsec:Method} and \ref{subsubsec:Spectral Fitting Code and EPS Models}). This pure continuum is constructed by subtracting the stacked emission-line component, provided by the MaNGA DAP model {\tt LOGCUBE}, from the stacked spectra within the same elliptical annular bins described above.

In the SAFI calculations, we fill in few masked fluxes within the index passbands at wavelengths longer than 4000\AA\ using the best-fitting pure continuum spectrum from the DAP {\tt LOGCUBE} (co-added in the same manner). This correction mitigates spurious absorption-like features that may arise from reduced spectral resolution and convolution effects.

Prior to pPXF fitting, the stacked pure continuum spectrum is cleaned by excluding masked fluxes and any outliers deviating by more than 5$\sigma$, ensuring robust and reliable fitting results. An example of stacked spectrum process and fitting is illustrated in Fig. \ref{fit_illustration}.

\begin{figure}[htb!]
\centering
\includegraphics[width=0.35\textwidth]{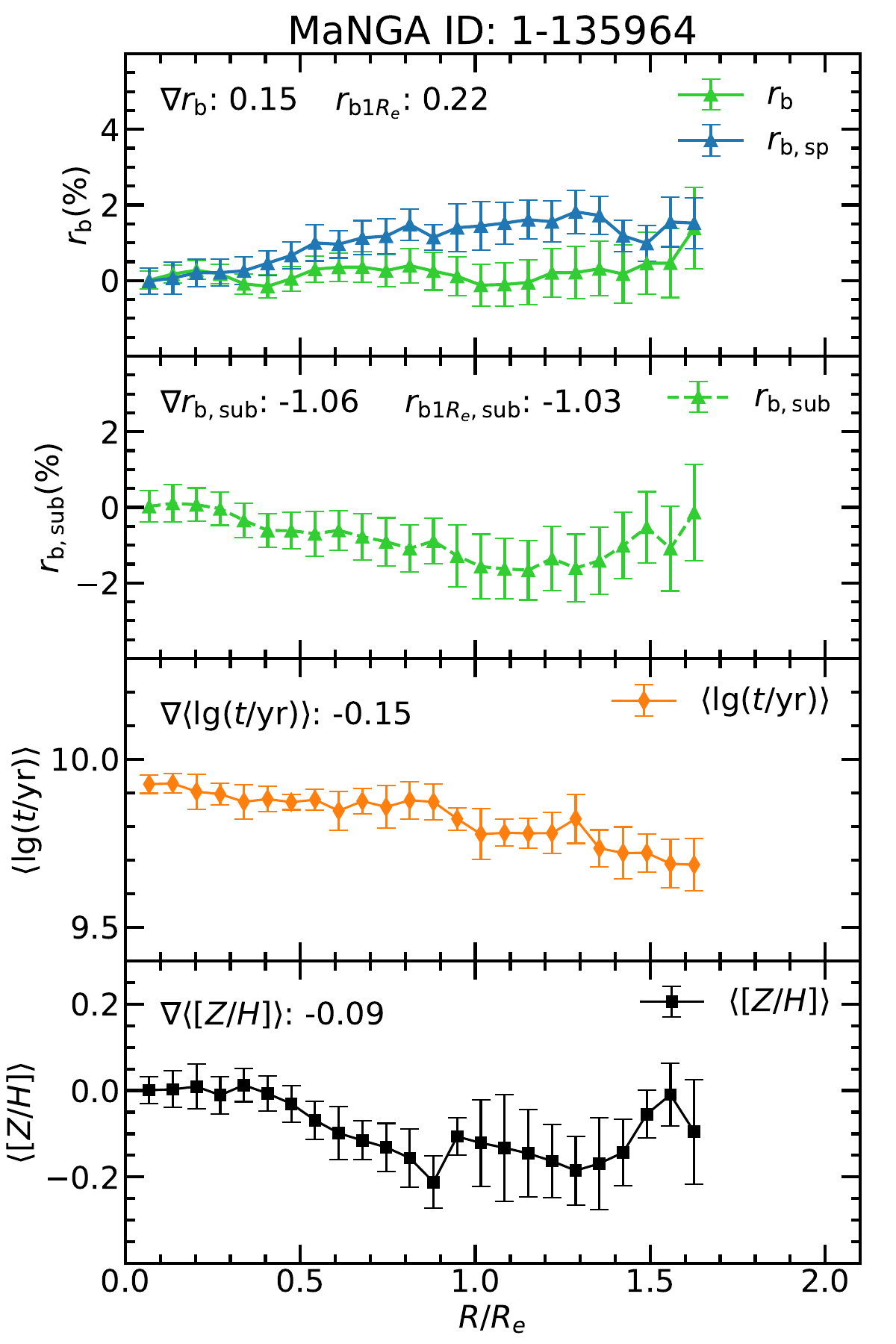}
\caption{An example galaxy with MaNGA ID: 1-135964. From top to bottom: The first panel is the radial profile of the derived binary fraction variation $r_{\rm b}$ (green solid line), and the SP-induced binary fraction variation $r_{\rm b,\,sp}$ (blue solid line). The second panel shows the corresponding SP-subtracted binary fraction variation $r_{\rm b,sub}$ (green dashed line). The bottom two panels show the radial profiles of light-weighted mean stellar age $\langle \lg (t/{\rm yr}) \rangle$ and metallicity $\langle [Z/H] \rangle$. In each panel, the gradient or values of binary fraction variation at $1R_e$ are labeled in the upper left, and the error bars indicate the $\pm1\sigma$ uncertainties.
\label{fig3}}
\end{figure}

\section{Binary Fraction Profiles}\label{sec:Binary Fraction Profiles}

We present the results of the binary fraction profile for our elliptical galaxy sample in Section~\ref{subsec:derived binary fraction Profile}. Given the potential impact of radial SP variation on binary fraction profile, the SP-subtracted binary fraction profiles are also given in Section~\ref{subsec:SP-Subtracted Binary Fraction Profile} for comparison.

\subsection{Binary Fraction Profile}\label{subsec:derived binary fraction Profile}

\begin{figure*}[htb!]
\plotone{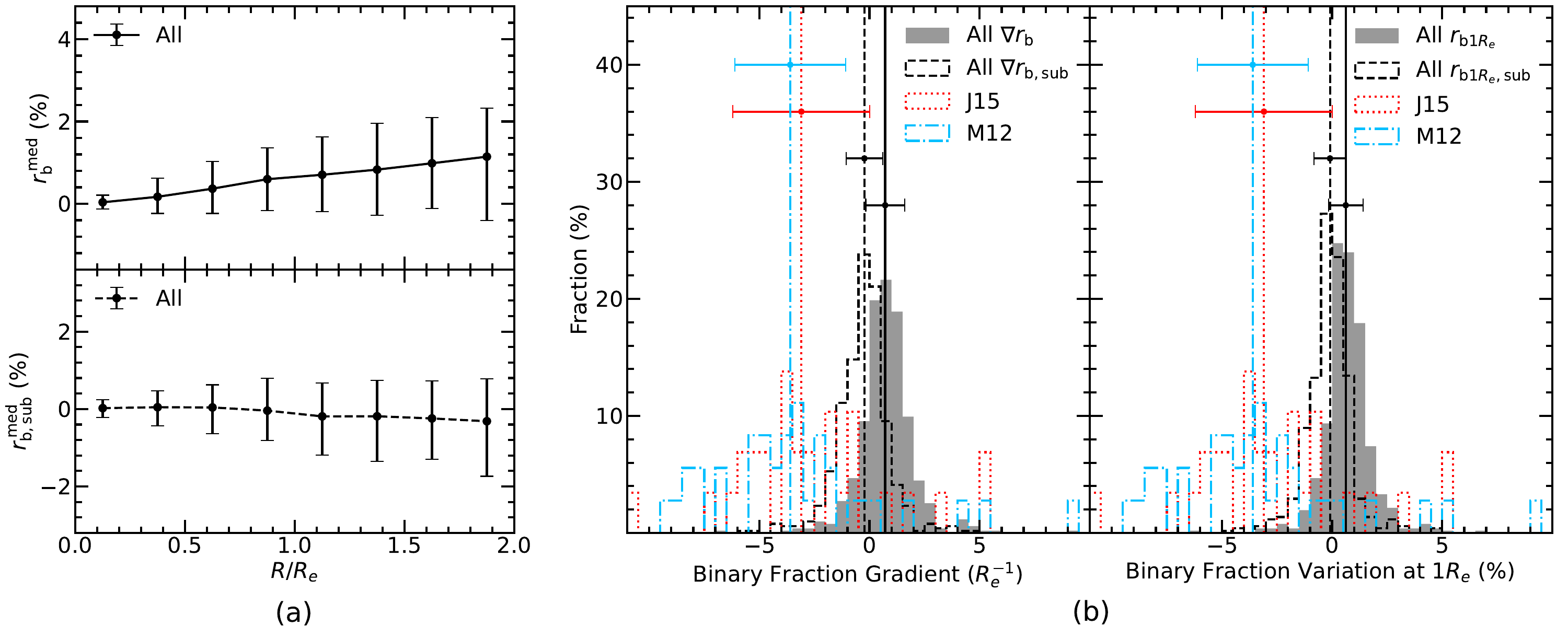}
\caption{(a): The median radial profile of binary fraction variation $r_{\rm b}^{\rm med}$ (solid line, top panel) and SP-subtracted binary fraction variation $r_{\rm b,\,sub}^{\rm med}$ (dashed line, bottom panel) for all 513 elliptical galaxies. The median value at each radius bin is shown as a filled black circle with error bars indicating the range from the 16th to the 84th percentiles. (b): Distributions of binary fraction gradients (left panel) and the values of binary fraction variation at $1R_e$ (right panel) for all 513 elliptical galaxies. The vertical lines indicate the corresponding median values of each histogram, with their dispersions denoted by the associated horizontal error bars. Results derived from GCs of \citet[][J15]{Ji2015ApJ...807...32J} and \citet[][M12]{Milone2012A&A...540A..16M} also ploted for illustration.
\label{fig4}}
\end{figure*}

An example of a galaxy with MaNGA ID 1-135964 is shown in Fig.~\ref{fig3}. As seen in the top panel, while the relative binary fraction variation $r_{\rm b}$ typically increases with radius, results vary across galaxies; therefore, we present statistical results for the entire sample. As shown in the top panel of Fig.~\ref{fig4}(a), the median radial profile of the binary fraction for all 513 elliptical galaxies, $r_{\rm b}^{\rm med}$, also increases slightly with radius. The increased dispersion at larger radii could be attributed to a more pronounced variation in the binary fraction relative to the galaxy center, compounded by fewer data points in the outer regions.

The distributions of the binary fraction gradient, $\nabla r_{\rm b}$ (left panel), and relative binary fraction variation at $1R_e$, $r_{{\rm b}1R_e}$ (right panel), for all 513 elliptical galaxies are shown in Fig.~\ref{fig4}(b). About 80\% of galaxies exhibit radially increasing binary fraction profiles as their $\nabla r_{\rm b}>0$. The distribution of $r_{{\rm b}1R_e}$ is centered around 1\%, with the highest (lowest) binary fraction variation at $1R_e$ is $\sim7 (-6)\%$. The median values of the binary fraction gradient, $\nabla r_{\rm b}^{\rm med}$, and the relative binary fraction variation at $1R_e$, $r_{{\rm b}1R_e}^{\rm med}$, for the full sample are 0.71\% and 0.62\%, respectively, as indicated by the vertical solid lines in each panel of Fig.~\ref{fig4}(b), also listed in the second and fourth columns of Table~\ref{tab4}.

Radial variation of binary fraction at $1R_e$ for almost all galaxies is lower than 5\%, which is comparable to that observed in the GCs, where the difference in binary fraction between the core and the region outside the half mass radius is typically a few percent for most of clusters \citep{Milone2012A&A...540A..16M, Ji2015ApJ...807...32J}.

\begin{deluxetable*}{ccccccc}
\tablenum{4}
\tablewidth{0pt}
\tablecaption{The Median Values and Dispersion of the Gradients of Binary Fraction and Binary Fraction Variations at $1R_e$. \label{tab4}}
\tablehead{\colhead{Sample (Num)} & \colhead{$\nabla r_{\rm b}^{\rm med}$} & \colhead{$\nabla r_{\rm b,\,sub}^{\rm med}$} & \colhead{$r_{{\rm b}1R_e}^{\rm med}$} & \colhead{$r_{{\rm b}1R_e{\rm ,\,sub}}^{\rm med}$}}
\startdata
All (513) & 0.71±0.88 & -0.23±0.83 & 0.62±0.79 & -0.09±0.73\\
UV (57) & 0.89±0.79 & -0.14±0.89  & 0.75±0.86 & -0.00±0.76\\
Non-UV (456) & 0.69±0.88 & -0.26±0.85  & 0.62±0.78 & -0.10±0.70\\
\enddata
\tablecomments{All values in the Table are given as percentages. The dispersion of a median value is estimated using 1.48 $\times$ MAD, where MAD is the median absolute deviation of the sample throughout the paper.}
\end{deluxetable*}

\subsection{SP-Subtracted Binary Fraction Profile} \label{subsec:SP-Subtracted Binary Fraction Profile}

The radial profile of the derived binary fraction could be attributed to the radial SP variation. As demonstrated in Fig.~\ref{fig3}, the light-weighted mean stellar age $\langle \lg (t/{\rm yr}) \rangle$ and metallicity $\langle [Z/H]\rangle$ decrease with radius, while the SP-induced binary fraction variation, $r_{\rm b,\,sp}$, increases with radius and even shows a steeper rising trend than the $r_{\rm b}$. After subtracting the contribution from SP variation, the SP-subtracted binary fraction variation, $r_{\rm b,\,sub}$, becomes flattened and even decreases slightly with radius. Therefore, the impact of the radial SP variation on the derived binary fraction profile should be taken into account.

As the bottom panel of Fig.~\ref{fig4}(a) shows, the median radial profile of SP-subtracted binary fraction variation for all 513 elliptical galaxies, $r_{\rm b,\,sub}^{\rm med}$, appears to flatter and decreases slightly along the radius compared to the $r_{\rm b}^{\rm med}$. However, we note that both the median radial profiles of $r_{\rm b}^{\rm med}$ and $r_{\rm b,sub}^{\rm med}$ are within the statistical noise of the data, therefore, the difference between these two profiles is not statistically significant.
The distribution of SP-subtracted binary fraction gradient, $\nabla r_{\rm b,\,sub}$, and SP-subtracted binary fraction variation at $1R_e$, $r_{{\rm b}1R_e{\rm ,\,sub}}$, also plotted in Fig.~\ref{fig4}(b), are systematically shift toward left compare to the $\nabla r_{\rm b}$ and $r_{{\rm b}1R_e}$. About 39\% of galaxies have their $\nabla r_{\rm b,\,sub}>0$, and the highest (lowest) binary fraction at $1R_e$ is $\sim5 (-5)\%$.

The median values of SP-subtracted binary fraction gradient, $\nabla r_{\rm b,\,sub}^{\rm med}$, and SP-subtracted binary fraction variation at $1R_e$, $r_{{\rm b}1R_e{\rm ,\,sub}}^{\rm med}$, for the full sample are $-0.23\%$ and $-0.09\%$, respectively, as illustrated by the vertical dashed line in each panel of Fig.~\ref{fig4}(b), which are also listed in the third and fifth columns of Table~\ref{tab4}.

A flattened radial profile of the SP-subtracted binary fraction indicates that the binary fraction variations derived from SAFIs variations are indeed influenced by radial variations in SP properties. But, we also note that 189 (89) galaxies exhibit both their binary fraction gradients and SP-subtracted binary fraction gradients are positive (negative), which suggests the binary fraction profile might increase slightly with radius for about 37\% of our elliptical galaxy sample. However, given the typical uncertainties listed in Table \ref{tab4}, we note that this fraction should be regarded as indicative and interpreted with caution.

Studies \citep{Mart2015MNRAS.447.1033M, van2017ApJ...841...68V} on the radial variations of the IMF in early-type galaxies have also shown that the absolute value of the IMF index decreases with radius, implying an increasing fraction of massive stars (a more top-heavy IMF) in the outer regions. Since the binary fraction typically increases with stellar mass \citep{Duch2013ARA&A..51..269D}, this trend may be linked with the behavior of the radially increasing binary fraction profile, which is also consistent with the hypothesis that younger or metal-poor stars tend to host higher binary fractions \citep{Sollima2007MNRAS.380..781S, Moe2019ApJ...875...61M}, as most of our galaxies show a radially decreasing stellar population age and metallicity (see Section~\ref{subsec:Connection with SP Properties}). And it might also be consistent with the hierarchical clustering scenario concerning elliptical galaxies \citep{Kang2017MNRAS.469.1636K}. 

However, this trend is opposite to that observed in GCs, where the binary fraction decreases with radius as a result of mass segregation \citep{Geller2012AJ....144...54G, Dib2018MNRAS.473..849D}. We compare the distribution of the binary fraction gradient in GCs with the results of our elliptical galaxies in Fig. \ref{fig4}(b). The data of binary fraction of GCs from \citet{Ji2015ApJ...807...32J} and \citet{Milone2012A&A...540A..16M}, and we excluded six GCs with negative binary fraction in \citet{Ji2015ApJ...807...32J}, as well as 19 GCs without binary fraction of core or half-mass radius and four GCs showing unusually large ($>$10\%) binary fraction variations in \citet{Milone2012A&A...540A..16M}. The core and half-mass radii were adopted from \citet{Harris1996AJ....112.1487H}. Since the binary fraction at different regions of GCs is measured over a radial range, we use the median radius of each bin to calculate the binary fraction gradient of GCs. We find that the distribution of the binary fraction gradient of GCs shows a wider dispersion and is dominated by a negative gradient, with median values of $-3.09\% \pm 3.11\%$ and $-3.60\% \pm 2.52\%$ for \citet{Ji2015ApJ...807...32J} and \citet{Milone2012A&A...540A..16M}, respectively, and approximately 83\% and 86\% of the GCs show negative gradients. However, we caution that the binary fraction gradient of GCs in the figure is only used to illustrate and might not be suitable for direct comparison with the binary fraction gradient of elliptical galaxies, as they are intrinsically different systems and have different radius scales.

\section{UV Upturn vs non-UV Upturn Galaxies}\label{sec:UV Upturn vs non-UV Upturn Galaxies}

\begin{figure*}[htb!]
\plotone{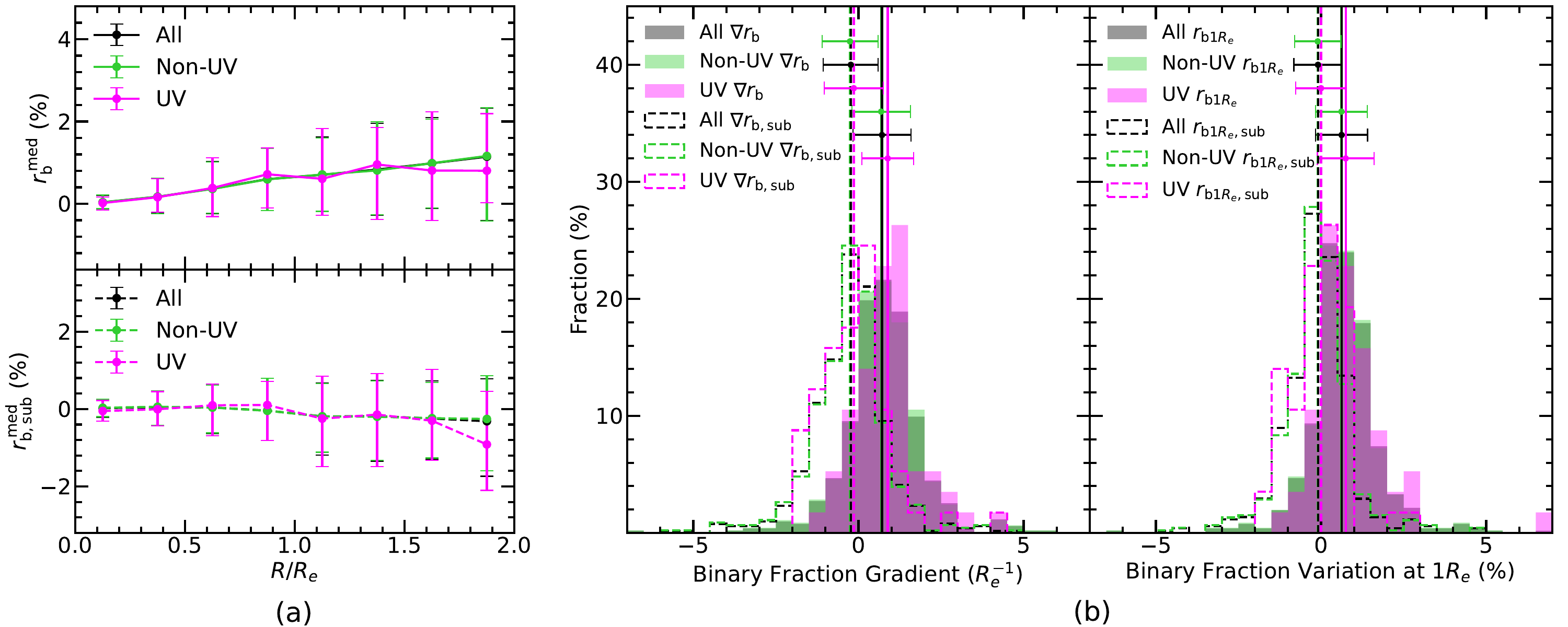}
\caption{Same as Fig.~\ref{fig4}, with the UV (magenta) and non-UV (green) subsamples also drawn for comparison.
\label{fig5}}
\end{figure*}

The origin of the UV upturn remains under debate. Either potential star formation activities or an old population of hot HB stars, which can be produced through binary interactions \citep{Han2002MNRAS.336..449H, Han2003MNRAS.341..669H}, give rise to this excessive UV emission. The binary fraction distribution and/or SP properties of UV upturn galaxies might differ from those of normal galaxies. Here, we compare the binary fraction profiles between UV and non-UV subsamples in Section~\ref{subsec:Comparison in Binary Fraction Profiles}, as well as their SP properties derived from EPS models without binaries (Section~\ref{subsubsec:Results From the GALAXEV}) and with binaries (Section~\ref{subsubsec:Results From Yunnan II EPS Model with Binaries}). A further discussion about the similarity between UV and non-UV subsamples is given in Section~\ref{subsubsec:Reasons of Similarity between UV and Non-UV Subsamples}

\subsection{Comparison in Binary Fraction Profiles}\label{subsec:Comparison in Binary Fraction Profiles}

The median radial profiles of binary fraction for the UV subsample (57 galaxies) and the non-UV subsample (456 galaxies) are shown in Fig.~\ref{fig5}(a). The binary fraction radial profile of the UV subsample is consistent with that of the non-UV subsample, where both of their $r_{\rm b}^{\rm med}$ increase slightly with the radius, while $r_{\rm b,\,sub}^{\rm med}$ becomes flattened.

As Fig.~\ref{fig5}(b) shows, the distributions of $\nabla r_{\rm b,\,sub}$ and $r_{{\rm b}1R_e{\rm ,\,sub}}$ for the UV subsample are also similar to those of the non-UV subsample, with their median values denoted by the vertical lines (also listed in Table~\ref{tab4}). The $p$-values of the Kolmogorov–Smirnov (KS) test between two subsamples on $\nabla r_{\rm b}$ ($\nabla r_{\rm b,\,sub}$) and $r_{{\rm b}1R_e}$ ($r_{{\rm b}1R_e{\rm ,\,sub}}$) are 0.43 (0.78) and 0.85 (0.24), respectively.
The subtle difference might raised due to the relatively higher ($>2\%$) binary fraction gradient of a few galaxies in the UV subsample, as can be seen in Fig.~\ref{fig5}(b). 

It should be noted that we derive only relative binary fractions (i.e., variations with respect to the galaxy center). Therefore, this result does not necessarily imply that UV upturn galaxies have absolute binary fractions similar to those of non-UV upturn galaxies. Instead, it suggests that the physical processes governing the dynamical evolution and spatial distribution of binaries in elliptical galaxies may be independent of whether the galaxy exhibits a UV upturn.

\begin{figure*}[htb!]
\plotone{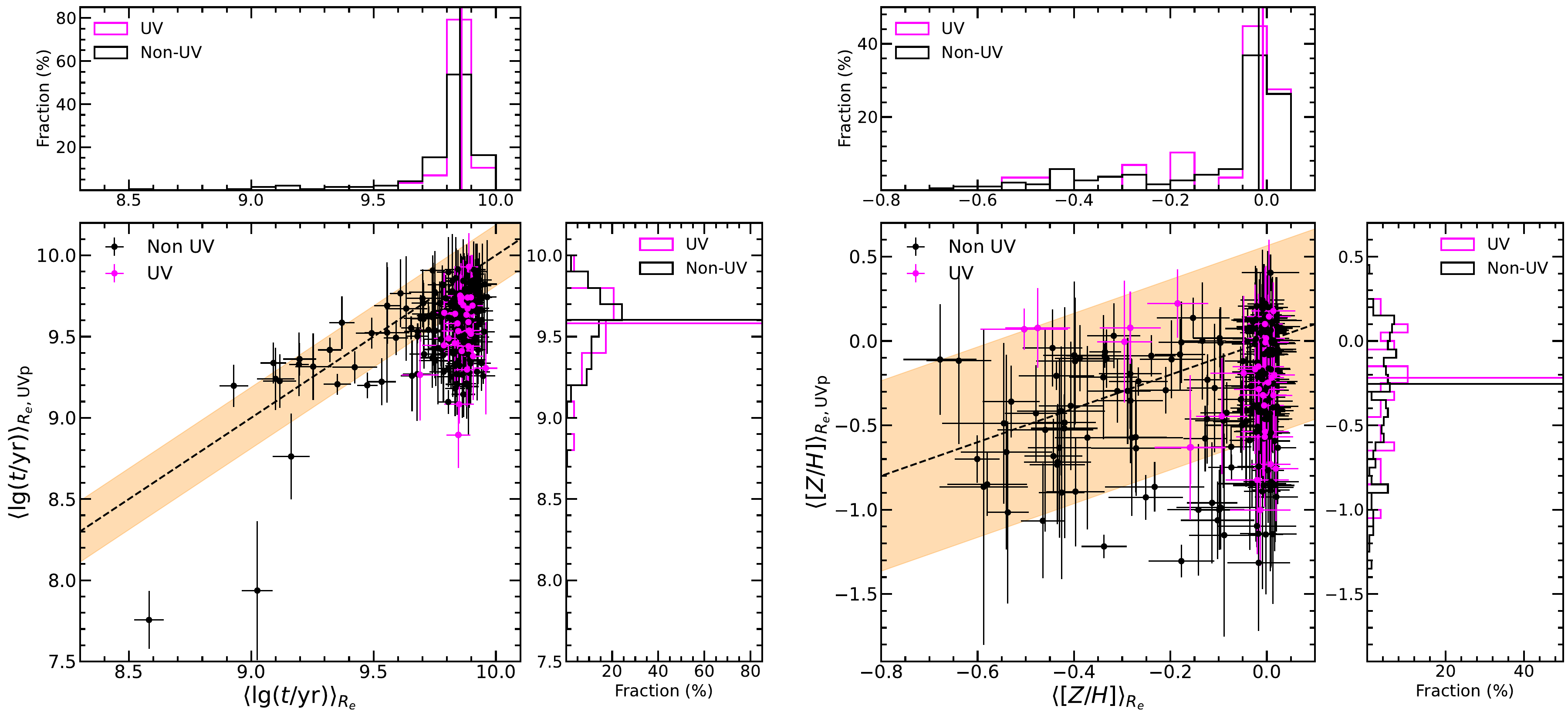}
\caption{SP properties within $1R_e$ derived with and without photometric fluxes. Left panel: The pure-continuum mean stellar age $\langle \lg (t/{\rm yr}) \rangle_{R_e}$ against the photometry-included mean stellar age $\langle \lg (t/{\rm yr}) \rangle_{R_e,\rm UVp}$. Right panel: The pure-continuum mean stellar metallicity $\langle [Z/H]\rangle_{R_e}$ against the photometry-included mean stellar mtallicity $\langle [Z/H]\rangle_{R_e,\rm UVp}$. (Non-) UV subsample is represented by filled magenta (black) circles, with error bars indicating $\pm1\sigma$ uncertainties. The orange error band in each panel gives the $\pm1\sigma$ uncertainty between two fitting methods. Distributions of different mean stellar age and metallicity for UV and non-UV subsamples are also plotted on the subpanels, with vertical lines indicating the corresponding median values.
\label{fig6}}
\end{figure*}

\subsection{Comparison in SP Properties}\label{subsec:Comparison in SP Properties}

The UV-upturn phenomenon is characterized by an excess of flux in the ultraviolet spectra of elliptical galaxies. A wider wavelength coverage, including the ultraviolet band, should be more reasonable for constraining their intrinsic stellar population properties in spectral fitting. Therefore, we derive photometry-included light-weighted mean stellar age $\langle \lg (t/{\rm yr}) \rangle_{R_e,\rm UVp}$ and metallicity $\langle [Z/H]\rangle_{R_e,\rm UVp}$ by fitting the pure continuum stacked within $1R_e$ simultaneously with the FUV, NUV, and $ugriz$ photometric fluxes provided by NSA to investigate further the differences between the UV and the non-UV subsample. In contrast, the pure continuum light-weighted mean stellar age $\langle \lg (t/{\rm yr}) \rangle_{R_e}$ and metallicity $\langle [Z/H]\rangle_{R_e}$ within $1R_e$ are also derived. 

The uncertainty between two methods is estimated from the difference between $\langle \lg (t/{\rm yr}) \rangle_{R_e}$ ($\langle [Z/H]\rangle_{R_e}$) and pure continuum plus $ugriz$ photometries fitting $\langle \lg (t/{\rm yr}) \rangle_{R_e,\rm p}$ ($\langle [Z/H]\rangle_{R_e,\rm p}$) within $1R_e$. All photometry-included fittings share the same configuration, and photometric fluxes are also Galactic extinction corrected as Section~\ref{subsec:Radial Spectra Construction}. We note that 294 sources (28 for the UV, and 266 for the non-UV subsample) are excluded in this Section due to the unavailability of FUV and NUV photometric flux errors, mismatched ($>3\sigma$) $griz$ photometric fluxes between the observed and synthetic photometries derived from spectra, and failed fitting with photometric measurements. Finally, the results of 219 (29 for UV and 190 for non-UV) galaxies are analyzed.

\begin{deluxetable*}{llcccccc}
\tablenum{5}
\tablewidth{0pt}
\tablecaption{The Median Values and Dispersion of SP Properties Derived with and without Photometric Fluxes within $1R_e$.
\label{tab5}}
\tablehead{\colhead{Model} & \colhead{Sample (Num)} & \colhead{$\langle \lg (t/{\rm yr}) \rangle_{R_e}^{\rm med}$} & \colhead{$\langle \lg (t/{\rm yr}) \rangle_{R_e,{\rm UVp}}^{\rm med}$} & \colhead{$\Delta t^{\rm med}$} & \colhead{$\langle [Z/H]\rangle_{R_e}^{\rm med}$} & \colhead{$\langle [Z/H]\rangle_{R_e,\rm UVp}^{\rm med}$} & \colhead{$\Delta Z^{\rm med}$}}
\startdata
GALAXEV & All (219) & 9.86±0.06 & 9.60±0.20 & 0.22±0.19 & -0.02±0.04 & -0.25±0.39 & 0.17±0.38\\
 -- & UV (29) & 9.86±0.04 & 9.58±0.20 & 0.30±0.21 & -0.01±0.02 & -0.22±0.43 & 0.23±0.36\\
 -- & Non-UV (190) & 9.85±0.07 & 9.60±0.20 & 0.21±0.19 & -0.02±0.04 & -0.25±0.38 & 0.16±0.37\\
Yunnan II-b & All (219) & 9.74±0.21 & 9.75±0.28 & -0.03±0.36 & -0.37±0.15 & -0.62±0.30 & 0.18±0.24\\
 -- & UV (29) & 9.76±0.16 & 9.86±0.22 & -0.05±0.40 & -0.31±0.07 & -0.59±0.18 & 0.24±0.24\\
 -- & Non-UV (190) & 9.73±0.22 & 9.73±0.27 & -0.02±0.36 & -0.39±0.17 & -0.62±0.30 & 0.17±0.24\\
\enddata
\tablecomments{$\Delta t^{\rm med}$ and $\Delta Z^{\rm med}$ are defined as the median difference of $\langle \lg (t/{\rm yr}) \rangle_{R_e} - \langle \lg (t/{\rm yr}) \rangle_{R_e,{\rm UVp}}$ and $\langle [Z/H]\rangle_{R_e} - \langle [Z/H]\rangle_{R_e,\rm UVp}$, respectively, for galaxies of different (sub)samples.}
\end{deluxetable*}

\subsubsection{Results From the GALAXEV}\label{subsubsec:Results From the GALAXEV}

The left panel of Fig.~\ref{fig6} plots the distribution of the $\langle \lg (t/{\rm yr}) \rangle_{R_e}$ against the $\langle \lg (t/{\rm yr}) \rangle_{R_e,\rm UVp}$ for different subsamples, based on the same GALAXEV model as above, and results are listed in the third and fourth columns of Table~\ref{tab5} (also see the vertical lines shown in the top and right subpanels). The median value of pure continuum light-weighted mean stellar age, $\langle \lg (t/{\rm yr}) \rangle_{R_e}^{\rm med}$, of the UV subsample is systematically about 0.01 dex older than that of the non-UV subsample. But, this difference is smaller than the reported uncertainty of 0.04 dex or 0.07 dex.
While the median value of photometry-included light-weighted mean stellar age, $\langle \lg (t/{\rm yr}) \rangle_{R_e,\rm UVp}^{\rm med}$, of the UV subsample is about 0.02 dex younger than that of the non-UV subsample. This difference is also not significant given the uncertainty of 0.2 dex.

The right panel of Fig.~\ref{fig6} plots the distribution of the $\langle [Z/H]\rangle_{R_e}$ against the $\langle [Z/H]\rangle_{R_e,\rm UVp}$. The median value of pure continuum light-weighted mean stellar metallicity, $\langle [Z/H]\rangle_{R_e}^{\rm med}$, of the UV subsample is slightly higher by 0.01 dex than that of the non-UV subsample. But this difference is smaller than the reported uncertainty of 0.02 dex or 0.04 dex.
The median value of photometry-included light-weighted mean stellar metallicity, $\langle [Z/H]\rangle_{R_e,\rm UVp}^{\rm med}$, of the UV subsample is 0.03 dex higher than that of the non-UV subsample. But, this difference is much smaller than the reported uncertainty of 0.43 dex or 0.38 dex.
Results are also listed in the sixth and seventh columns of Table~\ref{tab5} (also see the vertical lines shown in the top and right subpanels).

These minor differences are likely to be washed out by the fitting uncertainties of individual sources, or of the two fitting methods—typically about 0.2 dex for age and 0.6 dex for metallicity, as indicated by the orange error band in Fig. \ref{fig6}. The overall stellar population properties of the UV and the non-UV subsamples, therefore, are quite similar under the GALAXEV model.

Besides, photometry-included SP properties are not consistent with those derived from the pure-continuum fitting under the EPS model without binaries, as illustrated by the slanted dashed line in Fig.~\ref{fig6}. The results derived from photometry-included fitting seem to be more dispersed toward the younger or metal-poor side. 

For the UV subsample, the $\langle \lg (t/{\rm yr}) \rangle_{R_e,\rm UVp}^{\rm med}$ ($\langle [Z/H]\rangle_{R_e,\rm UVp}^{\rm med}$) is systematically younger (lower) by 0.30 (0.23) dex than the $\langle \lg (t/{\rm yr}) \rangle_{R_e}^{\rm med}$ ($\langle [Z/H]\rangle_{R_e}^{\rm med}$), as listed in the fifth and eighth columns of Table~\ref{tab5}, which are likely caused by the absence of UV bright stars (e.g., hot HB stars) in the GALAXEV model, for which younger and metal-poor SSPs are employed to reproduce the rising UV emission. The non-UV subsample exhibits similar trends with their $\langle \lg (t/{\rm yr}) \rangle_{R_e,\rm UVp}^{\rm med}$  ($\langle [Z/H]\rangle_{R_e,\rm UVp}^{\rm med}$) is also systematically younger (lower) by 0.21 (0.16) dex than the $\langle \lg (t/{\rm yr}) \rangle_{R_e}^{\rm med}$ ($\langle [Z/H]\rangle_{R_e}^{\rm med}$). Therefore, the EPS models without binaries are inadequate for describing populations where binary evolution plays an important role.

\begin{figure*}[htb!]
\plotone{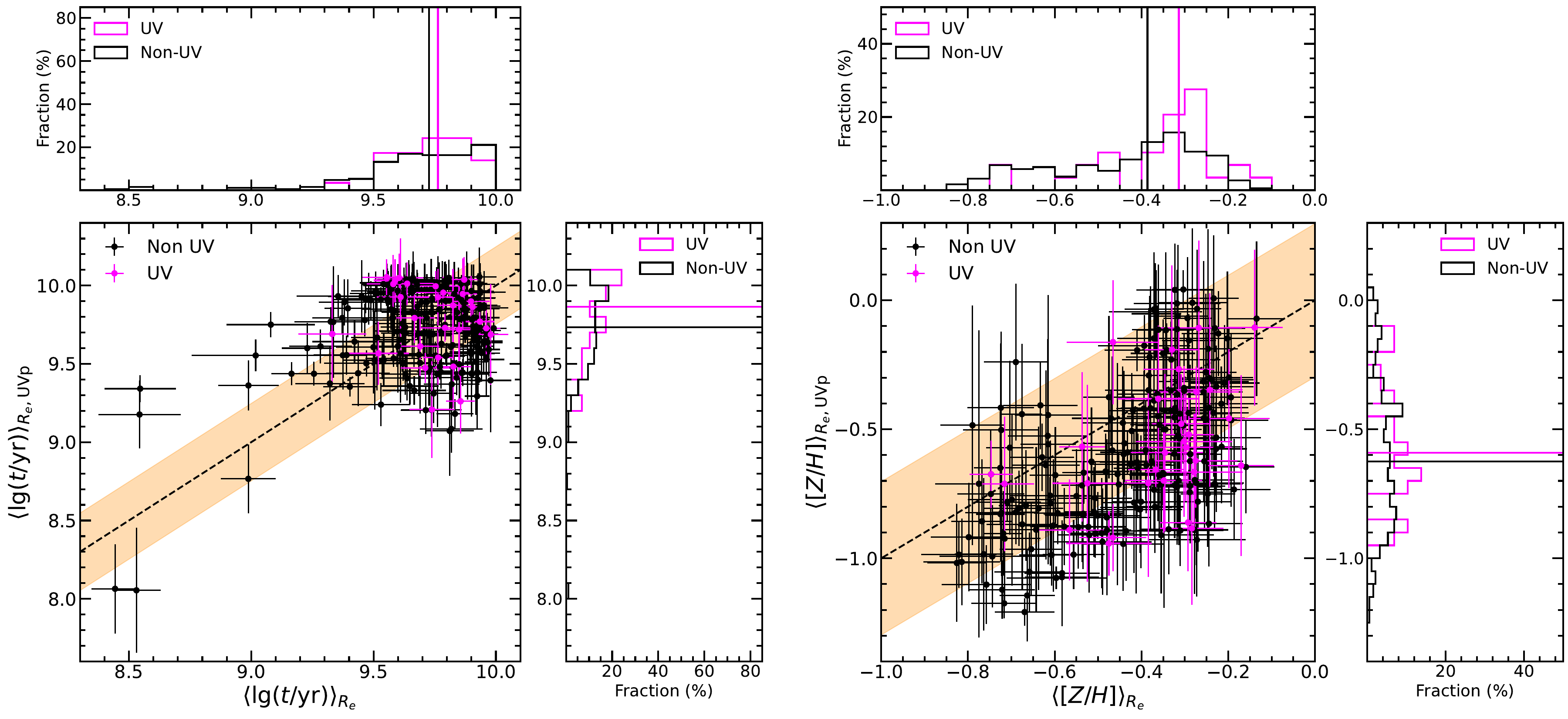}
\caption{Same as Fig.~\ref{fig6}, but using Yunnan II EPS model with binaries.
\label{fig7}}
\end{figure*}

\subsubsection{Results From Yunnan II EPS Model with Binaries}\label{subsubsec:Results From Yunnan II EPS Model with Binaries}

Considering the limitations of the GALAXEV, we performed an additional analysis using the Yunnan II EPS model with binaries (Yunnan II-b). 
As shown in Fig.~\ref{fig7}, for the mean stellar age (left panel), the $\langle \lg (t/{\rm yr}) \rangle_{R_e}^{\rm med}$ of the UV subsample is also slightly older by 0.03 dex than that of the non-UV subsample, similar to the result of GALAXEV, and this gap increases to 0.13 dex when photometry is included (see top and right subpanels, as listed in the third and fourth columns of Table~\ref{tab5}). But these differences are smaller than the typical age uncertainty of 0.2 dex, as the error bands shown in Fig.~\ref{fig7}, indicating no significant age difference.
For the mean stellar metallicity (right panel), the $\langle [Z/H]\rangle_{R_e}^{\rm med}$ of the UV subsample is 0.08 dex higher than that of the non-UV subsample, while the difference in $\langle [Z/H]\rangle_{R_e,\rm UVp}^{\rm med}$ between the UV and non-UV subsample reduces to 0.03 dex (see top and right subpanels, also listed in the sixth and seventh columns of Table~\ref{tab5}). These differences are much smaller than the typical metallicity uncertainty of 0.3 dex. Therefore, the differences in SP properties are not statistically significant between the UV and non-UV subsamples.

By contrast, under the Yunnan II-b model, the photometry-included SP properties of the UV subample seem more consistent with those derived from pure-continuum fitting than in the GALAXEV. 
As shown in Fig.~\ref{fig7} and its subpanels, for the UV subsample, the difference between the $\langle \lg (t/{\rm yr}) \rangle_{R_e,\rm UVp}^{\rm med}$  ($\langle [Z/H]\rangle_{R_e,\rm UVp}^{\rm med}$) and the $\langle \lg (t/{\rm yr}) \rangle_{R_e}^{\rm med}$ ($\langle [Z/H]\rangle_{R_e}^{\rm med}$) is 0.05 (0.24) dex, as listed in the fifth and last columns of Table~\ref{tab5}. For the non-UV subsample, the age difference is likewise reduced to 0.02 dex, while the metallicity difference is 0.17 dex.

We noted that the $\langle \lg (t/{\rm yr}) \rangle_{R_e,\rm UVp}^{\rm med}$  ($\langle [Z/H]\rangle_{R_e,\rm UVp}^{\rm med}$) of both UV and non-UV subsamples are slightly older (metal-poor) than the $\langle \lg (t/{\rm yr}) \rangle_{R_e}^{\rm med}$ ($\langle [Z/H]\rangle_{R_e}^{\rm med}$), especially for the metallicity, as can be seen clearly in the right panel of Fig~\ref{fig7}.
One possible explanation is that these galaxies are intrinsically old, and the observed UV emission remains stronger than predicted by the models even after accounting for the contribution from binary evolution. The SSPs generated by the EPS model with binaries are bluer than those without binaries \citep{Zhang2004A&A...415..117Z, Zhang2005MNRAS.357.1088Z}, which in turn leads to older inferred stellar ages. Consequently, older but more metal-poor SSPs may reproduce the excessive UV emission in photometry-included fitting.

For the Yunnan II-b model, it could also be a result of model dependency and likely affected by the age-metallicity degeneracy (as can be seen in Table~\ref{tab5}, where the result derived from pure-continuum fitting is young and metal-rich, while the photometry-included fitting is old and metal-poor). Given that more metal-poor photometry-included fitting is also present in the EPS model without binaries, suggesting that a hot, metal-poor star population may likewise contribute to the observed UV upturn.

\begin{figure*}[htb!]
\plotone{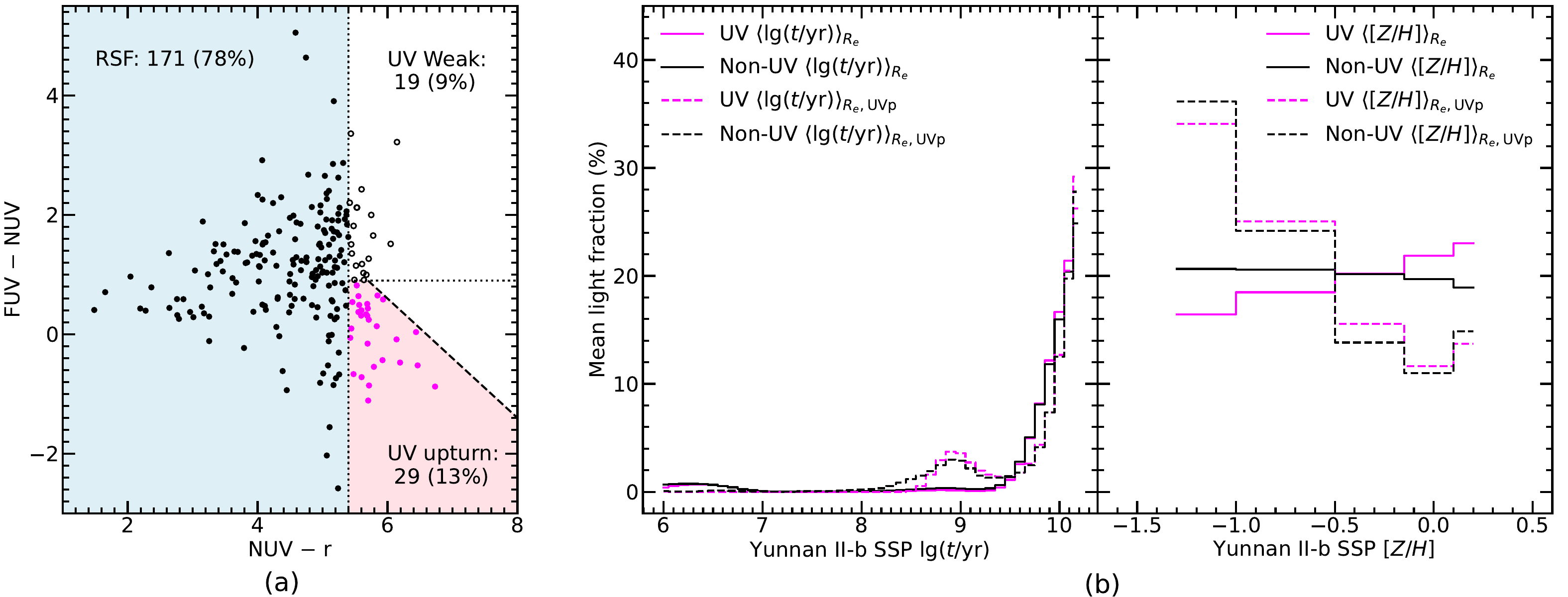}
\caption{(a): Two-color diagram for 29 UV (magenta dots) and 190 non-UV (black dots and circles) upturn galaxies following \citet{Yi2011ApJS..195...22Y} criteria (also see in Section~\ref{subsec:Data and Sample Selection}). The vertical and horizontal dotted lines denote the criteria for RSF and UV-weak regions, and the slanted dashed line is the demarcation for UV upturn galaxies. (b): Mean light-fractional contribution $x_j$ of pure-continuum (solid lines) and photometry-included (dashed lines) fitting as a function of SSP age (left panel) or metallicity (right panel) for different subsamples based on the Yunnan II EPS model with binaries. 
\label{fig8}}
\end{figure*}

\subsubsection{Reasons of Similarity between UV and Non-UV Subsamples}\label{subsubsec:Reasons of Similarity between UV and Non-UV Subsamples}

Although SP properties derived from the Yunnan II-b model show improved performance compared to those with GALAXEV, the non-UV subsample still cannot be efficiently distinguished from the UV subsample. The photometry-included age of the non-UV subsample is expected to be as old as those derived from pure-continuum fitting if they are completely retired/quenched systems, which is expected without any excess UV emission in spectra. However, their $\langle \lg (t/{\rm yr}) \rangle_{R_e,\rm UVp}$ is systematically lower than the $\langle \lg (t/{\rm yr}) \rangle_{R_e}$, similar to that of the UV subsample, under the GALAXEV model. It also follows the same trend of the UV subsample in the Yunnan II-b model, where the $\langle \lg (t/{\rm yr}) \rangle_{R_e,\rm UVp}$ is systematically larger than the $\langle \lg (t/{\rm yr}) \rangle_{R_e}$, as mentioned before and can be found in Table~\ref{tab5}. Since there are minor differences in binary fraction profiles between UV and non-UV subsamples (see Section~\ref{subsec:Comparison in Binary Fraction Profiles}), the fact that the non-UV subsample exhibits an age variation similar to that of the UV subsample suggests that these systems may host weak UV emission or RSF, which are common in low-redshift early-type galaxies \citep{Yi2005ApJ...619L.111Y, Schawinski2007ApJS..173..512S, Kaviraj2007ApJS..173..619K}.

For this reason, we examined the fitting results under the Yunnan II-b model for each galaxy in detail. We found that both FUV$-$NUV and NUV$-$r of non-UV subsample span a wide range and most (90\%) of them reside in the RSF region, as shown in Fig.~\ref{fig8}(a), indicating that their spectra may contain additional UV emission similar to that observed in UV upturn galaxies, thereby making their fitting results resemble those of the UV subsample. This can be supported by the mean light-fraction contribution of photometry-included fitting as shown in the left panel of Fig.~\ref{fig8}(b), where both the UV and non-UV subsamples show additional contribution from stellar populations younger than $10^9$ yr, and the non-UV subsample seems to have an more extended tail toward younger age compared to the UV subsample, suggesting the presence of possible star formation in some galaxies of the non-UV subsample. In fact, as described in Section~\ref{subsec:Data and Sample Selection} (also see Fig.~\ref{fig1}), we found that 152 out of 219 elliptical galaxies can be drawn on the BPT diagram, and 56 of them are located in the star formation region, implying at least 26\% of galaxies may host star formation activities.

Meanwhile, we also found that at least 14 galaxies of the whole sample are affected by the age-metallicity degeneracy (luminosity contribution dominated by old metal-poor stellar populations in pure-continuum fitting, while by young metal-rich stellar populations in photometry-included fitting, and vice versa), which is also reflected in the bimodal distributions of age and metallicity shown in Fig.~\ref{fig8}(b). 

Overall, the galaxies studied in this work are old, quenched systems, suggesting that they primarily consist of long-lived, low-mass stars, which typically have a low intrinsic binary fraction. The contribution from either binary populations or residual star formation to the integrated stellar population is therefore relatively limited. Additionally, the hot horizontal branch stars responsible for the UV upturn may be produced through single-star evolutionary channels, rather than being dominated by binary evolution. As a result, the global stellar population properties and binary fraction profiles of the two subsamples appear broadly similar within the current uncertainties.
Moreover, the UV properties of elliptical galaxies may vary continuously \citep{Smith2012MNRAS.421.2982S, Phillipps2020MNRAS.492.2128P, Jiang2025MNRAS.540.3770J}, which could further blur the distinctions in their SP properties, as evidenced by the similarity in stellar population properties between UV weak and UV upturn galaxies \citep{Dantas2021MNRAS.500.1870D}, and can be also seen in Fig. \ref{fig8}(a) where both UV and non-UV subsamples cover a wide range of FUV$-$NUV and NUV$-$r colors, and many galaxies distribute along the demarcation of criteria.

Corrections used to subtract the variations from the SP to obtain $r_{\rm b,sub}$ may not fully isolate the effects of intrinsic SP variation on the binary fraction, consequently, combined with the fitting uncertainties, age-metallicity degeneracy, and potential star formation, all these factors may lead to indistinguishable SP properties between the UV and non-UV subsamples in our elliptical galaxies. An appropriate EPS model combined with broader spectral coverage (including both UV and infrared), as well as additional diagnostics (e.g., H$\alpha$ emission as an indicator of recent star formation), may provide a more effective way to break the degeneracy, reduce the uncertainties in photometric fitting and better constrain the contribution from star formation.

\begin{figure*}[htb!]
\centering
\includegraphics[width=0.7\textwidth]{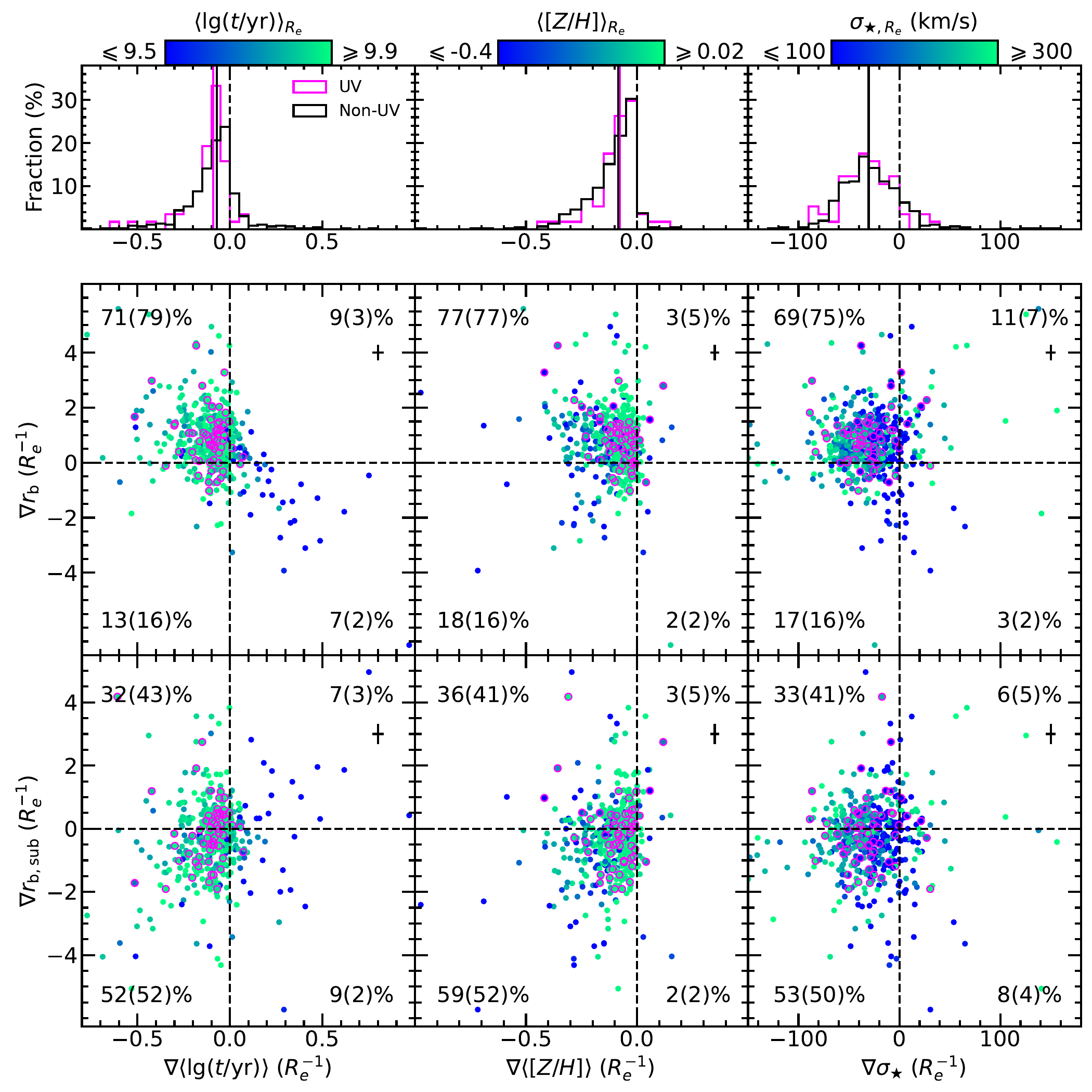}
\caption{The gradients of binary fraction against the gradients of mean stellar age $\nabla \langle \lg (t/{\rm yr}) \rangle$ (left column), metallicity $\nabla\langle [Z/H]\rangle$ (middle column), and stellar velocity dispersion $\nabla\sigma_{\star}$ (right column) for all 513 galaxies. The top panels show the distribution of corresponding SP property gradients, with vertical solid lines indicating the median values, for the UV and non-UV subsamples. Galaxies are colored by the corresponding SP properties within $1R_e$ derived from pure-continuum fitting, and the UV subsample is also marked by magenta circles. Percentages in the corners of each panel label the proportion of (UV upturn) galaxies that fall in different regions, which are divided by vertical and horizontal dashed lines, and the crosses in the upper right corner indicate the mean $\pm1\sigma$ uncertainties of gradients.
\label{fig9}}
\end{figure*}

\section{Discussion}\label{sec:Discussion}

\subsection{Connection with SP Properties} \label{subsec:Connection with SP Properties}

As mentioned in Section~\ref{sec:Binary Fraction Profiles}, binary fraction profiles are affected by variations in the radial SP. This might be owing to the dependence of the binary fraction on intrinsic stellar properties such as mass, age, and metallicity, although several aspects of these are still not well established. Generally, the younger population tends to have a higher binary fraction \citep{Duch2013ARA&A..51..269D}, which may also increase toward lower metallicity \citep{Moe2019ApJ...875...61M}. Furthermore, the radial dependence of the binary fraction in GCs is also a result of dynamical evolution (see Section~\ref{sec:Introduction}). Investigating these connections, therefore, provides valuable constraints on the formation mechanisms of binary systems and their dependence on the environment.

Since we only derived the relative binary fraction, here, we examine the potential correlations between binary fraction gradient and the gradients of mean stellar age, $\nabla \langle \lg (t/{\rm yr}) \rangle$, metallicity, $\nabla\langle [Z/H]\rangle$, and stellar velocity dispersion, $\nabla\sigma_{\star}$, as plotted in Fig.~\ref{fig9}. Most of our elliptical galaxies show mildly negative age and metallicity gradients (see top panels of Fig.~\ref{fig9}), with a median value of $\nabla \langle \lg (t/{\rm yr}) \rangle$ and $\nabla\langle [Z/H]\rangle$ are $-0.07$ and $-0.08$, respectively, consistent with previous studies of elliptical galaxies \citep{Sanchez2020ARA&A..58...99S, Gonzalez2015A&A...581A.103G, Li2015ApJ...804..125L, Zheng2017MNRAS.465.4572Z}, supporting an inside-out star-forming scenario in which galaxies assemble their stellar mass first in the center subsequently extend to the outer regions.

\begin{deluxetable*}{ccccccc}
\tablenum{6}
\tablewidth{0pt}
\tabletypesize{\footnotesize}
\tablecaption{Spearman Correlation Coefficients.\label{tab6}}
\tablehead{
\colhead{Sample (Num)} & \multicolumn{1}{c}{$\nabla\langle \lg (t/{\rm yr}) \rangle$-$\nabla r_{\rm b}$} & \multicolumn{1}{c}{$\nabla\langle \lg (t/{\rm yr}) \rangle$-$\nabla r_{\rm b,\,sub}$} & \multicolumn{1}{c}{$\nabla\langle [Z/H]\rangle$-$\nabla r_{\rm b}$} & \multicolumn{1}{c}{$\nabla\langle [Z/H]\rangle$-$\nabla r_{\rm b,\,sub}$} & \multicolumn{1}{c}{$\nabla\sigma_{\star}$-$\nabla r_{\rm b}$} & \multicolumn{1}{c}{$\nabla\sigma_{\star}$-$\nabla r_{\rm b,\,sub}$}\\
\colhead{} & \multicolumn{1}{c}{$\rho$ ($p$-value)} & \multicolumn{1}{c}{$\rho$ ($p$-value)} & \multicolumn{1}{c}{$\rho$ ($p$-value)} & \multicolumn{1}{c}{$\rho$ ($p$-value)} & \multicolumn{1}{c}{$\rho$ ($p$-value)} & \multicolumn{1}{c}{$\rho$ ($p$-value)}}
\startdata
All (513) & -0.26 (1.1E-09) & 0.16 (3.9E-04) & -0.18 (4.3E-05) & 0.23 (8.0E-08) & 0.09 (5.4E-02) & 0.06 (2.1E-01)\\
UV (57) & -0.09 (5.1E-01) & 0.27 (4.1E-02) & -0.37 (4.3E-03) & 0.06 (6.3E-01) & 0.11 (4.0E-01) & 0.13 (3.5E-01)\\
Non-UV (456) & -0.28 (1.4E-09) & 0.15 (1.2E-03) & -0.16 (7.2E-04) & 0.26 (2.7E-08) & 0.08 (8.1E-02) & 0.05 (3.0E-01)\\
$\langle \lg (t/{\rm yr}) \rangle_{R_e} \leqslant 9.5$ (34)& -0.74 (5.5E-07) & 0.47 (5.6E-03) & 0.20 (2.7E-01) & 0.17 (3.4E-01) & 0.16 (3.7E-01) & -0.12 (4.9E-01)\\
\enddata
\end{deluxetable*}

\textit{Correlation with age gradient.} In conjunction with the results from Section~\ref{sec:Binary Fraction Profiles}, as illustrated in the left column of Fig.~\ref{fig9}, most (71\%) galaxies fall in the upper-left ($\nabla \langle \lg (t/{\rm yr}) \rangle < 0$, $\nabla r_{\rm b} > 0$) region, indicating that their binary fractions tend to increase outward as the SP becomes younger. The SP-subtracted binary fraction of nearly half (52\%) of the galaxies increased with the larger stellar age, as they fall in the lower-left ($\nabla \langle \lg (t/{\rm yr}) \rangle<0$, $\nabla r_{\rm b,\,sub}<0$) region, nonetheless, about 32\% of galaxies are still located in the ($\nabla \langle \lg (t/{\rm yr}) \rangle<0$, $\nabla r_{\rm b,\,sub}>0$) region. Overall, neither the $\nabla r_{\rm b}$ nor $\nabla r_{\rm b,\,sub}$ shows a clear correlation with $\nabla \langle \lg (t/{\rm yr}) \rangle$.

However, we noted that some (34) relatively younger ($\langle \lg (t/{\rm yr}) \rangle_{R_e} \leqslant 9.5$, with deepest colored) galaxies, most (74\%) of which also have a positive age gradients, show the strongest Spearman correlation coefficients of $-0.74$ and 0.47 for $\nabla r_{\rm b}$ and $\nabla r_{\rm b,\,sub}$, respectively (see the second and third columns of Table~\ref{tab6}), suggesting that the binary fraction in younger ($\langle \lg (t/{\rm yr}) \rangle_{R_e} \leqslant 9.5$) galaxies is more susceptible to stellar age.

\textit{Correlation with metallicity gradient.} As the middle column of Fig.~\ref{fig9} shows, the binary fraction of most (77\%) galaxies increased with the lower metallicity, as they fall in the ($\nabla\langle [Z/H]\rangle<0$, $\nabla r_{\rm b}>0$) region. In contrast, the SP-subtracted binary fraction of 59\% galaxies decreased with the lower metallicity, as they fall in the ($\nabla\langle [Z/H]\rangle<0$, $\nabla r_{\rm b,\,sub}<0$) region, while about 36\% of galaxies remain in the ($\nabla\langle [Z/H]\rangle<0$, $\nabla r_{\rm b,\,sub}>0$) region. Neither the $\nabla r_{\rm b}$ nor $\nabla r_{\rm b,\,sub}$ shows a clear correlation with the metallicity gradient.

\textit{Correlation with velocity dispersion gradient.} While no correlation has been established between stellar velocity dispersion and the binary fraction, dynamical interaction could be another factor affecting the binary fraction and its spatial distribution. As the right column of Fig.~\ref{fig9} shows, most (86\%) of our elliptical galaxies also show a negative velocity dispersion gradient, with a median value of $\nabla\sigma_{\star}$ is about $-30$. The binary fraction of 69\% galaxies increased with the lower velocity dispersion, as they fall in the ($\nabla\sigma_{\star}<0$, $\nabla r_{\rm b}>0$) region. The SP-subtracted binary fraction of 53\% galaxies decreased with lower velocity dispersion, as they fall in the ($\nabla\sigma_{\star}<0$, $\nabla r_{\rm b,\,sub}<0$) region. The correlation between $\nabla\sigma_{\star}$ and the $\nabla r_{\rm b}$ ($\nabla r_{\rm b,\,sub}$) is the weakest with a Spearman correlation coefficient of 0.09 (0.06) for all galaxies, as listed in Table~\ref{tab6}. 

The lack of correlation between velocity dispersion and binary fraction suggests that dynamical interactions may not be the dominant factor governing the binary fraction and its radial distribution in elliptical galaxies. This could be because the process, such as capture or disruption of binaries by dynamical interactions, is complex and long-term, and it may be more effective in local, dense star environments (e.g., the cores of GCs). Given that our results represent galaxy-scale averages, any local dynamical effects may be diluted, making it difficult to identify a clear correlation with velocity dispersion.

The UV subsample is also marked by magenta circles in Fig.~\ref{fig9}. No significant gradient correlations are found for either the UV or the non-UV subsample, with the correlation coefficients listed in Table~\ref{tab6}. A further analysis based on the Yunnan II-b model also leads to the same conclusion as above.

Although some galaxies in our sample may show a potential correlation between binary fraction and SP properties, as they reside in the relevant gradient regions, no significant correlations are found between the binary fraction gradient and the gradients of stellar population age, metallicity, or stellar velocity dispersion. Nonetheless, this does not necessarily mean that the binary fraction is unrelated to stellar age or metallicity. In fact, the systematically higher correlation coefficients of $\nabla r_{\rm b}$ compared to $\nabla r_{\rm b,\,sub}$, as listed in Table~\ref{tab6}, along with their significant differences observed in younger galaxies ($\langle \lg (t/{\rm yr}) \rangle_{R_e} \leqslant 9.5$) are consistent with the conclusions in Section~\ref{sec:Binary Fraction Profiles}, suggesting that SP properties---especially stellar age---may play an important role in shaping the binary fraction. This result is also consistent with the theoretical expectation that older stellar populations are dominated by long-lived, low-mass stars, which tend to show lower binary fractions compared to short-lived, high-mass stars.

The weak correlations between the binary fraction gradient and SP gradients may be attributed to the relatively small ($<2\%$) radial variation in the derived binary fraction of our sample, compared to the significant ($>10\%$) changes where potential correlations with stellar age or metallicity can be observed \citep{Ji2015ApJ...807...32J, Moe2019ApJ...875...61M}. This may imply that the dependence of binary fraction on stellar age or metallicity may not follow a simple linear correlation, and UV and non-UV upturn galaxies may also share a similar dynamical evolution process concerning binary distribution.

\begin{figure}[tb!]
\centering
\includegraphics[width=0.35\textwidth]{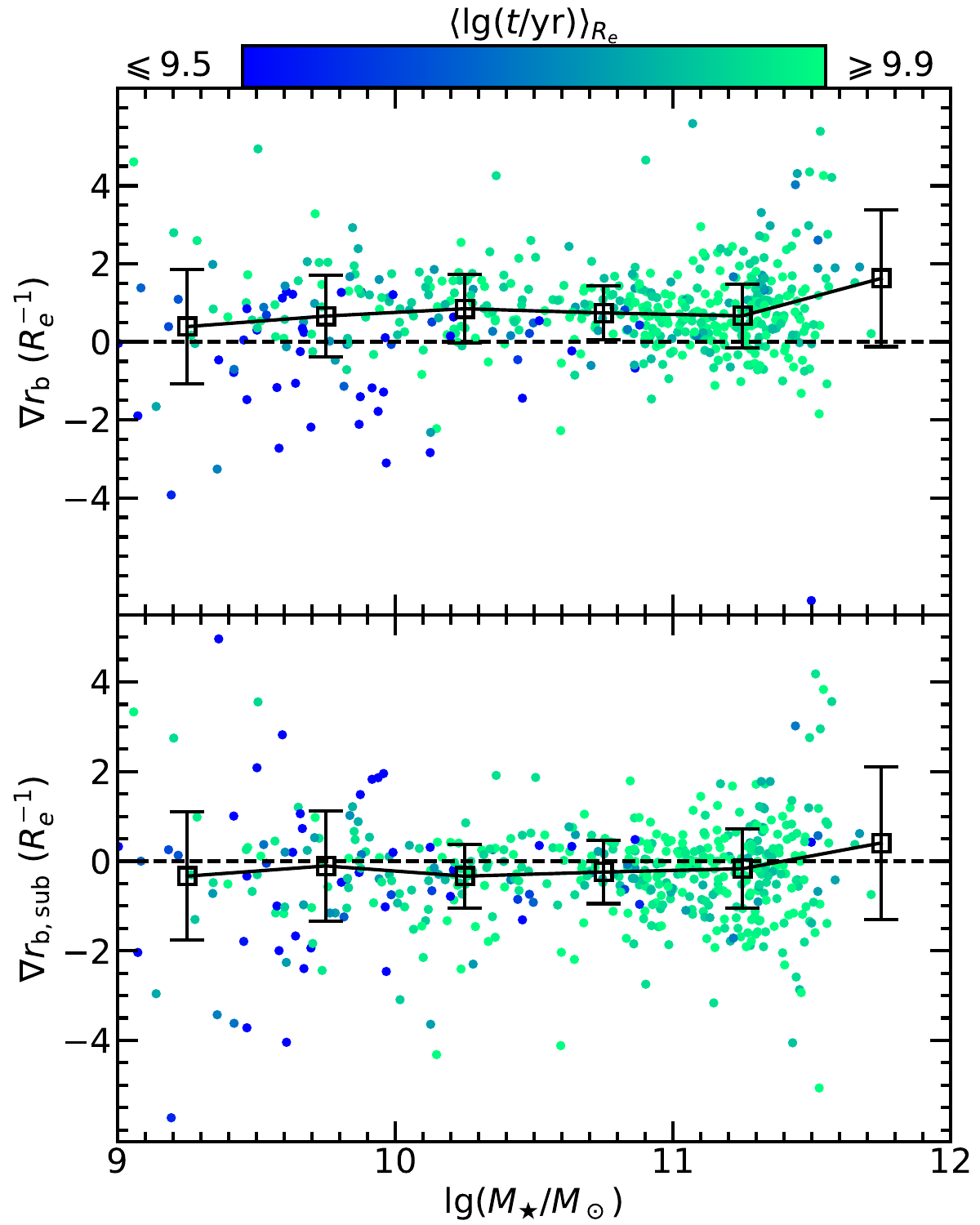}
\caption{The gradients of binary fraction as a function of stellar mass of galaxy for all 513 elliptical galaxies. Galaxies are colored by the corresponding mean stellar age derived from pure-continuum fitting within $1R_e$. The median value at each mass bin is shown as an open black square with error bars giving the corresponding dispersion.
\label{fig10}}
\end{figure}

\textit{Correlation with stellar mass of galaxies.} We also examine possible dependencies between binary fraction gradients and stellar masses of galaxies, using values from the NSA catalog with a Hubble constant of $H_0 \rm =73\,km\,s^{-1}Mpc^{-1}$, as mass is one of the most fundamental qualities in stellar and galaxy evolution, and galaxies of different masses exhibit different gradients in both SP properties and IMF index. Both \citet{Parikh2018MNRAS.477.3954P} and \citet{Zhou2019MNRAS.485.5256Z} found a negative IMF index gradient in massive ($\gtrsim 10^{10.5} M_{\odot}$) galaxies, and weaker/no gradient in low mass galaxies, which means for the massive galaxies (more top-heavy toward the outskirts), their binary fraction gradients might be more positive, while for the low mass galaxies (more bottom-heavy toward the outskirts), their binary fraction gradients might be flatter or more negative.

As shown in Fig.~\ref{fig10}, both the gradients of the binary fraction and SP-subtracted binary fraction seem to increase slightly with stellar mass of galaxy at the high-mass ($>10^{11} M_{\odot}$) end, although no significant correlation is found. It is noteworthy that most of the younger ($\langle \lg (t/{\rm yr}) \rangle_{R_e}\leqslant9.5$) galaxies are predominantly low-mass systems, and the binary fraction gradients seem to be more dispersed toward lower and higher mass sides.

For low mass galaxies, their binary fraction may be more sensitive to the SP properties (as discussed earlier), and they may still undergo mass growth. Processes such as gas inflow/outflow, stellar feedback (e.g., supernovae) or radial migration, and galaxy mergers could influence the formation and spatial distribution of binaries, leading to a wide diversity of binary fraction profiles. In contrast, the larger dispersion observed in massive galaxies is more likely driven by long-term intrinsic dynamical evolution. These factors might also explain the lack of gradient correlation in the aforementioned.

\subsection{Comparison of Different EPS Models}\label{subsec:Comparison of Different EPS Models}

As discussed in Section~\ref{sec:Binary Fraction Profiles}, the derived binary fraction profile could be affected by the radial variation of SP, which depends further on the adopted EPS models (see Section~\ref{subsec:Comparison in SP Properties}), the SP-subtracted binary fraction could also depend on the specific model. Therefore, we compare the SP-subtracted binary fraction profiles derived with the BPASS and Yunnan II models.

\begin{deluxetable}{ccccccc}
\tablenum{7}
\tablewidth{0pt}
\tablecaption{The Median Values and Dispersion of the SP-subtracted Binary Fraction Gradients and Binary Fraction Variations at $1R_e$ for Different EPS Models.
\label{tab7}}
\tablehead{\colhead{Model} & \colhead{$\nabla r_{\rm b,\,sub}^{\rm med}$} & \colhead{$r_{{\rm b}1R_e{\rm ,\,sub}}^{\rm med}$}}
\startdata
GALAXEV & -0.23±0.83 & -0.09±0.73 \\
Yunnan II-s & -0.19±1.33 & -0.09±1.19 \\
Yunnan II-b & -0.14±1.30 & -0.16±1.12 \\
BPASS-s & -0.71±1.03 & -0.65±0.82 \\
BPASS-b & -0.72±0.97 & -0.62±0.88
\enddata
\tablecomments{The suffix '-s' or '-b' in model names represents the EPS model without and with binaries.}
\end{deluxetable}

\begin{figure*}[htb!]
\plotone{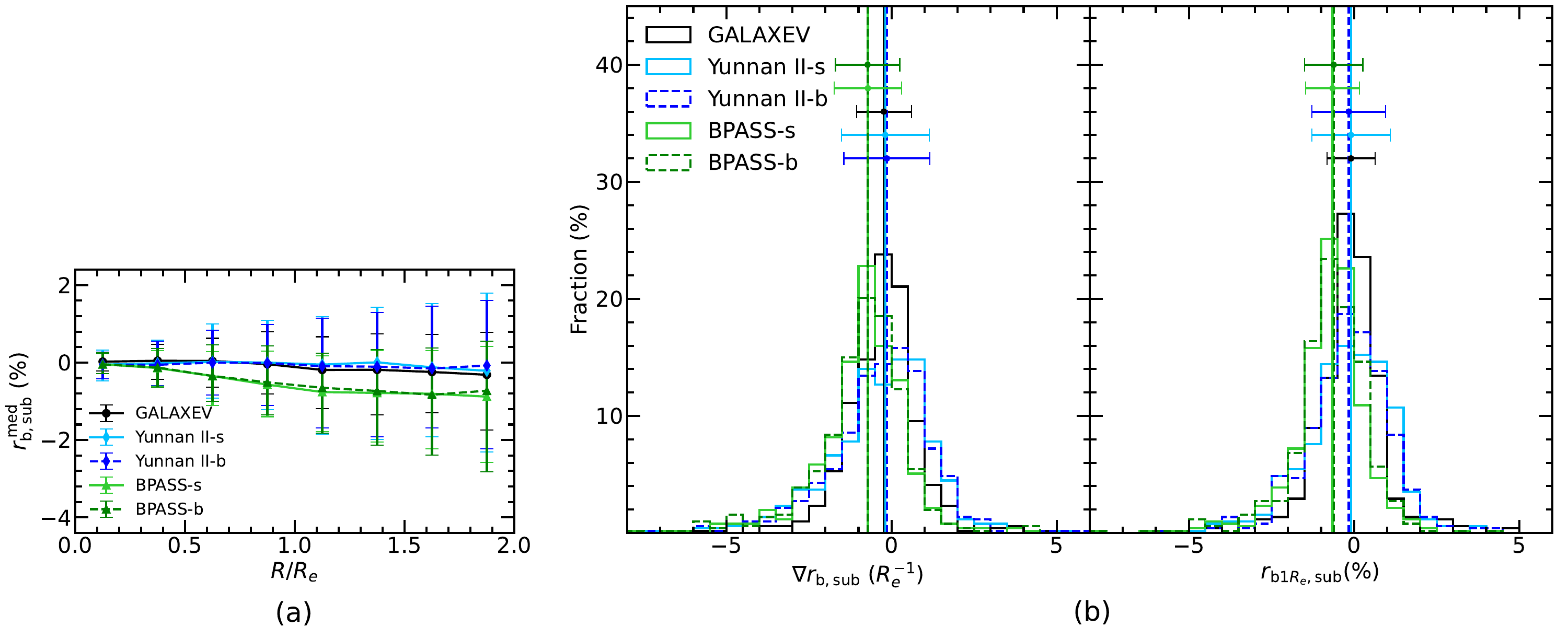}
\caption{Same as Fig.~\ref{fig4}, with results from the GALAXEV (black), Yunnan II-s (light blue), Yunnan II-b (blue), BPASS-s (green), and BPASS-b (dark green) EPS models are plotted for comparison.
\label{fig11}}
\end{figure*}

Results of median radial profile $r_{\rm b,\,sub}^{\rm med}$ for different models are plotted in Fig.~\ref{fig11}(a). For the Yunnan II EPS model without binaries (Yunnan II-s), the $r_{\rm b,\,sub}^{\rm med}$ is consistent with that of the GALAXEV, showing an almost flat radial profile. The $\nabla r_{\rm b,\,sub}^{\rm med}$ and $r_{{\rm b}1R_e{\rm ,\,sub}}^{\rm med}$ of the Yunnan II-s are $-0.19\%$ and $-0.09\%$, respectively, also close to those of the GALAXEV (see the second and third columns of Table~\ref{tab7}), but with a slightly larger dispersion as illustrated in Fig.~\ref{fig11}(b).
For the BPASS EPS model without binaries (BPASS-s), the $r_{\rm b,\,sub}^{\rm med}$ appears to decrease with radius and reaches about $-1\%$ at $2R_e$. The distributions of both $\nabla r_{\rm b,\,sub}$ and $r_{{\rm b}1R_e{\rm ,\,sub}}$ are systematically lower by $\sim0.5\%$ than the GALAXEV, with $\nabla r_{\rm b,\,sub}^{\rm med}$ and $r_{{\rm b}1R_e{\rm ,\,sub}}^{\rm med}$ are $-0.71\%$ and $-0.65\%$, respectively, and more than 78\% of galaxies have $\nabla r_{\rm b,\,sub}<0$, in contrast to 55\% for the Yunnan II-s and 61\% for GALAXEV. 

\begin{figure}[htb!]
\centering
\includegraphics[width=0.35\textwidth]{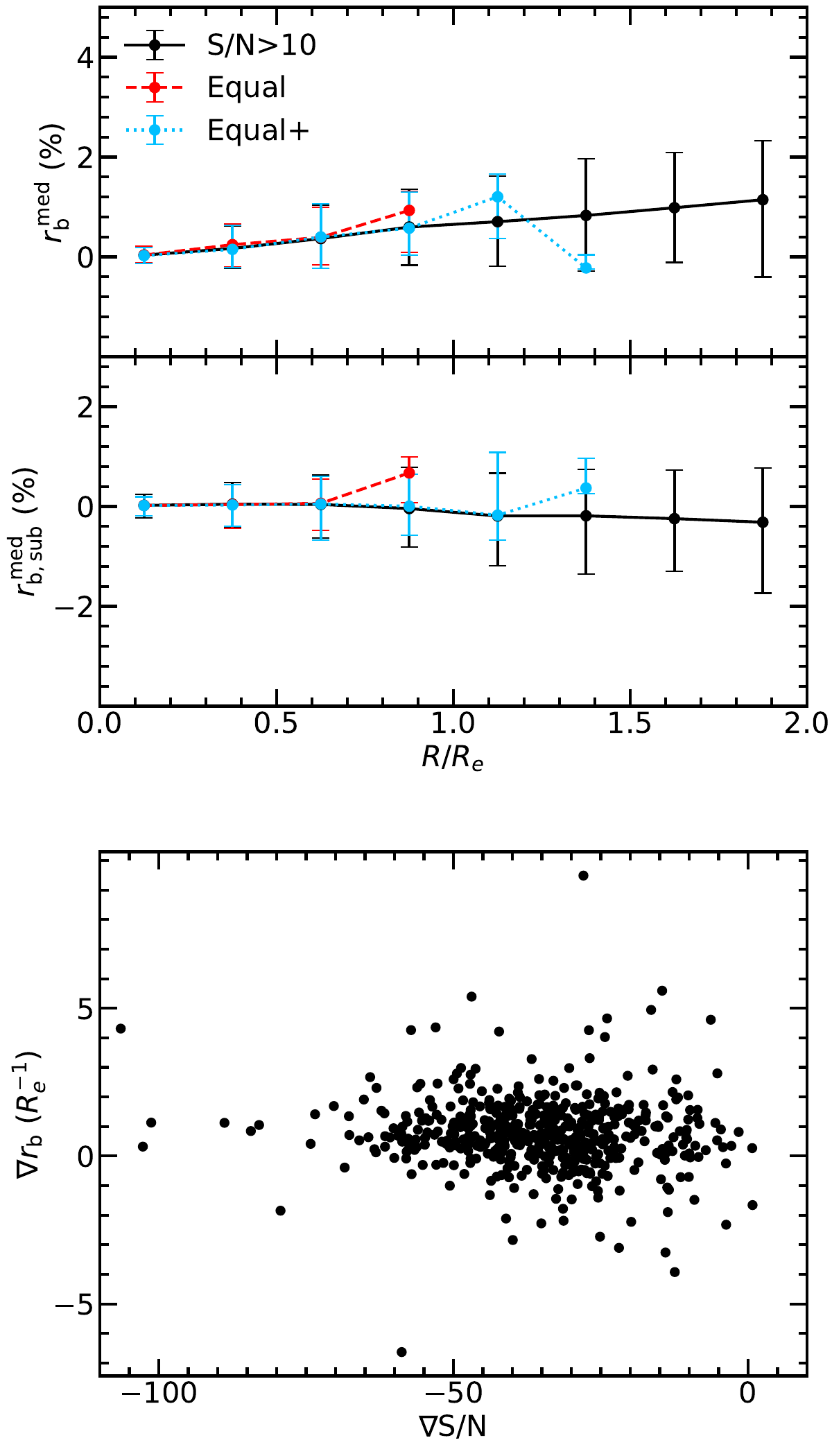}
\caption{Top panel: Radial profiles of the binary fraction (upper) and SP-subtracted binary fraction (lower) for subsamples using different S/N stacking methods. Bottom panel: Distribution of binary fraction gradient as a function of S/N gradient.
\label{fig12}}
\end{figure}

Radial profiles derived from the EPS model with binaries of BPASS (BPASS-b) and Yunnan II (Yunnan II-b) almost agree with their corresponding model without binaries, as can be seen in Fig.~\ref{fig11}(a). The distributions of $\nabla r_{\rm b,\,sub}$ and $r_{{\rm b}1R_e{\rm ,\,sub}}$ are also similar to corresponded model without binaries, with the $\nabla r_{\rm b,\,sub}^{\rm med}$ and $r_{{\rm b}1R_e{\rm ,\,sub}}^{\rm med}$ are $-0.14\%$ and $-0.16\%$, respectively, for the Yunnan II-b, and are $-0.72\%$ and $-0.62\%$, respectively, for the BPASS-b, as listed in the Table~\ref{tab7}. 

For a given EPS model set, the results obtained from models including binaries are consistent with those derived from the model without binaries. The major discrepancies only emerge when different EPS model sets are compared.
Overall, the SP-subtracted binary fraction profiles show some dependence on the adopted EPS model, but these minor ($<1\%$) differences do not significantly affect our main conclusions. Fitting using a linear combination of single and binary star templates could provide an alternative way to probe the binary fraction. However, given that the primary differences between single and binary star templates are most prominent in the ultraviolet, as shown in Fig. 1 of \citet{Zhang2020IAUS..341...35Z}, the results fitting with linear combinations of these templates would likely lie between the single and binary star templates, and might introduce additional degeneracies without additional constraints on the binary fraction.

\subsection{Effect of the S/N}\label{subsec:Effect of the S/N}

The S/N ratio of the stacked spectra in each annulus is greater than 10 but not uniform. The inner annulus generally has a higher S/N than the outer one. To assess whether the variation in S/N affects the derived binary fraction radial profiles, we adjust the width of each elliptical annulus to ensure that the S/N of the stacked spectra remains consistent across all annuli within the same galaxy. Since the increment of annulus width, the number of annuli decreases, therefore, we define two subsamples. The first subsample, known as the Equal subsample, includes 201 galaxies with more than three annuli and follows the same method outlined in Section~\ref{subsec:Measurement of Binary Fraction Profiles}. To preserve a larger dataset, any galaxy with more than one annulus is included in the Equal+ subsample, which contains 512 galaxies, and their outermost annulus is also kept in the analysis. 

As the top panel of Fig.~\ref{fig12} shows, both the Equal and Equal+ subsamples have a shorter radius coverage due to the decrease in the number of annuli. The median radial profiles of $r_{\rm b}^{\rm med}$ and $r_{\rm b,\,sub}^{\rm med}$ for Equal and Equal+ subsamples are consistent basically with the initial (S/N$>10$) result. The significant variation in the last bin can be attributed to the relatively smaller number of annuli and wider region stacked by the last annulus, especially for the Equal+ subsample. We also plot the distribution of the $\nabla r_{\rm b}$ as a function of the signal-to-noise ratio gradient $\nabla$S/N in the bottom panel of Fig.~\ref{fig12}. No significant correlation is found, with a Spearman correlation coefficient of $-0.11$ and a $p$-value of 0.02. Overall, the radial variation of S/N may not affect the radial profiles of derived binary fraction within $1R_e$ for our sample.

\section{Summary} \label{sec:Summary}

In this paper, we present detailed results on the radial profile of the binary fraction, as traced by the variation of binary fraction-sensitive SAFIs, for a sample of 513 elliptical galaxies using the IFU spectra from MaNGA DR17 and the full-spectrum fitting code pPXF. We compare the binary fraction profiles and the SP properties of UV upturn and non-UV upturn galaxies as derived from different EPS models, and investigate the possible connection between the binary fraction gradient and gradients of SP properties. Our main results can be summarized as follows:

1. The binary fraction derived from SAFIs increases slightly with the radius for our elliptical galaxy sample, with a median value of about 0.62\%$\pm$0.79\% at $1R_e$ relative to the galaxy center, which is not statistically significant, and almost all galaxies have their binary fraction variation less than 5\% at $1R_e$. After considering the profile induced by the radial SP variation, the SP-subtracted binary fraction profile for the whole sample is flattened, but there are still 39\% of galaxies exhibiting a radially increasing profile. Results derived from different EPS models are also similar.

2. About 11\% of our sample is classified as UV upturn galaxies. The EPS model with binaries appears to be more suitable for the SP analysis of UV upturn galaxies, as their SP properties derived from photometry-included fitting agree better with those of pure-continuum fitting than those inferred from the model without binaries. Within the current measurement uncertainties, our results do not show a statistically significant difference in binary fraction profiles and SP properties between our UV and non-UV subsamples. The indistinguishable SP properties between our UV and non-UV subsamples could be attributed to fitting uncertainties, age-metallicity degeneracy, and possible residual star formation. An improved sample data and broader wavelength coverage would be required to further test this conclusion.

3. No clear correlation is found between the binary fraction gradient and the gradients of mean stellar age, metallicity, or stellar velocity dispersion. However, the significant difference between binary fraction gradient and SP-subtracted binary fraction gradient in younger ($\leqslant 10^{9.5}$ yr) galaxies suggests that stellar age may be one of the key factors influencing the derived binary fraction profiles, particularly for younger systems.

One advantage of our spectral indices-based method used in this work to infer variations in the binary fraction is that it doesn't require resolved photometric data of individual stars, and can therefore be applied to more distant or unresolved stellar systems. However, this approach also has several limitations. It can not provide absolute binary fractions, but only relative variations. The measurement of spectral indices can be affected by many factors, such as contamination from emission lines, the age-metallicity degeneracy, and the dependency of stellar population templates if the contribution from stellar population is considered. Therefore, this method may be more applicable to systems with relatively simple populations and features, such as star clusters or local regions within galaxies.


\section{Acknowledgments}

We thank the anonymous referee for careful reading and thoughtful comments which improved the paper. The work is supported by the National Natural Science Foundation of China (NSFC; grant Nos. 12573039), the basic research program of Yunnan Province (No. 202401AT070142), and the Yunnan Revitalization Talent Support Program.

All the authors acknowledge the work of the Sloan Digital Sky Survey (SDSS) team. Funding for the Sloan Digital Sky Survey IV has been provided by the Alfred P. Sloan Foundation, the U.S. Department of Energy Office of Science, and the Participating Institutions. SDSS acknowledges support and resources from the Center for High-Performance Computing at the University of Utah. The SDSS website is www.sdss4.org.

SDSS is managed by the Astrophysical Research Consortium for the Participating Institutions of the SDSS Collaboration including the Brazilian Participation Group, the Carnegie Institution for Science, Carnegie Mellon University, Center for Astrophysics | Harvard \& Smithsonian (CfA), the Chilean Participation Group, the French Participation Group, Instituto de Astrofísica de Canarias, The Johns Hopkins University, Kavli Institute for the Physics and Mathematics of the Universe (IPMU) / University of Tokyo, the Korean Participation Group, Lawrence Berkeley National Laboratory, Leibniz Institut für Astrophysik Potsdam (AIP), Max-Planck-Institut für Astronomie (MPIA Heidelberg), Max-Planck-Institut für Astrophysik (MPA Garching), Max-Planck-Institut für Extraterrestrische Physik (MPE), National Astronomical Observatories of China, New Mexico State University, New York University, University of Notre Dame, Observatório Nacional / MCTI, The Ohio State University, Pennsylvania State University, Shanghai Astronomical Observatory, United Kingdom Participation Group, Universidad Nacional Autónoma de México, University of Arizona, University of Colorado Boulder, University of Oxford, University of Portsmouth, University of Utah, University of Virginia, University of Washington, University of Wisconsin, Vanderbilt University, and Yale University.


\bibliography{sample631}{}

@ARTICLE{Vazquez-Mata2022MNRAS.512.2222V,
       author = {{V{\'a}zquez-Mata}, J.~A. and {Hern{\'a}ndez-Toledo}, H.~M. and {Avila-Reese}, V. and {Herrera-Endoqui}, M. and {Rodr{\'\i}guez-Puebla}, A. and {Cano-D{\'\i}az}, M. and {Lacerna}, I. and {Mart{\'\i}nez-V{\'a}zquez}, L.~A. and {Lane}, R.},
        title = "{SDSS IV MaNGA: visual morphological and statistical characterization of the DR15 sample}",
      journal = {\mnras},
         year = 2022,
        month = may,
       volume = {512},
       number = {2},
        pages = {2222-2244},
          doi = {10.1093/mnras/stac635},
archivePrefix = {arXiv},
       eprint = {2203.02565},
 primaryClass = {astro-ph.GA},
       adsurl = {https://ui.adsabs.harvard.edu/abs/2022MNRAS.512.2222V}
}

@ARTICLE{Blanton2011AJ....142...31B,
       author = {{Blanton}, Michael R. and {Kazin}, Eyal and {Muna}, Demitri and {Weaver}, Benjamin A. and {Price-Whelan}, Adrian},
        title = "{Improved Background Subtraction for the Sloan Digital Sky Survey Images}",
      journal = {\aj},
         year = 2011,
        month = jul,
       volume = {142},
       number = {1},
          eid = {31},
        pages = {31},
          doi = {10.1088/0004-6256/142/1/31},
archivePrefix = {arXiv},
       eprint = {1105.1960},
 primaryClass = {astro-ph.IM},
       adsurl = {https://ui.adsabs.harvard.edu/abs/2011AJ....142...31B}
}

@ARTICLE{Yan2016AJ....152..197Y,
       author = {{Yan}, Renbin and {Bundy}, Kevin and {Law}, David R. and {Bershady}, Matthew A. and {Andrews}, Brett and {Cherinka}, Brian and {Diamond-Stanic}, Aleksandar M. and {Drory}, Niv and {MacDonald}, Nicholas and {S{\'a}nchez-Gallego}, Jos{\'e} R. and {Thomas}, Daniel and {Wake}, David A. and {Weijmans}, Anne-Marie and {Westfall}, Kyle B. and {Zhang}, Kai and {Arag{\'o}n-Salamanca}, Alfonso and {Belfiore}, Francesco and {Bizyaev}, Dmitry and {Blanc}, Guillermo A. and {Blanton}, Michael R. and {Brownstein}, Joel and {Cappellari}, Michele and {D'Souza}, Richard and {Emsellem}, Eric and {Fu}, Hai and {Gaulme}, Patrick and {Graham}, Mark T. and {Goddard}, Daniel and {Gunn}, James E. and {Harding}, Paul and {Jones}, Amy and {Kinemuchi}, Karen and {Li}, Cheng and {Li}, Hongyu and {Maiolino}, Roberto and {Mao}, Shude and {Maraston}, Claudia and {Masters}, Karen and {Merrifield}, Michael R. and {Oravetz}, Daniel and {Pan}, Kaike and {Parejko}, John K. and {Sanchez}, Sebastian F. and {Schlegel}, David and {Simmons}, Audrey and {Thanjavur}, Karun and {Tinker}, Jeremy and {Tremonti}, Christy and {van den Bosch}, Remco and {Zheng}, Zheng},
        title = "{SDSS-IV MaNGA IFS Galaxy Survey{\textemdash}Survey Design, Execution, and Initial Data Quality}",
      journal = {\aj},
         year = 2016,
        month = dec,
       volume = {152},
       number = {6},
          eid = {197},
        pages = {197},
          doi = {10.3847/0004-6256/152/6/197},
archivePrefix = {arXiv},
       eprint = {1607.08613},
 primaryClass = {astro-ph.GA},
       adsurl = {https://ui.adsabs.harvard.edu/abs/2016AJ....152..197Y}
}

@ARTICLE{Bundy2015ApJ...798....7B,
       author = {{Bundy}, Kevin and {Bershady}, Matthew A. and {Law}, David R. and {Yan}, Renbin and {Drory}, Niv and {MacDonald}, Nicholas and {Wake}, David A. and {Cherinka}, Brian and {S{\'a}nchez-Gallego}, Jos{\'e} R. and {Weijmans}, Anne-Marie and {Thomas}, Daniel and {Tremonti}, Christy and {Masters}, Karen and {Coccato}, Lodovico and {Diamond-Stanic}, Aleksandar M. and {Arag{\'o}n-Salamanca}, Alfonso and {Avila-Reese}, Vladimir and {Badenes}, Carles and {Falc{\'o}n-Barroso}, J{\'e}sus and {Belfiore}, Francesco and {Bizyaev}, Dmitry and {Blanc}, Guillermo A. and {Bland-Hawthorn}, Joss and {Blanton}, Michael R. and {Brownstein}, Joel R. and {Byler}, Nell and {Cappellari}, Michele and {Conroy}, Charlie and {Dutton}, Aaron A. and {Emsellem}, Eric and {Etherington}, James and {Frinchaboy}, Peter M. and {Fu}, Hai and {Gunn}, James E. and {Harding}, Paul and {Johnston}, Evelyn J. and {Kauffmann}, Guinevere and {Kinemuchi}, Karen and {Klaene}, Mark A. and {Knapen}, Johan H. and {Leauthaud}, Alexie and {Li}, Cheng and {Lin}, Lihwai and {Maiolino}, Roberto and {Malanushenko}, Viktor and {Malanushenko}, Elena and {Mao}, Shude and {Maraston}, Claudia and {McDermid}, Richard M. and {Merrifield}, Michael R. and {Nichol}, Robert C. and {Oravetz}, Daniel and {Pan}, Kaike and {Parejko}, John K. and {Sanchez}, Sebastian F. and {Schlegel}, David and {Simmons}, Audrey and {Steele}, Oliver and {Steinmetz}, Matthias and {Thanjavur}, Karun and {Thompson}, Benjamin A. and {Tinker}, Jeremy L. and {van den Bosch}, Remco C.~E. and {Westfall}, Kyle B. and {Wilkinson}, David and {Wright}, Shelley and {Xiao}, Ting and {Zhang}, Kai},
        title = "{Overview of the SDSS-IV MaNGA Survey: Mapping nearby Galaxies at Apache Point Observatory}",
      journal = {\apj},
         year = 2015,
        month = jan,
       volume = {798},
       number = {1},
          eid = {7},
        pages = {7},
          doi = {10.1088/0004-637X/798/1/7},
archivePrefix = {arXiv},
       eprint = {1412.1482},
 primaryClass = {astro-ph.GA},
       adsurl = {https://ui.adsabs.harvard.edu/abs/2015ApJ...798....7B}
}

@article{Kewley2001ApJ...556..121K,
   author = {Kewley, L. J. and Dopita, M. A. and Sutherland, R. S. and Heisler, C. A. and Trevena, J.},
   title = {Theoretical Modeling of Starburst Galaxies},
   journal = {The Astrophysical Journal},
   volume = {556},
   pages = {121-140},
   ISSN = {0004-637X},
   DOI = {10.1086/321545},
   url = {https://ui.adsabs.harvard.edu/abs/2001ApJ...556..121K},
   year = {2001},
   type = {Journal Article}
}

@ARTICLE{Yi2011ApJS..195...22Y,
       author = {{Yi}, Sukyoung K. and {Lee}, Jihye and {Sheen}, Yun-Kyeong and {Jeong}, Hyunjin and {Suh}, Hyewon and {Oh}, Kyuseok},
        title = "{The Ultraviolet Upturn in Elliptical Galaxies and Environmental Effects}",
      journal = {\apjs},
         year = 2011,
        month = aug,
       volume = {195},
       number = {2},
          eid = {22},
        pages = {22},
          doi = {10.1088/0067-0049/195/2/22},
archivePrefix = {arXiv},
       eprint = {1107.0005},
 primaryClass = {astro-ph.GA},
       adsurl = {https://ui.adsabs.harvard.edu/abs/2011ApJS..195...22Y}
}

@ARTICLE{Cappellari2004PASP..116..138C,
       author = {{Cappellari}, Michele and {Emsellem}, Eric},
        title = "{Parametric Recovery of Line-of-Sight Velocity Distributions from Absorption-Line Spectra of Galaxies via Penalized Likelihood}",
      journal = {\pasp},
         year = 2004,
        month = feb,
       volume = {116},
       number = {816},
        pages = {138-147},
          doi = {10.1086/381875},
archivePrefix = {arXiv},
       eprint = {astro-ph/0312201},
 primaryClass = {astro-ph},
       adsurl = {https://ui.adsabs.harvard.edu/abs/2004PASP..116..138C}
}

@ARTICLE{Cappellari2017MNRAS.466..798C,
       author = {{Cappellari}, Michele},
        title = "{Improving the full spectrum fitting method: accurate convolution with Gauss-Hermite functions}",
      journal = {\mnras},
         year = 2017,
        month = apr,
       volume = {466},
       number = {1},
        pages = {798-811},
          doi = {10.1093/mnras/stw3020},
archivePrefix = {arXiv},
       eprint = {1607.08538},
 primaryClass = {astro-ph.GA},
       adsurl = {https://ui.adsabs.harvard.edu/abs/2017MNRAS.466..798C}
}

@ARTICLE{Cappellari2023MNRAS.526.3273C,
       author = {{Cappellari}, Michele},
        title = "{Full spectrum fitting with photometry in PPXF: stellar population versus dynamical masses, non-parametric star formation history and metallicity for 3200 LEGA-C galaxies at redshift z {\ensuremath{\approx}} 0.8}",
      journal = {\mnras},
         year = 2023,
        month = dec,
       volume = {526},
       number = {3},
        pages = {3273-3300},
          doi = {10.1093/mnras/stad2597},
archivePrefix = {arXiv},
       eprint = {2208.14974},
 primaryClass = {astro-ph.GA},
       adsurl = {https://ui.adsabs.harvard.edu/abs/2023MNRAS.526.3273C}
}

@ARTICLE{Woo2024MNRAS.530.4260W,
       author = {{Woo}, Joanna and {Walters}, Dan and {Archinuk}, Finn and {Faber}, S.~M. and {Ellison}, Sara L. and {Teimoorinia}, Hossen and {Iyer}, Kartheik},
        title = "{Stellar populations with optical spectra: deep learning versus popular spectrum fitting codes}",
      journal = {\mnras},
         year = 2024,
        month = jun,
       volume = {530},
       number = {4},
        pages = {4260-4276},
          doi = {10.1093/mnras/stae1114},
archivePrefix = {arXiv},
       eprint = {2401.12300},
 primaryClass = {astro-ph.GA},
       adsurl = {https://ui.adsabs.harvard.edu/abs/2024MNRAS.530.4260W}
}

@ARTICLE{Zhang2023A&A...679A..27Z,
       author = {{Zhang}, F. and {Li}, L. and {Han}, Z. and {Gong}, X.},
        title = "{Relation between spectral indices and binary fractions in globular clusters}",
      journal = {\aap},
         year = 2023,
        month = nov,
       volume = {679},
          eid = {A27},
        pages = {A27},
          doi = {10.1051/0004-6361/202245212},
archivePrefix = {arXiv},
       eprint = {2309.05223},
 primaryClass = {astro-ph.SR},
       adsurl = {https://ui.adsabs.harvard.edu/abs/2023A&A...679A..27Z}
}

@ARTICLE{Zhang2024MNRAS.531.3468Z,
       author = {{Zhang}, F. and {Li}, L. and {Han}, Z. and {Wang}, X.},
        title = "{Determination method for binary fractions using the integrated spectrum}",
      journal = {\mnras},
         year = 2024,
        month = jul,
       volume = {531},
       number = {3},
        pages = {3468-3478},
          doi = {10.1093/mnras/stae1346},
archivePrefix = {arXiv},
       eprint = {2406.06951},
 primaryClass = {astro-ph.SR},
       adsurl = {https://ui.adsabs.harvard.edu/abs/2024MNRAS.531.3468Z}
}

@ARTICLE{Conroy2018ApJ...854..139C,
       author = {{Conroy}, Charlie and {Villaume}, Alexa and {van Dokkum}, Pieter G. and {Lind}, Karin},
        title = "{Metal-rich, Metal-poor: Updated Stellar Population Models for Old Stellar Systems}",
      journal = {\apj},
         year = 2018,
        month = feb,
       volume = {854},
       number = {2},
          eid = {139},
        pages = {139},
          doi = {10.3847/1538-4357/aaab49},
archivePrefix = {arXiv},
       eprint = {1801.10185},
 primaryClass = {astro-ph.GA},
       adsurl = {https://ui.adsabs.harvard.edu/abs/2018ApJ...854..139C}
}

@ARTICLE{Choi2016ApJ...823..102C,
       author = {{Choi}, Jieun and {Dotter}, Aaron and {Conroy}, Charlie and {Cantiello}, Matteo and {Paxton}, Bill and {Johnson}, Benjamin D.},
        title = "{Mesa Isochrones and Stellar Tracks (MIST). I. Solar-scaled Models}",
      journal = {\apj},
         year = 2016,
        month = jun,
       volume = {823},
       number = {2},
          eid = {102},
        pages = {102},
          doi = {10.3847/0004-637X/823/2/102},
archivePrefix = {arXiv},
       eprint = {1604.08592},
 primaryClass = {astro-ph.SR},
       adsurl = {https://ui.adsabs.harvard.edu/abs/2016ApJ...823..102C}
}

@ARTICLE{Eggleton1971MNRAS.151..351E,
       author = {{Eggleton}, Peter P.},
        title = "{The evolution of low mass stars}",
      journal = {\mnras},
         year = 1971,
        month = jan,
       volume = {151},
        pages = {351},
          doi = {10.1093/mnras/151.3.351},
       adsurl = {https://ui.adsabs.harvard.edu/abs/1971MNRAS.151..351E}
}

@ARTICLE{Eldridge2008MNRAS.384.1109E,
       author = {{Eldridge}, John J. and {Izzard}, Robert G. and {Tout}, Christopher A.},
        title = "{The effect of massive binaries on stellar populations and supernova progenitors}",
      journal = {\mnras},
         year = 2008,
        month = mar,
       volume = {384},
       number = {3},
        pages = {1109-1118},
          doi = {10.1111/j.1365-2966.2007.12738.x},
archivePrefix = {arXiv},
       eprint = {0711.3079},
 primaryClass = {astro-ph},
       adsurl = {https://ui.adsabs.harvard.edu/abs/2008MNRAS.384.1109E}
}

@ARTICLE{Byrne2022MNRAS.512.5329B,
       author = {{Byrne}, C.~M. and {Stanway}, E.~R. and {Eldridge}, J.~J. and {McSwiney}, L. and {Townsend}, O.~T.},
        title = "{The dependence of theoretical synthetic spectra on {\ensuremath{\alpha}}-enhancement in young, binary stellar populations}",
      journal = {\mnras},
         year = 2022,
        month = jun,
       volume = {512},
       number = {4},
        pages = {5329-5338},
          doi = {10.1093/mnras/stac807},
archivePrefix = {arXiv},
       eprint = {2203.13275},
 primaryClass = {astro-ph.SR},
       adsurl = {https://ui.adsabs.harvard.edu/abs/2022MNRAS.512.5329B}
}

@ARTICLE{Moe2017ApJS..230...15M,
       author = {{Moe}, Maxwell and {Di Stefano}, Rosanne},
        title = "{Mind Your Ps and Qs: The Interrelation between Period (P) and Mass-ratio (Q) Distributions of Binary Stars}",
      journal = {\apjs},
         year = 2017,
        month = jun,
       volume = {230},
       number = {2},
          eid = {15},
        pages = {15},
          doi = {10.3847/1538-4365/aa6fb6},
archivePrefix = {arXiv},
       eprint = {1606.05347},
 primaryClass = {astro-ph.SR},
       adsurl = {https://ui.adsabs.harvard.edu/abs/2017ApJS..230...15M}
}

@ARTICLE{Sollima2007MNRAS.380..781S,
       author = {{Sollima}, A. and {Beccari}, G. and {Ferraro}, F.~R. and {Fusi Pecci}, F. and {Sarajedini}, A.},
        title = "{The fraction of binary systems in the core of 13 low-density Galactic globular clusters}",
      journal = {\mnras},
         year = 2007,
        month = sep,
       volume = {380},
       number = {2},
        pages = {781-791},
          doi = {10.1111/j.1365-2966.2007.12116.x},
archivePrefix = {arXiv},
       eprint = {0706.2288},
 primaryClass = {astro-ph},
       adsurl = {https://ui.adsabs.harvard.edu/abs/2007MNRAS.380..781S}
}

@ARTICLE{Ji2015ApJ...807...32J,
       author = {{Ji}, Jun and {Bregman}, Joel N.},
        title = "{Binary Frequencies in a Sample of Globular Clusters. II. Sample Analysis and Comparison to Models}",
      journal = {\apj},
         year = 2015,
        month = jul,
       volume = {807},
       number = {1},
          eid = {32},
        pages = {32},
          doi = {10.1088/0004-637X/807/1/32},
       adsurl = {https://ui.adsabs.harvard.edu/abs/2015ApJ...807...32J}
}

@ARTICLE{Duch2013ARA&A..51..269D,
       author = {{Duch{\^e}ne}, Gaspard and {Kraus}, Adam},
        title = "{Stellar Multiplicity}",
      journal = {\araa},
         year = 2013,
        month = aug,
       volume = {51},
       number = {1},
        pages = {269-310},
          doi = {10.1146/annurev-astro-081710-102602},
archivePrefix = {arXiv},
       eprint = {1303.3028},
 primaryClass = {astro-ph.SR},
       adsurl = {https://ui.adsabs.harvard.edu/abs/2013ARA&A..51..269D}
}

@ARTICLE{Gao2014ApJ...788L..37G,
       author = {{Gao}, Shuang and {Liu}, Chao and {Zhang}, Xiaobin and {Justham}, Stephen and {Deng}, Licai and {Yang}, Ming},
        title = "{The Binarity of Milky Way F,G,K Stars as a Function of Effective Temperature and Metallicity}",
      journal = {\apjl},
         year = 2014,
        month = jun,
       volume = {788},
       number = {2},
          eid = {L37},
        pages = {L37},
          doi = {10.1088/2041-8205/788/2/L37},
archivePrefix = {arXiv},
       eprint = {1405.7105},
 primaryClass = {astro-ph.GA},
       adsurl = {https://ui.adsabs.harvard.edu/abs/2014ApJ...788L..37G}
}

@ARTICLE{Liu2019MNRAS.490..550L,
       author = {{Liu}, Chao},
        title = "{Smoking gun of the dynamical processing of solar-type field binary stars}",
      journal = {\mnras},
         year = 2019,
        month = nov,
       volume = {490},
       number = {1},
        pages = {550-565},
          doi = {10.1093/mnras/stz2274},
archivePrefix = {arXiv},
       eprint = {1907.02250},
 primaryClass = {astro-ph.SR},
       adsurl = {https://ui.adsabs.harvard.edu/abs/2019MNRAS.490..550L}
}

@ARTICLE{Raghavan2010ApJS..190....1R,
       author = {{Raghavan}, Deepak and {McAlister}, Harold A. and {Henry}, Todd J. and {Latham}, David W. and {Marcy}, Geoffrey W. and {Mason}, Brian D. and {Gies}, Douglas R. and {White}, Russel J. and {ten Brummelaar}, Theo A.},
        title = "{A Survey of Stellar Families: Multiplicity of Solar-type Stars}",
      journal = {\apjs},
         year = 2010,
        month = sep,
       volume = {190},
       number = {1},
        pages = {1-42},
          doi = {10.1088/0067-0049/190/1/1},
archivePrefix = {arXiv},
       eprint = {1007.0414},
 primaryClass = {astro-ph.SR},
       adsurl = {https://ui.adsabs.harvard.edu/abs/2010ApJS..190....1R}
}

@ARTICLE{Moe2019ApJ...875...61M,
       author = {{Moe}, Maxwell and {Kratter}, Kaitlin M. and {Badenes}, Carles},
        title = "{The Close Binary Fraction of Solar-type Stars Is Strongly Anticorrelated with Metallicity}",
      journal = {\apj},
         year = 2019,
        month = apr,
       volume = {875},
       number = {1},
          eid = {61},
        pages = {61},
          doi = {10.3847/1538-4357/ab0d88},
archivePrefix = {arXiv},
       eprint = {1808.02116},
 primaryClass = {astro-ph.SR},
       adsurl = {https://ui.adsabs.harvard.edu/abs/2019ApJ...875...61M}
}

@ARTICLE{Hurley2002MNRAS.329..897H,
       author = {{Hurley}, Jarrod R. and {Tout}, Christopher A. and {Pols}, Onno R.},
        title = "{Evolution of binary stars and the effect of tides on binary populations}",
      journal = {\mnras},
         year = 2002,
        month = feb,
       volume = {329},
       number = {4},
        pages = {897-928},
          doi = {10.1046/j.1365-8711.2002.05038.x},
archivePrefix = {arXiv},
       eprint = {astro-ph/0201220},
 primaryClass = {astro-ph},
       adsurl = {https://ui.adsabs.harvard.edu/abs/2002MNRAS.329..897H}
}

@ARTICLE{Milone2012A&A...540A..16M,
       author = {{Milone}, A.~P. and {Piotto}, G. and {Bedin}, L.~R. and {Aparicio}, A. and {Anderson}, J. and {Sarajedini}, A. and {Marino}, A.~F. and {Moretti}, A. and {Davies}, M.~B. and {Chaboyer}, B. and {Dotter}, A. and {Hempel}, M. and {Mar{\'\i}n-Franch}, A. and {Majewski}, S. and {Paust}, N.~E.~Q. and {Reid}, I.~N. and {Rosenberg}, A. and {Siegel}, M.},
        title = "{The ACS survey of Galactic globular clusters. XII. Photometric binaries along the main sequence}",
      journal = {\aap},
         year = 2012,
        month = apr,
       volume = {540},
          eid = {A16},
        pages = {A16},
          doi = {10.1051/0004-6361/201016384},
archivePrefix = {arXiv},
       eprint = {1111.0552},
 primaryClass = {astro-ph.SR},
       adsurl = {https://ui.adsabs.harvard.edu/abs/2012A&A...540A..16M}
}

@ARTICLE{Han2002MNRAS.336..449H,
       author = {{Han}, Z. and {Podsiadlowski}, Ph. and {Maxted}, P.~F.~L. and {Marsh}, T.~R. and {Ivanova}, N.},
        title = "{The origin of subdwarf B stars - I. The formation channels}",
      journal = {\mnras},
         year = 2002,
        month = oct,
       volume = {336},
       number = {2},
        pages = {449-466},
          doi = {10.1046/j.1365-8711.2002.05752.x},
archivePrefix = {arXiv},
       eprint = {astro-ph/0206130},
 primaryClass = {astro-ph},
       adsurl = {https://ui.adsabs.harvard.edu/abs/2002MNRAS.336..449H}
}

@ARTICLE{Han2003MNRAS.341..669H,
       author = {{Han}, Z. and {Podsiadlowski}, Ph. and {Maxted}, P.~F.~L. and {Marsh}, T.~R.},
        title = "{The origin of subdwarf B stars - II}",
      journal = {\mnras},
         year = 2003,
        month = may,
       volume = {341},
       number = {2},
        pages = {669-691},
          doi = {10.1046/j.1365-8711.2003.06451.x},
archivePrefix = {arXiv},
       eprint = {astro-ph/0301380},
 primaryClass = {astro-ph},
       adsurl = {https://ui.adsabs.harvard.edu/abs/2003MNRAS.341..669H}
}

@ARTICLE{Han2007MNRAS.380.1098H,
       author = {{Han}, Z. and {Podsiadlowski}, Ph. and {Lynas-Gray}, A.~E.},
        title = "{A binary model for the UV-upturn of elliptical galaxies}",
      journal = {\mnras},
         year = 2007,
        month = sep,
       volume = {380},
       number = {3},
        pages = {1098-1118},
          doi = {10.1111/j.1365-2966.2007.12151.x},
archivePrefix = {arXiv},
       eprint = {0704.0863},
 primaryClass = {astro-ph},
       adsurl = {https://ui.adsabs.harvard.edu/abs/2007MNRAS.380.1098H}
}

@ARTICLE{Jiang2010MNRAS.401..977J,
       author = {{Jiang}, Yan-Fei and {Tremaine}, Scott},
        title = "{The evolution of wide binary stars}",
      journal = {\mnras},
         year = 2010,
        month = jan,
       volume = {401},
       number = {2},
        pages = {977-994},
          doi = {10.1111/j.1365-2966.2009.15744.x},
archivePrefix = {arXiv},
       eprint = {0907.2952},
 primaryClass = {astro-ph.GA},
       adsurl = {https://ui.adsabs.harvard.edu/abs/2010MNRAS.401..977J}
}

@ARTICLE{Pe2021MNRAS.501.3670P,
       author = {{Pe{\~n}arrubia}, Jorge},
        title = "{Creation/destruction of ultra-wide binaries in tidal streams}",
      journal = {\mnras},
         year = 2021,
        month = mar,
       volume = {501},
       number = {3},
        pages = {3670-3686},
          doi = {10.1093/mnras/staa3700},
archivePrefix = {arXiv},
       eprint = {2012.06180},
 primaryClass = {astro-ph.GA},
       adsurl = {https://ui.adsabs.harvard.edu/abs/2021MNRAS.501.3670P}
}

@ARTICLE{Grishin2022MNRAS.512.4993G,
       author = {{Grishin}, Evgeni and {Perets}, Hagai B.},
        title = "{Chaotic dynamics of wide triples induced by galactic tides: a novel channel for producing compact binaries, mergers, and collisions}",
      journal = {\mnras},
         year = 2022,
        month = jun,
       volume = {512},
       number = {4},
        pages = {4993-5009},
          doi = {10.1093/mnras/stac706},
archivePrefix = {arXiv},
       eprint = {2112.11475},
 primaryClass = {astro-ph.SR},
       adsurl = {https://ui.adsabs.harvard.edu/abs/2022MNRAS.512.4993G}
}

@ARTICLE{Albrow2024MNRAS.528.6211A,
       author = {{Albrow}, Michael D.},
        title = "{The frequency and mass-ratio distribution of binaries in clusters II: radial segregation in the nearby dissolving open clusters Hyades and Praesepe}",
      journal = {\mnras},
         year = 2024,
        month = mar,
       volume = {528},
       number = {4},
        pages = {6211-6220},
          doi = {10.1093/mnras/stae425},
archivePrefix = {arXiv},
       eprint = {2402.05177},
 primaryClass = {astro-ph.SR},
       adsurl = {https://ui.adsabs.harvard.edu/abs/2024MNRAS.528.6211A}
}

@ARTICLE{Geller2012AJ....144...54G,
       author = {{Geller}, Aaron M. and {Mathieu}, Robert D.},
        title = "{WIYN Open Cluster Study. XLVIII. The Hard-binary Population of NGC 188}",
      journal = {\aj},
         year = 2012,
        month = aug,
       volume = {144},
       number = {2},
          eid = {54},
        pages = {54},
          doi = {10.1088/0004-6256/144/2/54},
archivePrefix = {arXiv},
       eprint = {1111.3950},
 primaryClass = {astro-ph.SR},
       adsurl = {https://ui.adsabs.harvard.edu/abs/2012AJ....144...54G}
}

@ARTICLE{Geller2013AJ....145....8G,
       author = {{Geller}, Aaron M. and {Hurley}, Jarrod R. and {Mathieu}, Robert D.},
        title = "{Direct N-body Modeling of the Old Open Cluster NGC 188: A Detailed Comparison of Theoretical and Observed Binary Star and Blue Straggler Populations}",
      journal = {\aj},
         year = 2013,
        month = jan,
       volume = {145},
       number = {1},
          eid = {8},
        pages = {8},
          doi = {10.1088/0004-6256/145/1/8},
archivePrefix = {arXiv},
       eprint = {1210.1575},
 primaryClass = {astro-ph.SR},
       adsurl = {https://ui.adsabs.harvard.edu/abs/2013AJ....145....8G}
}

@ARTICLE{Giesers2019A&A...632A...3G,
       author = {{Giesers}, Benjamin and {Kamann}, Sebastian and {Dreizler}, Stefan and {Husser}, Tim-Oliver and {Askar}, Abbas and {G{\"o}ttgens}, Fabian and {Brinchmann}, Jarle and {Latour}, Marilyn and {Weilbacher}, Peter M. and {Wendt}, Martin and {Roth}, Martin M.},
        title = "{A stellar census in globular clusters with MUSE: Binaries in NGC 3201}",
      journal = {\aap},
         year = 2019,
        month = dec,
       volume = {632},
          eid = {A3},
        pages = {A3},
          doi = {10.1051/0004-6361/201936203},
archivePrefix = {arXiv},
       eprint = {1909.04050},
 primaryClass = {astro-ph.SR},
       adsurl = {https://ui.adsabs.harvard.edu/abs/2019A&A...632A...3G}
}

@ARTICLE{Fregeau2009ApJ...707.1533F,
       author = {{Fregeau}, John M. and {Ivanova}, Natalia and {Rasio}, Frederic A.},
        title = "{Evolution of the Binary Fraction in Dense Stellar Systems}",
      journal = {\apj},
         year = 2009,
        month = dec,
       volume = {707},
       number = {2},
        pages = {1533-1540},
          doi = {10.1088/0004-637X/707/2/1533},
archivePrefix = {arXiv},
       eprint = {0907.4196},
 primaryClass = {astro-ph.GA},
       adsurl = {https://ui.adsabs.harvard.edu/abs/2009ApJ...707.1533F}
}

@ARTICLE{Hurley2007ApJ...665..707H,
       author = {{Hurley}, Jarrod R. and {Aarseth}, Sverre J. and {Shara}, Michael M.},
        title = "{The Core Binary Fractions of Star Clusters from Realistic Simulations}",
      journal = {\apj},
         year = 2007,
        month = aug,
       volume = {665},
       number = {1},
        pages = {707-718},
          doi = {10.1086/517879},
archivePrefix = {arXiv},
       eprint = {0704.0290},
 primaryClass = {astro-ph},
       adsurl = {https://ui.adsabs.harvard.edu/abs/2007ApJ...665..707H}
}

@ARTICLE{Gautam2024ApJ...964..164G,
       author = {{Gautam}, Abhimat K. and {Do}, Tuan and {Ghez}, Andrea M. and {Chu}, Devin S. and {Hosek}, Matthew W. and {Sakai}, Shoko and {Naoz}, Smadar and {Morris}, Mark R. and {Ciurlo}, Anna and {Haggard}, Zo{\"e} and {Lu}, Jessica R.},
        title = "{An Estimate of the Binary Star Fraction among Young Stars at the Galactic Center: Possible Evidence of a Radial Dependence}",
      journal = {\apj},
         year = 2024,
        month = apr,
       volume = {964},
       number = {2},
          eid = {164},
        pages = {164},
          doi = {10.3847/1538-4357/ad26e6},
archivePrefix = {arXiv},
       eprint = {2401.12555},
 primaryClass = {astro-ph.GA},
       adsurl = {https://ui.adsabs.harvard.edu/abs/2024ApJ...964..164G}
}

@ARTICLE{Law2015AJ....150...19L,
       author = {{Law}, David R. and {Yan}, Renbin and {Bershady}, Matthew A. and {Bundy}, Kevin and {Cherinka}, Brian and {Drory}, Niv and {MacDonald}, Nicholas and {S{\'a}nchez-Gallego}, Jos{\'e} R. and {Wake}, David A. and {Weijmans}, Anne-Marie and {Blanton}, Michael R. and {Klaene}, Mark A. and {Moran}, Sean M. and {Sanchez}, Sebastian F. and {Zhang}, Kai},
        title = "{Observing Strategy for the SDSS-IV/MaNGA IFU Galaxy Survey}",
      journal = {\aj},
         year = 2015,
        month = jul,
       volume = {150},
       number = {1},
          eid = {19},
        pages = {19},
          doi = {10.1088/0004-6256/150/1/19},
archivePrefix = {arXiv},
       eprint = {1505.04285},
 primaryClass = {astro-ph.IM},
       adsurl = {https://ui.adsabs.harvard.edu/abs/2015AJ....150...19L}
}

@ARTICLE{Wake2017AJ....154...86W,
       author = {{Wake}, David A. and {Bundy}, Kevin and {Diamond-Stanic}, Aleksandar M. and {Yan}, Renbin and {Blanton}, Michael R. and {Bershady}, Matthew A. and {S{\'a}nchez-Gallego}, Jos{\'e} R. and {Drory}, Niv and {Jones}, Amy and {Kauffmann}, Guinevere and {Law}, David R. and {Li}, Cheng and {MacDonald}, Nicholas and {Masters}, Karen and {Thomas}, Daniel and {Tinker}, Jeremy and {Weijmans}, Anne-Marie and {Brownstein}, Joel R.},
        title = "{The SDSS-IV MaNGA Sample: Design, Optimization, and Usage Considerations}",
      journal = {\aj},
         year = 2017,
        month = sep,
       volume = {154},
       number = {3},
          eid = {86},
        pages = {86},
          doi = {10.3847/1538-3881/aa7ecc},
archivePrefix = {arXiv},
       eprint = {1707.02989},
 primaryClass = {astro-ph.GA},
       adsurl = {https://ui.adsabs.harvard.edu/abs/2017AJ....154...86W}
}

@ARTICLE{Drory2015AJ....149...77D,
       author = {{Drory}, N. and {MacDonald}, N. and {Bershady}, M.~A. and {Bundy}, K. and {Gunn}, J. and {Law}, D.~R. and {Smith}, M. and {Stoll}, R. and {Tremonti}, C.~A. and {Wake}, D.~A. and {Yan}, R. and {Weijmans}, A.~M. and {Byler}, N. and {Cherinka}, B. and {Cope}, F. and {Eigenbrot}, A. and {Harding}, P. and {Holder}, D. and {Huehnerhoff}, J. and {Jaehnig}, K. and {Jansen}, T.~C. and {Klaene}, M. and {Paat}, A.~M. and {Percival}, J. and {Sayres}, C.},
        title = "{The MaNGA Integral Field Unit Fiber Feed System for the Sloan 2.5 m Telescope}",
      journal = {\aj},
         year = 2015,
        month = feb,
       volume = {149},
       number = {2},
          eid = {77},
        pages = {77},
          doi = {10.1088/0004-6256/149/2/77},
archivePrefix = {arXiv},
       eprint = {1412.1535},
 primaryClass = {astro-ph.IM},
       adsurl = {https://ui.adsabs.harvard.edu/abs/2015AJ....149...77D}
}

@ARTICLE{Cai2021RAA....21..204C,
       author = {{Cai}, Wei and {Zhao}, Ying-He and {Bai}, Jin-Ming},
        title = "{Radial stellar populations of AGN-host dwarf galaxies in SDSS-IV MaNGA survey}",
      journal = {Research in Astronomy and Astrophysics},
         year = 2021,
        month = oct,
       volume = {21},
       number = {8},
          eid = {204},
        pages = {204},
          doi = {10.1088/1674-4527/21/8/204},
archivePrefix = {arXiv},
       eprint = {2103.16826},
 primaryClass = {astro-ph.GA},
       adsurl = {https://ui.adsabs.harvard.edu/abs/2021RAA....21..204C}
}

@ARTICLE{Chabrier2003PASP..115..763C,
       author = {{Chabrier}, Gilles},
        title = "{Galactic Stellar and Substellar Initial Mass Function}",
      journal = {\pasp},
         year = 2003,
        month = jul,
       volume = {115},
       number = {809},
        pages = {763-795},
          doi = {10.1086/376392},
archivePrefix = {arXiv},
       eprint = {astro-ph/0304382},
 primaryClass = {astro-ph},
       adsurl = {https://ui.adsabs.harvard.edu/abs/2003PASP..115..763C}
}

@ARTICLE{Bertelli1994AAS..106..275B,
       author = {{Bertelli}, G. and {Bressan}, A. and {Chiosi}, C. and {Fagotto}, F. and {Nasi}, E.},
        title = "{Theoretical isochrones from models with new radiative opacities}",
      journal = {\aaps},
         year = 1994,
        month = aug,
       volume = {106},
        pages = {275-302},
       adsurl = {https://ui.adsabs.harvard.edu/abs/1994A&AS..106..275B}
}

@ARTICLE{Bruzual2003MNRAS.344.1000B,
       author = {{Bruzual}, G. and {Charlot}, S.},
        title = "{Stellar population synthesis at the resolution of 2003}",
      journal = {\mnras},
         year = 2003,
        month = oct,
       volume = {344},
       number = {4},
        pages = {1000-1028},
          doi = {10.1046/j.1365-8711.2003.06897.x},
archivePrefix = {arXiv},
       eprint = {astro-ph/0309134},
 primaryClass = {astro-ph},
       adsurl = {https://ui.adsabs.harvard.edu/abs/2003MNRAS.344.1000B}
}

@ARTICLE{Falcon2011AA...532A..95F,
       author = {{Falc{\'o}n-Barroso}, J. and {S{\'a}nchez-Bl{\'a}zquez}, P. and {Vazdekis}, A. and {Ricciardelli}, E. and {Cardiel}, N. and {Cenarro}, A.~J. and {Gorgas}, J. and {Peletier}, R.~F.},
        title = "{An updated MILES stellar library and stellar population models}",
      journal = {\aap},
         year = 2011,
        month = aug,
       volume = {532},
          eid = {A95},
        pages = {A95},
          doi = {10.1051/0004-6361/201116842},
archivePrefix = {arXiv},
       eprint = {1107.2303},
 primaryClass = {astro-ph.CO},
       adsurl = {https://ui.adsabs.harvard.edu/abs/2011A&A...532A..95F}
}

@ARTICLE{Sanchez2006MNRAS.371..703S,
       author = {{S{\'a}nchez-Bl{\'a}zquez}, P. and {Peletier}, R.~F. and {Jim{\'e}nez-Vicente}, J. and {Cardiel}, N. and {Cenarro}, A.~J. and {Falc{\'o}n-Barroso}, J. and {Gorgas}, J. and {Selam}, S. and {Vazdekis}, A.},
        title = "{Medium-resolution Isaac Newton Telescope library of empirical spectra}",
      journal = {\mnras},
         year = 2006,
        month = sep,
       volume = {371},
       number = {2},
        pages = {703-718},
          doi = {10.1111/j.1365-2966.2006.10699.x},
archivePrefix = {arXiv},
       eprint = {astro-ph/0607009},
 primaryClass = {astro-ph},
       adsurl = {https://ui.adsabs.harvard.edu/abs/2006MNRAS.371..703S}
}

@ARTICLE{Zhang2005MNRAS.357.1088Z,
       author = {{Zhang}, Fenghui and {Han}, Zhanwen and {Li}, Lifang and {Hurley}, Jarrod R.},
        title = "{Inclusion of binaries in evolutionary population synthesis}",
      journal = {\mnras},
         year = 2005,
        month = mar,
       volume = {357},
       number = {3},
        pages = {1088-1103},
          doi = {10.1111/j.1365-2966.2005.08739.x},
archivePrefix = {arXiv},
       eprint = {astro-ph/0412008},
 primaryClass = {astro-ph},
       adsurl = {https://ui.adsabs.harvard.edu/abs/2005MNRAS.357.1088Z}
}

@ARTICLE{Zhang2004A&A...415..117Z,
       author = {{Zhang}, F. and {Han}, Z. and {Li}, L. and {Hurley}, J.~R.},
        title = "{Evolutionary population synthesis for binary stellar populations}",
      journal = {\aap},
         year = 2004,
        month = feb,
       volume = {415},
        pages = {117-122},
          doi = {10.1051/0004-6361:20031268},
archivePrefix = {arXiv},
       eprint = {astro-ph/0403297},
 primaryClass = {astro-ph},
       adsurl = {https://ui.adsabs.harvard.edu/abs/2004A&A...415..117Z}
}

@ARTICLE{Hurley2000MNRAS.315..543H,
       author = {{Hurley}, Jarrod R. and {Pols}, Onno R. and {Tout}, Christopher A.},
        title = "{Comprehensive analytic formulae for stellar evolution as a function of mass and metallicity}",
      journal = {\mnras},
         year = 2000,
        month = jul,
       volume = {315},
       number = {3},
        pages = {543-569},
          doi = {10.1046/j.1365-8711.2000.03426.x},
archivePrefix = {arXiv},
       eprint = {astro-ph/0001295},
 primaryClass = {astro-ph},
       adsurl = {https://ui.adsabs.harvard.edu/abs/2000MNRAS.315..543H}
}

@ARTICLE{Miller1979ApJS...41..513M,
       author = {{Miller}, G.~E. and {Scalo}, J.~M.},
        title = "{The Initial Mass Function and Stellar Birthrate in the Solar Neighborhood}",
      journal = {\apjs},
         year = 1979,
        month = nov,
       volume = {41},
        pages = {513},
          doi = {10.1086/190629},
       adsurl = {https://ui.adsabs.harvard.edu/abs/1979ApJS...41..513M}
}

@ARTICLE{Eggleton1989ApJ...347..998E,
       author = {{Eggleton}, Peter P. and {Fitchett}, Michael J. and {Tout}, Christopher A.},
        title = "{The Distribution of Visual Binaries with Two Bright Components}",
      journal = {\apj},
         year = 1989,
        month = dec,
       volume = {347},
        pages = {998},
          doi = {10.1086/168190},
       adsurl = {https://ui.adsabs.harvard.edu/abs/1989ApJ...347..998E}
}

@ARTICLE{Bertone2008AA...485..823B,
       author = {{Bertone}, E. and {Buzzoni}, A. and {Ch{\'a}vez}, M. and {Rodr{\'\i}guez-Merino}, L.~H.},
        title = "{Probing Atlas model atmospheres at high spectral resolution. Stellar synthesis and reference template validation}",
      journal = {\aap},
         year = 2008,
        month = jul,
       volume = {485},
       number = {3},
        pages = {823-835},
          doi = {10.1051/0004-6361:20078923},
       adsurl = {https://ui.adsabs.harvard.edu/abs/2008A&A...485..823B}
}

@ARTICLE{Lejeune1997AAS..125..229L,
       author = {{Lejeune}, Th. and {Cuisinier}, F. and {Buser}, R.},
        title = "{Standard stellar library for evolutionary synthesis. I. Calibration of theoretical spectra}",
      journal = {\aaps},
         year = 1997,
        month = oct,
       volume = {125},
        pages = {229-246},
          doi = {10.1051/aas:1997373},
archivePrefix = {arXiv},
       eprint = {astro-ph/9701019},
 primaryClass = {astro-ph},
       adsurl = {https://ui.adsabs.harvard.edu/abs/1997A&AS..125..229L}
}

@ARTICLE{Lejeune1998AAS..130...65L,
       author = {{Lejeune}, T. and {Cuisinier}, F. and {Buser}, R.},
        title = "{A standard stellar library for evolutionary synthesis. II. The M dwarf extension}",
      journal = {\aaps},
         year = 1998,
        month = may,
       volume = {130},
        pages = {65-75},
          doi = {10.1051/aas:1998405},
archivePrefix = {arXiv},
       eprint = {astro-ph/9710350},
 primaryClass = {astro-ph},
       adsurl = {https://ui.adsabs.harvard.edu/abs/1998A&AS..130...65L}
}

@ARTICLE{Trager1998ApJS..116....1T,
       author = {{Trager}, S.~C. and {Worthey}, Guy and {Faber}, S.~M. and {Burstein}, David and {Gonz{\'a}lez}, J. Jes{\'u}s},
        title = "{Old Stellar Populations. VI. Absorption-Line Spectra of Galaxy Nuclei and Globular Clusters}",
      journal = {\apjs},
         year = 1998,
        month = jan,
       volume = {116},
       number = {1},
        pages = {1-28},
          doi = {10.1086/313099},
archivePrefix = {arXiv},
       eprint = {astro-ph/9712258},
 primaryClass = {astro-ph},
       adsurl = {https://ui.adsabs.harvard.edu/abs/1998ApJS..116....1T}
}

@ARTICLE{Worthey1994ApJS...94..687W,
       author = {{Worthey}, Guy and {Faber}, S.~M. and {Gonzalez}, J. Jesus and {Burstein}, D.},
        title = "{Old Stellar Populations. V. Absorption Feature Indices for the Complete Lick/IDS Sample of Stars}",
      journal = {\apjs},
         year = 1994,
        month = oct,
       volume = {94},
        pages = {687},
          doi = {10.1086/192087},
       adsurl = {https://ui.adsabs.harvard.edu/abs/1994ApJS...94..687W}
}

@ARTICLE{Worthey1994ApJS...95..107W,
       author = {{Worthey}, Guy},
        title = "{Comprehensive Stellar Population Models and the Disentanglement of Age and Metallicity Effects}",
      journal = {\apjs},
         year = 1994,
        month = nov,
       volume = {95},
        pages = {107},
          doi = {10.1086/192096},
       adsurl = {https://ui.adsabs.harvard.edu/abs/1994ApJS...95..107W}
}

@ARTICLE{van2017ApJ...841...68V,
       author = {{van Dokkum}, Pieter and {Conroy}, Charlie and {Villaume}, Alexa and {Brodie}, Jean and {Romanowsky}, Aaron J.},
        title = "{The Stellar Initial Mass Function in Early-type Galaxies from Absorption Line Spectroscopy. III. Radial Gradients}",
      journal = {\apj},
         year = 2017,
        month = jun,
       volume = {841},
       number = {2},
          eid = {68},
        pages = {68},
          doi = {10.3847/1538-4357/aa7135},
archivePrefix = {arXiv},
       eprint = {1611.09859},
 primaryClass = {astro-ph.GA},
       adsurl = {https://ui.adsabs.harvard.edu/abs/2017ApJ...841...68V}
}

@ARTICLE{Mart2015MNRAS.447.1033M,
       author = {{Mart{\'\i}n-Navarro}, Ignacio and {La Barbera}, Francesco and {Vazdekis}, Alexandre and {Falc{\'o}n-Barroso}, Jes{\'u}s and {Ferreras}, Ignacio},
        title = "{Radial variations in the stellar initial mass function of early-type galaxies}",
      journal = {\mnras},
         year = 2015,
        month = feb,
       volume = {447},
       number = {2},
        pages = {1033-1048},
          doi = {10.1093/mnras/stu2480},
archivePrefix = {arXiv},
       eprint = {1404.6533},
 primaryClass = {astro-ph.GA},
       adsurl = {https://ui.adsabs.harvard.edu/abs/2015MNRAS.447.1033M}
}

@ARTICLE{OConnell1999ARA&A..37..603O,
       author = {{O'Connell}, Robert W.},
        title = "{Far-Ultraviolet Radiation from Elliptical Galaxies}",
      journal = {\araa},
         year = 1999,
        month = jan,
       volume = {37},
        pages = {603-648},
          doi = {10.1146/annurev.astro.37.1.603},
archivePrefix = {arXiv},
       eprint = {astro-ph/9906068},
 primaryClass = {astro-ph},
       adsurl = {https://ui.adsabs.harvard.edu/abs/1999ARA&A..37..603O}
}

@ARTICLE{Dantas2021MNRAS.500.1870D,
       author = {{Dantas}, M.~L.~L. and {Coelho}, P.~R.~T. and {S{\'a}nchez-Bl{\'a}zquez}, P.},
        title = "{UV upturn versus UV weak galaxies: differences and similarities of their stellar populations unveiled by a de-biased sample}",
      journal = {\mnras},
         year = 2021,
        month = jan,
       volume = {500},
       number = {2},
        pages = {1870-1883},
          doi = {10.1093/mnras/staa3447},
archivePrefix = {arXiv},
       eprint = {2009.03915},
 primaryClass = {astro-ph.GA},
       adsurl = {https://ui.adsabs.harvard.edu/abs/2021MNRAS.500.1870D}
}

@ARTICLE{Kaviraj2007ApJS..173..619K,
       author = {{Kaviraj}, S. and {Schawinski}, K. and {Devriendt}, J.~E.~G. and {Ferreras}, I. and {Khochfar}, S. and {Yoon}, S. -J. and {Yi}, S.~K. and {Deharveng}, J. -M. and {Boselli}, A. and {Barlow}, T. and {Conrow}, T. and {Forster}, K. and {Friedman}, P.~G. and {Martin}, D.~C. and {Morrissey}, P. and {Neff}, S. and {Schiminovich}, D. and {Seibert}, M. and {Small}, T. and {Wyder}, T. and {Bianchi}, L. and {Donas}, J. and {Heckman}, T. and {Lee}, Y. -W. and {Madore}, B. and {Milliard}, B. and {Rich}, R.~M. and {Szalay}, A.},
        title = "{UV-Optical Colors As Probes of Early-Type Galaxy Evolution}",
      journal = {\apjs},
         year = 2007,
        month = dec,
       volume = {173},
       number = {2},
        pages = {619-642},
          doi = {10.1086/516633},
archivePrefix = {arXiv},
       eprint = {astro-ph/0601029},
 primaryClass = {astro-ph},
       adsurl = {https://ui.adsabs.harvard.edu/abs/2007ApJS..173..619K}
}

@ARTICLE{Schawinski2007ApJS..173..512S,
       author = {{Schawinski}, K. and {Kaviraj}, S. and {Khochfar}, S. and {Yoon}, S. -J. and {Yi}, S.~K. and {Deharveng}, J. -M. and {Boselli}, A. and {Barlow}, T. and {Conrow}, T. and {Forster}, K. and {Friedman}, P.~G. and {Martin}, D.~C. and {Morrissey}, P. and {Neff}, S. and {Schiminovich}, D. and {Seibert}, M. and {Small}, T. and {Wyder}, T. and {Bianchi}, L. and {Donas}, J. and {Heckman}, T. and {Lee}, Y. -W. and {Madore}, B. and {Milliard}, B. and {Rich}, R.~M. and {Szalay}, A.},
        title = "{The Effect of Environment on the Ultraviolet Color-Magnitude Relation of Early-Type Galaxies}",
      journal = {\apjs},
         year = 2007,
        month = dec,
       volume = {173},
       number = {2},
        pages = {512-523},
          doi = {10.1086/516631},
archivePrefix = {arXiv},
       eprint = {astro-ph/0601036},
 primaryClass = {astro-ph},
       adsurl = {https://ui.adsabs.harvard.edu/abs/2007ApJS..173..512S}
}

@ARTICLE{Yi2005ApJ...619L.111Y,
       author = {{Yi}, S.~K. and {Yoon}, S. -J. and {Kaviraj}, S. and {Deharveng}, J. -M. and {Rich}, R.~M. and {Salim}, S. and {Boselli}, A. and {Lee}, Y. -W. and {Ree}, C.~H. and {Sohn}, Y. -J. and {Rey}, S. -C. and {Lee}, J. -W. and {Rhee}, J. and {Bianchi}, L. and {Byun}, Y. -I. and {Donas}, J. and {Friedman}, P.~G. and {Heckman}, T.~M. and {Jelinsky}, P. and {Madore}, B.~F. and {Malina}, R. and {Martin}, D.~C. and {Milliard}, B. and {Morrissey}, P. and {Neff}, S. and {Schiminovich}, D. and {Siegmund}, O. and {Small}, T. and {Szalay}, A.~S. and {Jee}, M.~J. and {Kim}, S. -W. and {Barlow}, T. and {Forster}, K. and {Welsh}, B. and {Wyder}, T.~K.},
        title = "{Galaxy Evolution Explorer Ultraviolet Color-Magnitude Relations and Evidence of Recent Star Formation in Early-Type Galaxies}",
      journal = {\apjl},
         year = 2005,
        month = jan,
       volume = {619},
       number = {1},
        pages = {L111-L114},
          doi = {10.1086/422811},
archivePrefix = {arXiv},
       eprint = {astro-ph/0411327},
 primaryClass = {astro-ph},
       adsurl = {https://ui.adsabs.harvard.edu/abs/2005ApJ...619L.111Y}
}

@ARTICLE{Sanchez2020ARA&A..58...99S,
       author = {{S{\'a}nchez}, Sebasti{\'a}n F.},
        title = "{Spatially Resolved Spectroscopic Properties of Low-Redshift Star-Forming Galaxies}",
      journal = {\araa},
         year = 2020,
        month = aug,
       volume = {58},
        pages = {99-155},
          doi = {10.1146/annurev-astro-012120-013326},
archivePrefix = {arXiv},
       eprint = {1911.06925},
 primaryClass = {astro-ph.GA},
       adsurl = {https://ui.adsabs.harvard.edu/abs/2020ARA&A..58...99S}
}

@ARTICLE{Gonzalez2015A&A...581A.103G,
       author = {{Gonz{\'a}lez Delgado}, R.~M. and {Garc{\'\i}a-Benito}, R. and {P{\'e}rez}, E. and {Cid Fernandes}, R. and {de Amorim}, A.~L. and {Cortijo-Ferrero}, C. and {Lacerda}, E.~A.~D. and {L{\'o}pez Fern{\'a}ndez}, R. and {Vale-Asari}, N. and {S{\'a}nchez}, S.~F. and {Moll{\'a}}, M. and {Ruiz-Lara}, T. and {S{\'a}nchez-Bl{\'a}zquez}, P. and {Walcher}, C.~J. and {Alves}, J. and {Aguerri}, J.~A.~L. and {Bekerait{\'e}}, S. and {Bland-Hawthorn}, J. and {Galbany}, L. and {Gallazzi}, A. and {Husemann}, B. and {Iglesias-P{\'a}ramo}, J. and {Kalinova}, V. and {L{\'o}pez-S{\'a}nchez}, A.~R. and {Marino}, R.~A. and {M{\'a}rquez}, I. and {Masegosa}, J. and {Mast}, D. and {M{\'e}ndez-Abreu}, J. and {Mendoza}, A. and {del Olmo}, A. and {P{\'e}rez}, I. and {Quirrenbach}, A. and {Zibetti}, S.},
        title = "{The CALIFA survey across the Hubble sequence. Spatially resolved stellar population properties in galaxies}",
      journal = {\aap},
         year = 2015,
        month = sep,
       volume = {581},
          eid = {A103},
        pages = {A103},
          doi = {10.1051/0004-6361/201525938},
archivePrefix = {arXiv},
       eprint = {1506.04157},
 primaryClass = {astro-ph.GA},
       adsurl = {https://ui.adsabs.harvard.edu/abs/2015A&A...581A.103G}
}

@ARTICLE{Li2015ApJ...804..125L,
       author = {{Li}, Cheng and {Wang}, Enci and {Lin}, Lin and {Bershady}, Matthew A. and {Bundy}, Kevin and {Tremonti}, Christy A. and {Xiao}, Ting and {Yan}, Renbin and {Bizyaev}, Dmitry and {Blanton}, Michael and {Cales}, Sabrina and {Cherinka}, Brian and {Cheung}, Edmond and {Drory}, Niv and {Emsellem}, Eric and {Fu}, Hai and {Gelfand}, Joseph and {Law}, David R. and {Lin}, Lihwai and {MacDonald}, Nick and {Maraston}, Claudia and {Masters}, Karen L. and {Merrifield}, Michael R. and {Pan}, Kaike and {S{\'a}nchez}, S.~F. and {Schneider}, Donald P. and {Thomas}, Daniel and {Wake}, David and {Wang}, Lixin and {Weijmans}, Anne-Marie and {Wilkinson}, David and {Yoachim}, Peter and {Zhang}, Kai and {Zheng}, Tiantian},
        title = "{P-MaNGA: Gradients in Recent Star Formation Histories as Diagnostics for Galaxy Growth and Death}",
      journal = {\apj},
         year = 2015,
        month = may,
       volume = {804},
       number = {2},
          eid = {125},
        pages = {125},
          doi = {10.1088/0004-637X/804/2/125},
archivePrefix = {arXiv},
       eprint = {1502.07040},
 primaryClass = {astro-ph.GA},
       adsurl = {https://ui.adsabs.harvard.edu/abs/2015ApJ...804..125L}
}

@ARTICLE{Zheng2017MNRAS.465.4572Z,
       author = {{Zheng}, Zheng and {Wang}, Huiyuan and {Ge}, Junqiang and {Mao}, Shude and {Li}, Cheng and {Li}, Ran and {Mo}, Houjun and {Goddard}, Daniel and {Bundy}, Kevin and {Li}, Hongyu and {Nair}, Preethi and {Lin}, Lihwai and {Long}, R.~J. and {Riffel}, Rog{\'e}rio and {Thomas}, Daniel and {Masters}, Karen and {Bizyaev}, Dmitry and {Brownstein}, Joel R. and {Zhang}, Kai and {Law}, David R. and {Drory}, Niv and {Roman Lopes}, Alexandre and {Malanushenko}, Olena},
        title = "{SDSS-IV MaNGA: environmental dependence of stellar age and metallicity gradients in nearby galaxies}",
      journal = {\mnras},
         year = 2017,
        month = mar,
       volume = {465},
       number = {4},
        pages = {4572-4588},
          doi = {10.1093/mnras/stw3030},
archivePrefix = {arXiv},
       eprint = {1612.01523},
 primaryClass = {astro-ph.GA},
       adsurl = {https://ui.adsabs.harvard.edu/abs/2017MNRAS.465.4572Z}
}

@ARTICLE{Baldwin1981PASP...93....5B,
       author = {{Baldwin}, J.~A. and {Phillips}, M.~M. and {Terlevich}, R.},
        title = "{Classification parameters for the emission-line spectra of extragalactic objects.}",
      journal = {\pasp},
         year = 1981,
        month = feb,
       volume = {93},
        pages = {5-19},
          doi = {10.1086/130766},
       adsurl = {https://ui.adsabs.harvard.edu/abs/1981PASP...93....5B}
}

@ARTICLE{Veilleux1987ApJS...63..295V,
       author = {{Veilleux}, Sylvain and {Osterbrock}, Donald E.},
        title = "{Spectral Classification of Emission-Line Galaxies}",
      journal = {\apjs},
         year = 1987,
        month = feb,
       volume = {63},
        pages = {295},
          doi = {10.1086/191166},
       adsurl = {https://ui.adsabs.harvard.edu/abs/1987ApJS...63..295V}
}

@ARTICLE{Kauffmann2003MNRAS.346.1055K,
       author = {{Kauffmann}, Guinevere and {Heckman}, Timothy M. and {Tremonti}, Christy and {Brinchmann}, Jarle and {Charlot}, St{\'e}phane and {White}, Simon D.~M. and {Ridgway}, Susan E. and {Brinkmann}, Jon and {Fukugita}, Masataka and {Hall}, Patrick B. and {Ivezi{\'c}}, {\v{Z}}eljko and {Richards}, Gordon T. and {Schneider}, Donald P.},
        title = "{The host galaxies of active galactic nuclei}",
      journal = {\mnras},
         year = 2003,
        month = dec,
       volume = {346},
       number = {4},
        pages = {1055-1077},
          doi = {10.1111/j.1365-2966.2003.07154.x},
archivePrefix = {arXiv},
       eprint = {astro-ph/0304239},
 primaryClass = {astro-ph},
       adsurl = {https://ui.adsabs.harvard.edu/abs/2003MNRAS.346.1055K}
}

@ARTICLE{Cardelli1989ApJ...345..245C,
       author = {{Cardelli}, Jason A. and {Clayton}, Geoffrey C. and {Mathis}, John S.},
        title = "{The Relationship between Infrared, Optical, and Ultraviolet Extinction}",
      journal = {\apj},
         year = 1989,
        month = oct,
       volume = {345},
        pages = {245},
          doi = {10.1086/167900},
       adsurl = {https://ui.adsabs.harvard.edu/abs/1989ApJ...345..245C}
}

@ARTICLE{ODonnell1994ApJ...422..158O,
       author = {{O'Donnell}, James E.},
        title = "{R v-dependent Optical and Near-Ultraviolet Extinction}",
      journal = {\apj},
         year = 1994,
        month = feb,
       volume = {422},
        pages = {158},
          doi = {10.1086/173713},
       adsurl = {https://ui.adsabs.harvard.edu/abs/1994ApJ...422..158O}
}

@ARTICLE{Schlegel1998ApJ...500..525S,
       author = {{Schlegel}, David J. and {Finkbeiner}, Douglas P. and {Davis}, Marc},
        title = "{Maps of Dust Infrared Emission for Use in Estimation of Reddening and Cosmic Microwave Background Radiation Foregrounds}",
      journal = {\apj},
         year = 1998,
        month = jun,
       volume = {500},
       number = {2},
        pages = {525-553},
          doi = {10.1086/305772},
archivePrefix = {arXiv},
       eprint = {astro-ph/9710327},
 primaryClass = {astro-ph},
       adsurl = {https://ui.adsabs.harvard.edu/abs/1998ApJ...500..525S}
}

@ARTICLE{Li2013MNRAS.436.1497L,
       author = {{Li}, Chengyuan and {de Grijs}, Richard and {Deng}, Licai},
        title = "{The binary fractions in the massive young Large Magellanic Cloud star clusters NGC 1805 and NGC 1818}",
      journal = {\mnras},
         year = 2013,
        month = dec,
       volume = {436},
       number = {2},
        pages = {1497-1512},
          doi = {10.1093/mnras/stt1669},
archivePrefix = {arXiv},
       eprint = {1309.0929},
 primaryClass = {astro-ph.SR},
       adsurl = {https://ui.adsabs.harvard.edu/abs/2013MNRAS.436.1497L}
}

@ARTICLE{deGrijs2013ApJ...765....4D,
       author = {{de Grijs}, Richard and {Li}, Chengyuan and {Zheng}, Yong and {Deng}, Licai and {Hu}, Yi and {Kouwenhoven}, M.~B.~N. and {Wicker}, James E.},
        title = "{Gravitational Conundrum? Dynamical Mass Segregation versus Disruption of Binary Stars in Dense Stellar Systems}",
      journal = {\apj},
         year = 2013,
        month = mar,
       volume = {765},
       number = {1},
          eid = {4},
        pages = {4},
          doi = {10.1088/0004-637X/765/1/4},
archivePrefix = {arXiv},
       eprint = {1301.1926},
 primaryClass = {astro-ph.SR},
       adsurl = {https://ui.adsabs.harvard.edu/abs/2013ApJ...765....4D}
}

@ARTICLE{Geller2015ApJ...805...11G,
       author = {{Geller}, Aaron M. and {de Grijs}, Richard and {Li}, Chengyuan and {Hurley}, Jarrod R.},
        title = "{Different Dynamical Ages for the Two Young and Coeval LMC Star Clusters, NGC 1805 and NGC 1818, Imprinted on Their Binary Populations}",
      journal = {\apj},
         year = 2015,
        month = may,
       volume = {805},
       number = {1},
          eid = {11},
        pages = {11},
          doi = {10.1088/0004-637X/805/1/11},
archivePrefix = {arXiv},
       eprint = {1503.05198},
 primaryClass = {astro-ph.SR},
       adsurl = {https://ui.adsabs.harvard.edu/abs/2015ApJ...805...11G}
}

@ARTICLE{Pipino2009MNRAS.395..462P,
       author = {{Pipino}, A. and {Kaviraj}, S. and {Bildfell}, C. and {Babul}, A. and {Hoekstra}, H. and {Silk}, J.},
        title = "{Evidence for recent star formation in BCGs: a correspondence between blue cores and UV excess}",
      journal = {\mnras},
         year = 2009,
        month = may,
       volume = {395},
       number = {1},
        pages = {462-471},
          doi = {10.1111/j.1365-2966.2009.14534.x},
archivePrefix = {arXiv},
       eprint = {0807.2760},
 primaryClass = {astro-ph},
       adsurl = {https://ui.adsabs.harvard.edu/abs/2009MNRAS.395..462P}
}

@ARTICLE{Salim2010ApJ...714L.290S,
       author = {{Salim}, Samir and {Rich}, R. Michael},
        title = "{Star Formation Signatures in Optically Quiescent Early-type Galaxies}",
      journal = {\apjl},
         year = 2010,
        month = may,
       volume = {714},
       number = {2},
        pages = {L290-L294},
          doi = {10.1088/2041-8205/714/2/L290},
archivePrefix = {arXiv},
       eprint = {1004.2041},
 primaryClass = {astro-ph.CO},
       adsurl = {https://ui.adsabs.harvard.edu/abs/2010ApJ...714L.290S}
}

@INPROCEEDINGS{Yi2008ASPC..392....3Y,
       author = {{Yi}, S.~K.},
        title = "{The Current Understanding on the UV Upturn}",
    booktitle = {Hot Subdwarf Stars and Related Objects},
         year = 2008,
       editor = {{Heber}, U. and {Jeffery}, C.~S. and {Napiwotzki}, R.},
       series = {Astronomical Society of the Pacific Conference Series},
       volume = {392},
        month = jan,
        pages = {3},
          doi = {10.48550/arXiv.0808.0254},
archivePrefix = {arXiv},
       eprint = {0808.0254},
 primaryClass = {astro-ph},
       adsurl = {https://ui.adsabs.harvard.edu/abs/2008ASPC..392....3Y}
}

@ARTICLE{Chung2013ApJS..204....3C,
       author = {{Chung}, Chul and {Yoon}, Suk-Jin and {Lee}, Sang-Yoon and {Lee}, Young-Wook},
        title = "{Yonsei Evolutionary Population Synthesis (YEPS) Model. I. Spectroscopic Evolution of Simple Stellar Populations}",
      journal = {\apjs},
         year = 2013,
        month = jan,
       volume = {204},
       number = {1},
          eid = {3},
        pages = {3},
          doi = {10.1088/0067-0049/204/1/3},
archivePrefix = {arXiv},
       eprint = {1210.6032},
 primaryClass = {astro-ph.GA},
       adsurl = {https://ui.adsabs.harvard.edu/abs/2013ApJS..204....3C}
}

@ARTICLE{Chung2017ApJ...842...91C,
       author = {{Chung}, Chul and {Yoon}, Suk-Jin and {Lee}, Young-Wook},
        title = "{Yonsei Evolutionary Population Synthesis (YEPS). II. Spectro-photometric Evolution of Helium-enhanced Stellar Populations}",
      journal = {\apj},
         year = 2017,
        month = jun,
       volume = {842},
       number = {2},
          eid = {91},
        pages = {91},
          doi = {10.3847/1538-4357/aa6f19},
archivePrefix = {arXiv},
       eprint = {1704.07382},
 primaryClass = {astro-ph.GA},
       adsurl = {https://ui.adsabs.harvard.edu/abs/2017ApJ...842...91C}
}

@ARTICLE{Lee2005ApJ...621L..57L,
       author = {{Lee}, Young-Wook and {Joo}, Seok-Joo and {Han}, Sang-Il and {Chung}, Chul and {Ree}, Chang H. and {Sohn}, Young-Jong and {Kim}, Yong-Cheol and {Yoon}, Suk-Jin and {Yi}, Sukyoung K. and {Demarque}, Pierre},
        title = "{Super-Helium-rich Populations and the Origin of Extreme Horizontal-Branch Stars in Globular Clusters}",
      journal = {\apjl},
         year = 2005,
        month = mar,
       volume = {621},
       number = {1},
        pages = {L57-L60},
          doi = {10.1086/428944},
archivePrefix = {arXiv},
       eprint = {astro-ph/0501500},
 primaryClass = {astro-ph},
       adsurl = {https://ui.adsabs.harvard.edu/abs/2005ApJ...621L..57L}
}

@ARTICLE{Jiang2025MNRAS.540.3770J,
       author = {{Jiang}, Zhen and {Li}, Cheng and {Zhang}, Fenghui and {Zhou}, Shuang},
        title = "{Two categories of UV-upturn galaxies revealed by semi-analytic models}",
      journal = {\mnras},
         year = 2025,
        month = jul,
       volume = {540},
       number = {4},
        pages = {3770-3788},
          doi = {10.1093/mnras/staf923},
archivePrefix = {arXiv},
       eprint = {2502.14263},
 primaryClass = {astro-ph.GA},
       adsurl = {https://ui.adsabs.harvard.edu/abs/2025MNRAS.540.3770J}
}

@ARTICLE{Dib2018MNRAS.473..849D,
       author = {{Dib}, Sami and {Schmeja}, Stefan and {Parker}, Richard J.},
        title = "{Structure and mass segregation in Galactic stellar clusters}",
      journal = {\mnras},
         year = 2018,
        month = jan,
       volume = {473},
       number = {1},
        pages = {849-859},
          doi = {10.1093/mnras/stx2413},
archivePrefix = {arXiv},
       eprint = {1707.00744},
 primaryClass = {astro-ph.GA},
       adsurl = {https://ui.adsabs.harvard.edu/abs/2018MNRAS.473..849D}
}

@ARTICLE{Kang2017MNRAS.469.1636K,
       author = {{Kang}, Xiaoyu and {Zhang}, Fenghui and {Chang}, Ruixiang},
        title = "{The role of environment on the star formation history of disc galaxies}",
      journal = {\mnras},
         year = 2017,
        month = aug,
       volume = {469},
       number = {2},
        pages = {1636-1646},
          doi = {10.1093/mnras/stx1001},
archivePrefix = {arXiv},
       eprint = {1704.07697},
 primaryClass = {astro-ph.GA},
       adsurl = {https://ui.adsabs.harvard.edu/abs/2017MNRAS.469.1636K}
}

@ARTICLE{Parikh2018MNRAS.477.3954P,
       author = {{Parikh}, Taniya and {Thomas}, Daniel and {Maraston}, Claudia and {Westfall}, Kyle B. and {Goddard}, Daniel and {Lian}, Jianhui and {Meneses-Goytia}, Sofia and {Jones}, Amy and {Vaughan}, Sam and {Andrews}, Brett H. and {Bershady}, Matthew and {Bizyaev}, Dmitry and {Brinkmann}, Jonathan and {Brownstein}, Joel R. and {Bundy}, Kevin and {Drory}, Niv and {Emsellem}, Eric and {Law}, David R. and {Newman}, Jeffrey A. and {Roman-Lopes}, Alexandre and {Wake}, David and {Yan}, Renbin and {Zheng}, Zheng},
        title = "{SDSS-IV MaNGA: the spatially resolved stellar initial mass function in {\ensuremath{\sim}}400 early-type galaxies}",
      journal = {\mnras},
         year = 2018,
        month = jul,
       volume = {477},
       number = {3},
        pages = {3954-3982},
          doi = {10.1093/mnras/sty785},
archivePrefix = {arXiv},
       eprint = {1803.08515},
 primaryClass = {astro-ph.GA},
       adsurl = {https://ui.adsabs.harvard.edu/abs/2018MNRAS.477.3954P}
}

@ARTICLE{Zhou2019MNRAS.485.5256Z,
       author = {{Zhou}, Shuang and {Mo}, H.~J. and {Li}, Cheng and {Zheng}, Zheng and {Li}, Niu and {Du}, Cheng and {Mao}, Shude and {Parikh}, Taniya and {Lane}, Richard R. and {Thomas}, Daniel},
        title = "{SDSS-IV MaNGA: stellar initial mass function variation inferred from Bayesian analysis of the integral field spectroscopy of early-type galaxies}",
      journal = {\mnras},
         year = 2019,
        month = jun,
       volume = {485},
       number = {4},
        pages = {5256-5275},
          doi = {10.1093/mnras/stz764},
archivePrefix = {arXiv},
       eprint = {1811.09799},
 primaryClass = {astro-ph.GA},
       adsurl = {https://ui.adsabs.harvard.edu/abs/2019MNRAS.485.5256Z}
}

@ARTICLE{Li2018ApJ...859...36L,
       author = {{Li}, Zhongmu and {Mao}, Caiyan},
        title = "{Evolution of Optical Binary Fraction in Sparse Stellar Systems}",
      journal = {\apj},
         year = 2018,
        month = may,
       volume = {859},
       number = {1},
          eid = {36},
        pages = {36},
          doi = {10.3847/1538-4357/aabc09},
       adsurl = {https://ui.adsabs.harvard.edu/abs/2018ApJ...859...36L}
}

@ARTICLE{Liu2012MNRAS.425.2144L,
       author = {{Liu}, Chao and {van de Ven}, Glenn},
        title = "{Chemo-orbital evidence from SDSS/SEGUE G-type dwarf stars for a mixed origin of the Milky Way's thick disc}",
      journal = {\mnras},
         year = 2012,
        month = sep,
       volume = {425},
       number = {3},
        pages = {2144-2156},
          doi = {10.1111/j.1365-2966.2012.21551.x},
archivePrefix = {arXiv},
       eprint = {1201.1635},
 primaryClass = {astro-ph.GA},
       adsurl = {https://ui.adsabs.harvard.edu/abs/2012MNRAS.425.2144L}
}

@ARTICLE{Hwang2022MNRAS.513..754H,
       author = {{Hwang}, Hsiang-Chih and {Ting}, Yuan-Sen and {Conroy}, Charlie and {Zakamska}, Nadia L. and {El-Badry}, Kareem and {Cargile}, Phillip and {Zaritsky}, Dennis and {Chandra}, Vedant and {Han}, Jiwon Jesse and {Speagle}, Joshua S. and {Bonaca}, Ana},
        title = "{Wide binaries from the H3 survey: the thick disc and halo have similar wide binary fractions}",
      journal = {\mnras},
         year = 2022,
        month = jun,
       volume = {513},
       number = {1},
        pages = {754-767},
          doi = {10.1093/mnras/stac650},
archivePrefix = {arXiv},
       eprint = {2111.01788},
 primaryClass = {astro-ph.GA},
       adsurl = {https://ui.adsabs.harvard.edu/abs/2022MNRAS.513..754H}
}

@ARTICLE{Sellwood2002MNRAS.336..785S,
       author = {{Sellwood}, J.~A. and {Binney}, J.~J.},
        title = "{Radial mixing in galactic discs}",
      journal = {\mnras},
         year = 2002,
        month = nov,
       volume = {336},
       number = {3},
        pages = {785-796},
          doi = {10.1046/j.1365-8711.2002.05806.x},
archivePrefix = {arXiv},
       eprint = {astro-ph/0203510},
 primaryClass = {astro-ph},
       adsurl = {https://ui.adsabs.harvard.edu/abs/2002MNRAS.336..785S}
}

@ARTICLE{Li2023ChPhB..32c9801L,
       author = {{Li}, Niu and {Li}, Cheng},
        title = "{Measuring stellar populations, dust attenuation and ionized gas at kpc scales in 10010 nearby galaxies using the integral field spectroscopy from MaNGA}",
      journal = {Chinese Physics B},
         year = 2023,
        month = mar,
       volume = {32},
       number = {3},
          eid = {039801},
        pages = {039801},
          doi = {10.1088/1674-1056/acb0ba},
       adsurl = {https://ui.adsabs.harvard.edu/abs/2023ChPhB..32c9801L}
}

@ARTICLE{Phillipps2020MNRAS.492.2128P,
       author = {{Phillipps}, S. and {Ali}, S.~S. and {Bremer}, M.~N. and {De Propris}, R. and {Sansom}, A.~E. and {Cluver}, M.~E. and {Alpaslan}, M. and {Brough}, S. and {Brown}, M.~J.~I. and {Davies}, L.~J.~M. and {Driver}, S.~P. and {Grootes}, M.~W. and {Holwerda}, B.~W. and {Hopkins}, A.~M. and {James}, P.~A. and {Pimbblet}, K. and {Robotham}, A.~S.~G. and {Taylor}, E.~N. and {Wang}, L.},
        title = "{Galaxy And Mass Assembly (GAMA): Defining passive galaxy samples and searching for the UV upturn}",
      journal = {\mnras},
         year = 2020,
        month = feb,
       volume = {492},
       number = {2},
        pages = {2128-2139},
          doi = {10.1093/mnras/stz3552},
archivePrefix = {arXiv},
       eprint = {2001.02465},
 primaryClass = {astro-ph.GA},
       adsurl = {https://ui.adsabs.harvard.edu/abs/2020MNRAS.492.2128P}
}

@ARTICLE{Smith2012MNRAS.421.2982S,
       author = {{Smith}, Russell J. and {Lucey}, John R. and {Carter}, David},
        title = "{What drives the ultraviolet colours of passive galaxies?}",
      journal = {\mnras},
         year = 2012,
        month = apr,
       volume = {421},
       number = {4},
        pages = {2982-2997},
          doi = {10.1111/j.1365-2966.2012.20524.x},
archivePrefix = {arXiv},
       eprint = {1201.1907},
 primaryClass = {astro-ph.CO},
       adsurl = {https://ui.adsabs.harvard.edu/abs/2012MNRAS.421.2982S}
}

@ARTICLE{Harris1996AJ....112.1487H,
       author = {{Harris}, William E.},
        title = "{A Catalog of Parameters for Globular Clusters in the Milky Way}",
      journal = {\aj},
         year = 1996,
        month = oct,
       volume = {112},
        pages = {1487},
          doi = {10.1086/118116},
       adsurl = {https://ui.adsabs.harvard.edu/abs/1996AJ....112.1487H}
}

@INPROCEEDINGS{Zhang2020IAUS..341...35Z,
       author = {{Zhang}, F. and {Han}, Z. and {Li}, L.},
        title = "{Evolutionary Population Synthesis model with binary stars - Yunnan-II model}",
    booktitle = {Panchromatic Modelling with Next Generation Facilities},
         year = 2020,
       editor = {{Boquien}, M{\'e}d{\'e}ric and {Lusso}, Elisabeta and {Gruppioni}, Carlotta and {Tissera}, Patricia},
       series = {IAU Symposium},
       volume = {341},
        month = jan,
        pages = {35-38},
          doi = {10.1017/S174392131900262X},
       adsurl = {https://ui.adsabs.harvard.edu/abs/2020IAUS..341...35Z}
}

@ARTICLE{Pacifici2023ApJ...944..141P,
       author = {{Pacifici}, Camilla and {Iyer}, Kartheik G. and {Mobasher}, Bahram and {da Cunha}, Elisabete and {Acquaviva}, Viviana and {Burgarella}, Denis and {Calistro Rivera}, Gabriela and {Carnall}, Adam C. and {Chang}, Yu-Yen and {Chartab}, Nima and {Cooke}, Kevin C. and {Fairhurst}, Ciaran and {Kartaltepe}, Jeyhan and {Leja}, Joel and {Ma{\l}ek}, Katarzyna and {Salmon}, Brett and {Torelli}, Marianna and {Vidal-Garc{\'\i}a}, Alba and {Boquien}, M{\'e}d{\'e}ric and {Brammer}, Gabriel G. and {Brown}, Michael J.~I. and {Capak}, Peter L. and {Chevallard}, Jacopo and {Circosta}, Chiara and {Croton}, Darren and {Davidzon}, Iary and {Dickinson}, Mark and {Duncan}, Kenneth J. and {Faber}, Sandra M. and {Ferguson}, Harry C. and {Fontana}, Adriano and {Guo}, Yicheng and {Haeussler}, Boris and {Hemmati}, Shoubaneh and {Jafariyazani}, Marziye and {Kassin}, Susan A. and {Larson}, Rebecca L. and {Lee}, Bomee and {Mantha}, Kameswara Bharadwaj and {Marchi}, Francesca and {Nayyeri}, Hooshang and {Newman}, Jeffrey A. and {Pandya}, Viraj and {Pforr}, Janine and {Reddy}, Naveen and {Sanders}, Ryan and {Shah}, Ekta and {Shahidi}, Abtin and {Stevans}, Matthew L. and {Triani}, Dian Puspita and {Tyler}, Krystal D. and {Vanderhoof}, Brittany N. and {de la Vega}, Alexander and {Wang}, Weichen and {Weston}, Madalyn E.},
        title = "{The Art of Measuring Physical Parameters in Galaxies: A Critical Assessment of Spectral Energy Distribution Fitting Techniques}",
      journal = {\apj},
         year = 2023,
        month = feb,
       volume = {944},
       number = {2},
          eid = {141},
        pages = {141},
          doi = {10.3847/1538-4357/acacff},
archivePrefix = {arXiv},
       eprint = {2212.01915},
 primaryClass = {astro-ph.GA},
       adsurl = {https://ui.adsabs.harvard.edu/abs/2023ApJ...944..141P}
}

@ARTICLE{Jones2025MNRAS.543..167J,
       author = {{Jones}, Gareth T. and {Byrne}, Conor M. and {Stanway}, Elizabeth R.},
        title = "{Impact of uncertainties in spectral energy distribution modelling on inferred galaxy properties}",
      journal = {\mnras},
         year = 2025,
        month = oct,
       volume = {543},
       number = {1},
        pages = {167-189},
          doi = {10.1093/mnras/staf1462},
archivePrefix = {arXiv},
       eprint = {2509.02741},
 primaryClass = {astro-ph.GA},
       adsurl = {https://ui.adsabs.harvard.edu/abs/2025MNRAS.543..167J}
}
\bibliographystyle{aasjournal}



\end{document}